\documentclass{ws-ijmpd}
\usepackage{latexsym,revsymb}
\usepackage{amsmath,amssymb,amsfonts}
\usepackage[super,compress]{cite}
\usepackage{hyperref}
\usepackage{color}

\newcommand{\GN}{G}
\newcommand{\ac}{\bar{a}_0}

\begin{document}

\markboth{Edmonds, Farrah, Minic, Ng, Takeuchi}
{MDM: Relating Dark Energy, Dark Matter and Baryonic Matter}

%
\catchline{}{}{}{}{}
%

\title{Modified Dark Matter: Relating Dark Energy, Dark Matter and Baryonic Matter}

\author{Douglas Edmonds${}^{1*}$, ${}^{2}$,
Duncan Farrah${}^{2\dagger}$, 
Djordje Minic${}^{2\ddagger}$, 
Y.\ Jack Ng${}^{3\S}$, and
Tatsu Takeuchi${}^{2\P}$}
\address{%
${}^{1}$Department of Physics, The Pennsylvania State University, Hazleton, PA 18202 USA,\\
${}^{2}$Department of Physics, Center for Neutrino Physics, Virginia Tech,\\Blacksburg, VA 24061 USA,\\
${}^{3}$Institute of Field Physics, Department of Physics and Astronomy,\\
University of North Carolina, Chapel Hill, NC 27599 USA\\
${}^{*}$bde12@psu.edu, 
${}^{\dagger}$farrah@vt.edu, 
${}^{\ddagger}$dminic@vt.edu, 
${}^{\S}$yjng@physics.unc.edu, 
${}^{\P}$takeuchi@vt.edu
}

\maketitle

\begin{history}
\received{Day Month Year}
\revised{Day Month Year}
\end{history}

\begin{abstract}
Modified dark matter (MDM) is a phenomenological model of dark matter, inspired by gravitational thermodynamics. For an accelerating Universe with positive cosmological constant ($\Lambda$), such phenomenological considerations lead to the emergence of a 
critical acceleration parameter related to $\Lambda$.  Such a critical acceleration is
an effective phenomenological manifestation of MDM, and it is found in  correlations between dark matter and baryonic matter in galaxy rotation curves. The resulting MDM mass profiles, which are sensitive to $\Lambda$, are consistent with observational data at both the galactic and cluster scales. In particular, the same critical acceleration appears both in the galactic and cluster data fits based on MDM. Furthermore, using some robust qualitative arguments, MDM appears to work well on cosmological scales, even though quantitative studies are still lacking. Finally, we comment on certain non-local aspects of the quanta of modified dark matter,
which may lead to novel non-particle phenomenology and which may explain why, so far, 
dark matter detection experiments have failed to detect dark matter particles.
\end{abstract}


\ccode{PACS numbers:}

\section{Introduction and Overview}

The `missing mass' problem is one of the fundamental puzzles in contemporary physics and astronomy.\cite{Bertone:2016nfn}  Since the pioneering work of Oort \cite{Oort}, Zwicky \cite{Zwicky:1933gu}, and Rubin and Ford \cite{Rubin:1970zza,Rubin:1978kmz,Rubin:1980zd}, observational evidence for substantial mass discrepancies between dynamical studies and observations of visible (baryonic) matter have become overwhelming on all scales spanning from the galactic to the cosmological. 

From the perspective of Einstein's equations
\begin{equation}
G_{ab} + \Lambda\,g_{ab}\;=\; 8\pi\GN\, T_{ab} \;,
\label{EinsteinEquation}
\end{equation}
where $G_{ab}$ is the Einstein tensor, $T_{ab}$ is the energy-momentum tensor, and $\Lambda$ is the cosmological constant, the problem amounts to a mismatch between the left- and right-hand sides. The curvature of spacetime (on the left-hand side) which determines the dynamics is larger than what is expected from contributions of baryonic matter to the energy-momentum tensor on the right-hand side.  Note that we will be treating the dark energy as the cosmological constant or vacuum energy \cite{Peebles:2002gy} in this review, which meshes well with our view of 
the modified dark matter, that aims to relate the dark matter, dark energy and baryonic matter sectors of Einstein's equations.

This mismatch can be alleviated by the modification of, or addition of extra terms to, either the left- or the right-hand sides of the equation. The problem with modifications to the geometric/gravitational (left-hand) side\cite{Bekenstein:2004ne,Mannheim:2012qw,Moffat:2013,Moffat:2014} is that they are difficult to motivate from general physical principles as elegantly as the original formulation of general relativity (GR) by Einstein. Moreover, such modifications are necessarily classical, and thus their quantum nature is obscure and almost certainly more opaque than the quantum nature of Einstein's gravity, which itself remains controversial.

Attention has thus focused mainly on modifications to the source (right-hand) side of Eq.~(\ref{EinsteinEquation}), the simplest of which is to add contributions to the energy-momentum tensor from heretofore unknown and unobserved degrees of freedom, \textit{i.e.} Dark Matter (DM). An obvious candidate for DM is baryons that are not easily detectable by observations of photons, either in absorption or emission. This led to Massive Compact Halo Object (MACHO) models, in which the DM consists of brown dwarfs, neutron stars, black holes and/or other collapsed objects. However, exhaustive searches for microlensing events that would signify the presence of such objects in our Milky Way's halo turned up far too few events to make MACHOs a significant source of DM\cite{alco96,alco97,Alcock:2000ph,calchi05,tisser07}. 

Instead, observational evidence favors that DM is largely non-baryonic in nature. In particular, comparisons of the observed deuterium to hydrogen ratio to that expected from Big Bang nucleosynthesis shows that the bulk of matter cannot be baryonic\cite{cyburt04}. Observations of the power spectrum of anisotropies in the cosmic microwave background (CMB) are consistent with at least most DM being non-relativistic (\textit{i.e.} cold) 
and diffuse\cite{smoot92,debernar00,spergel07,komat11,planck16}. The evolution of large-scale structures (LSS) of galaxies across the history of the Universe are also consistent with 
this idea\cite{davis85,whfr91,cole05,eisen05,sprin05}. Other evidence includes the observations of colliding galaxy clusters, which are straightforward to explain with non-baryonic cold DM\cite{clowe06} but convoluted to explain in competing models\cite{angus06,lililin13}. 

This paradigm of a cold, diffuse, non-relativistic DM is known as the Cold Dark Matter (CDM) paradigm 
\cite{Peebles:1982ff, Bond:1982uy, Blumenthal:1982mv, Blumenthal:1984bp}. 
It is however notable that CDM is not particularly restrictive with models that fit this framework; all that is required is a Weakly Interacting Massive Particle (WIMP) in which extra and independent (from the baryonic matter) degrees of freedom are described by new weakly-interacting quantum fields \cite{Jungman:1995df}, but in which (at least from the astrophysical perspective) the mass and interaction cross section of individual WIMPs are barely constrained\footnote{In this review we are not going to concentrate on alternative dark matter scenarios that include warm dark matter models \cite{Bode:2000gq}, axions \cite{Asztalos:2009yp}, and hidden sectors \cite{Strassler:2006im}.}. Note that the standard cosmological model, $\Lambda CDM$, assumes that the dark energy sector is modeled with the cosmological constant (i.e. vacuum energy) and this will remain the case in our discussion of the MDM proposal.

The popularity of the WIMP models is in part due to the possibility of directly detecting the WIMPs using laboratory-based experiments. Starting in the 1980s, increasingly sensitive searches have been made for 
direct signatures of DM particles \cite{Goodman:1984dc}. 
However, no direct detections have been made, and even the latest results of experiements searching for recoil events \cite{Klasen:2015uma} only set limits on the mass of the assumed DM particle and its cross section with baryonic matter.

In the absence of any direct detection of DM, it is important to look for other constraints that we should place on the nature and properties of DM in order to narrow down the list of possibilities. For this, we note that hints may be discernible in the tensions between observations and CDM models. First, there is a set of `problems' between $N$-body simulations of galaxy/cluster evolution and observations:

\begin{itemize}
\item \textbf{Missing Satellite Problem:}\\
Simulations predict a much larger number of satellite galaxies in CDM haloes than is typically observed \cite{Moore:1998zn,Moore:1999jv,Klypin:1999uc,Tollerud:2008ze,Springel:2008cc}. Also, there are tensions between the dispersion in mass predicted by simulations, and the dispersion that is observed \cite{Kravtsov:2004cm,Geha:2008zr}. These discrepancies can be at least partly resolved by suppression of dwarf galaxy formation via the UVB (ultraviolet background) heating of the IGM (intergalactic medium) gas (e.g. Refs.~\citen{Crain:2006sb,Tassis:2006zt}) or supernova feedback \cite{wadepuh11,brooks13}. Moreover, recent simulations at high spatial resolution do reproduce the observed Milky Way satellite number \cite{Maccio:2009aek,Wetzel:2016wro}. 

\item \textbf{Core/Cusp Problem:}\\
Simulations predict that the central CDM distribution in galaxy clusters should be sharply peaked, but observations instead favor a much flatter central density profile \cite{Dubinski:1991bm,Governato:2009bg,Walker:2011zu,Oh:2015xoa,Navarro:2016bfs}.  

\item \textbf{Too-Big-to-Fail Problem:}\\
Observations infer that luminous dwarf galaxies inhabit lower mass CDM halos than those predicted by simulations \cite{Kroupa:2010hf,BoylanKolchin:2011de,BoylanKolchin:2011dk,Tollerud:2014zha,Papastergis:2014aba}. 

\item \textbf{Satellite Planes Problem:}\\
Observations find that satellite galaxies are distributed much more anisotropically than is predicted by simulations \cite{Kroupa:2004pt,Metz:2006zc,Ibata:2014pja,Gillet:2015}.

\end{itemize}

\noindent Summaries of these observational tensions can be found in e.g. Refs.~\citen{Kroupa:2012qj,Famaey:2013ty,McGaugh:2014nsa,Weinberg:2013aya,Walker:2014isa,Pawlowski:2015qta,Kroupa:2014ria,Schaye:2014tpa}.

These problems, however, may not pose insurmountable challenges for CDM. For example, some studies suggest that all of these problems are due to the limitations of our current simulation capabilities, not to our limited understanding of the nature of CDM, and that they can be (at least partly) overcome by including a sufficiently complete suite of baryonic physical processes in simulations \cite{Koposov:2009ru,Pontzen:2011ty,Governato:2012fa,Ogiya:2012jq,Garrison-Kimmel:2017zes}. Moreover, modifications to cold dark matter which introduce relativistic degrees of freedom, such as warm dark matter (sterile neutrinos), or other popular relativistic degrees of freedom, such as axions, do solve some of these problems\cite{Elbert:2014bma}.

The second, and in our view more serious, tension is the set of observations at the galactic scale which strongly suggests that DM may not simply be extra, independent degrees of freedom. These observations include:
\begin{enumerate}
\item \textbf{Presence of a Universal Acceleration Scale:}\\
A recent work by McGaugh and collaborators claims a precise correlation between the mass profiles of dark and baryonic matter in disk galaxies spanning an extremely wide ranges in scale, mass, and age\cite{Milgrom:2007br,McGaugh:2016leg,Lelli:2017vgz}. This correlation is expressed as a relation between the observed acceleration $a_\mathrm{obs}$ and 
the expected acceleration $a_\mathrm{bar}$ from baryonic matter only as
\begin{equation}
a_\mathrm{obs} \;=\; 
\dfrac{a_\mathrm{bar}}{1-e^{-\sqrt{a_\mathrm{bar}/\ac}}},
\label{McGaughRelation}
\end{equation}
where $\ac$ is a universal constant which has been fit to $\ac = (1.20\pm 0.02)\times 10^{-10}\mathrm{m/s^2}$.
%
(Here we have written $\ac$ rather than $a_0\approx cH_o$, which will be introduced later.
Note that $\ac = a_0/2\pi$, in parallel with $\hbar = h/2\pi$ and $\lambdabar=\lambda/2\pi$.
)
Given that the Hubble parameter is
$H_0  =  (67.74\pm 0.46)\,\mathrm{km/s/Mpc}$ (see Table 4 of Ref.~\citen{Ade:2015xua})\footnote{%
That is, $H_0  =  \left[(6.581\pm 0.045)\times 10^{-8}\mathrm{cm/s^2}\right]/c$.
} 
it has been noted that\cite{Milgrom:1998sy}
\begin{equation}
\ac \;\approx\; \dfrac{cH_0}{2\pi}\;,
\label{coincidence}
\end{equation}
which suggests that the constant $a_0$ may be cosmological in origin.
\\

Notice that the above correlation (Eq.2) is difficult to motivate within purely collision-less CDM, since some dispersion might be expected between the collision-less CDM particles, and the free baryons that act as fuel for the star formation. Nevertheless, it may be possible to reconcile this result with CDM models\cite{Ludlow:2016qzh,Navarro:2016bfs} by relying on dissipative baryonic dynamics.
On the other hand, this correlation could be pointing to a yet unknown property of DM which
connects its distribution to that of baryonic matter.
Perhaps the more mysterious part of this relation is the appearance of the universial acceleration constant $a_0$.
If this acceleration $a_0$  is cosmological in origin and related to $H_0$, how can the DM mass profile be sensitive to it?
\\

\item \textbf{Baryonic Tully-Fisher Relation:}\\
The Tully-Fisher Relation\cite{Tully:1977fu} is a universal relation between the total observed baryonic mass (stars + gas) of a galaxy $M_\mathrm{bar,total}$ and the asymptote of the galactic rotation curve $v_\infty$:
\begin{equation}
M_\mathrm{bar,total} \;=\; Av_\infty^4\;,\qquad
A \;=\; (47\pm 6) \,M_\odot\,\mathrm{s^4/km^4}\;.
\label{TullyFisher}
\end{equation}
This relation holds regardless of the value of $M_\mathrm{bar,total}$, or how it is distributed.
Even when the shapes of the rotation curves are different, the asymptote is always the same for galaxies with the
same baryonic mass.
This is remarkable when one recognizes that the rotation velocity $v(r)$ 
at a distance $r$ from the center of the galaxy is determined by the distribution of the
sum of baryonic and dark matter, and not by baryonic matter alone.
Nevertheless, the asymptotic velocity depends only on the total baryonic mass,
again suggesting a correlation between the baryonic and DM mass distributions.
\\


\end{enumerate}

While it is not clear that current DM models  naturally explain these relations, 
they are natural consequences of Milgrom's
MOdified Newtonican Dynamics (MOND).\cite{Milgrom:1983ca,Milgrom:1983pn,Milgrom:1983zz} (Note that one of the aims of our work is
to provide a DM model that explains the baryonic Tully-Fisher relation and accounts for the accelation scale $a_0$.)

In MOND, it is postulated that Newton's equation of motion $F=ma$ should be modified to
\begin{equation}
F \;=\; 
\begin{cases}
ma            & (a \gg \ac) \\
ma^2/\ac  & (a \ll \ac)
\end{cases}\;.
\end{equation}
More specifically,
\begin{equation}
F \;=\; ma\,\mu(a/\ac)\;,
\label{MOND-EQM}
\end{equation}
where $\mu(x) = 1$ for $x \gg 1$ and $\mu(x) = x$ for $x \ll 1$.
The choice of interpolating functions $\mu(x)$ is arbitrary.
This implies
\begin{equation}
a_\mathrm{obs} \;=\; 
\begin{cases}
a_\mathrm{bar} & (a_\mathrm{bar} \gg \ac) \\
\sqrt{\ac a_\mathrm{bar}} & (a_\mathrm{bar} \ll \ac)
\end{cases}\;,
\end{equation}
\textit{i.e.} the same relation implied by Eq.~(\ref{McGaughRelation}).
Far away from the galactic center, we can expect the following baryonic acceleration
\begin{equation}
a_\mathrm{bar}(r) \;=\; \dfrac{\GN M_\mathrm{bar,total}}{r^2}\;,
\end{equation}
and thus
\begin{equation}
v^2(r) \;=\; r\,a_\mathrm{obs}(r) \;\xrightarrow{r\rightarrow\infty}\; r\sqrt{\ac a_\mathrm{bar}(r)}
\;=\; \sqrt{\ac\GN M_\mathrm{bar,total}} 
\;\equiv\;v_\infty^2
\;,
\end{equation}
which gives us flat rotation curves\footnote{%
In reality, rotation curves are not all flat; they display a variety of 
properties. See, e.g. Ref.~\citen{Persic:1991}.
}
and
\begin{equation}
M_\mathrm{bar,total} \;=\;
\dfrac{v_\infty^4}{\ac\GN}
\;=\; (63\,M_\odot\,\mathrm{s^4/km^4})\,v_\infty^4
\;,
\end{equation}
cf. Eq.~(\ref{TullyFisher}).
Other studies of MOND\footnote{There are also the relativistic versions
AQUAL, RAQUEL and TeVeS; but they tend to be more limited in their
predictive power.  See Ref.~\citen{Famaey:2011kh} and references therein.} 
in the context of rotation curves include
Ref.~\citen{Sanders:1996ua,Sanders:1998gr}.


We note, however, that MOND can also be interpreted as the introduction of a very specific type of DM.
Consider a spherically symmetric distribution of baryonic matter where
$M_\mathrm{bar}(r)$ is the total baryonic mass enclosed in a sphere of radius $r$.
Then, the gravitational force on a test mass $m$ placed at $r$ due to this distribution will be given by
\begin{equation}
F(r) \;=\; \dfrac{\GN M_\mathrm{bar}(r)m}{r^2}\;.
\end{equation}
Eq.~(\ref{MOND-EQM}) in this case can be rewritten as:
\begin{equation}
a(r)
\;=\;
\dfrac{1}{\mu(a(r)/\ac)}\dfrac{\GN M_\mathrm{bar}(r)}{r^2}
\;\equiv\;
\dfrac{\GN[M_\mathrm{bar}(r)+M_\mathrm{DM}(r)]}{r^2}
\;,
\end{equation}
where we identify
\begin{equation}
M_\mathrm{DM}(r) \;=\; \left[\dfrac{1}{\mu(a(r)/\ac)}-1\right]M_\mathrm{bar}(r)\;,
\end{equation}
as the total DM mass within a radius of $r$ form the center.
Thus, to reproduce the success of MOND at galactic scales, we need a DM model which 
predicts such a mass distribution. (Note that such a dark matter model is {\it not} going to be an
inversion of Milgrom's MOND, {\it i.e.} it is {\it not} going to be a  ``phantom'' dark matter, because it will
have to work on all scales: galactic, cluster and cosmological.)
We still have to account for the impressive successes of the canonical
$\Lambda CDM$ model on cluster and galactic scales.

However, there are problems with MOND at the cluster and cosmological scales.  
\footnote{The reason may be due to
the lack of a fundamental relativisitic quantum theory of MOND.}
In general MOND fails to address the dynamics of galactic clusters 
(more later in Section 3.2) and other cosmological measurements, 
in particular, it cannot explain the third
and higher CMB peaks, and the shape of matter power spectrum.
Nevertheless, the fundamental acceleration parameter, also known as Milgrom's scaling, appears to be both real and a potentially important signpost towards a deeper understanding of the dark sector. CDM, for example, can only reproduce Milgrom's scaling through a somewhat convoluted argument of Kaplinghat and Turner \cite{Kaplinghat:2001me}. The fundamental meaning (if there is any) of Milgrom's scaling is obscure in this argument.\footnote{We should note that MOND has been
formulated in a relativistic context in Ref.~\citen{Bekenstein:2004ne} and even argued to be a consequence of quantum gravity in Ref.~\citen{Smolin:2017kkb}.} \\


Thus the question regarding the relation between this fundamental acceleration parameter $a_0$ and dark matter is still outstanding. The nature of this question motivated us to examine a new model for non-baryonic dark matter, which we term modified (or ``Mondian'') dark matter, or simply, MDM. 
The idea here is that by taking into account the existence of the fundamental acceleration as well as of the baryonic Tully-Fisher 
relation, without modifying the Einstein equations, and thus Newtonian dynamics in the non-relativistic regimes, and by combining it with the non-baryonic dark matter paradigm, we should be able to sharpen the CDM proposal, and point towards a more focused origin of dark matter quanta.
(At the moment, the nature of dark matter quanta is not constrained at all, and these can span enormous energy scales.)

The defining feature of the MDM proposal is that the modified dark matter profile should be sensitive to the fundamental acceleration $a_0$, or alternatively, to the cosmological constant, at all scales (galactic, cluster and cosmological) and
that on galactic scales the modified dark matter mass profile should be correlated to the baryonic mass profile. The question really is, whether we might be able to modify the energy momentum tensor in such a way so that this modification depends both on the original baryonic source, and on the inertial properties, such as the acceleration, associated with the geometric side of Einstein's equation.

The canonical formulation of General Relativity is ignorant of any modified inertial properties like the ones suggested by MOND. Moreover, the effective field theory (which is used to model dark matter particles, as in WIMP models) does not know about inertial properties at all.
So, whatever one does to implement the dependence of the dark matter profile on some fundamental acceleration, it has to go beyond the classical Einstein theory and the usual local effective quantum field theory without violating these two
pillars of modern physics in their domains of validity.

One way to do so is by appealing to quantum gravity, which should reproduce Einstein's gravity and
the energy momentum tensor of the sources described by effective field theory. We will comment on the role of quantum gravity in the conclusion, when we talk about the recent new formulation of string theory and quantum gravity in terms of metastring theory.

Regardless of the lofty origins of MDM, our working proposal is much simpler and more practical: we propose to look at the thermodynamic reformulation of Einstein's theory of gravity and search for a consistent modification of the energy momentum tensor in that thermodynamics approach, so that the fundamental acceleration is included ab initio. The reason for this is that gravitational thermodynamics (the prototype of which is black hole thermodynamics) is the only sure place where quantum theory and physics in accelerating frames are precisely related. Thus, whatever quantum gravity is, it should be consistent with gravitational thermodynamics, and the proposal for modified dark matter, MDM, should be based in robust 
features of gravitational thermodynamics which include the sensitivity of the dark matter profile on the fundamental acceleration.


The main result of our investigation can be summarized in the following formula for the mass profile of non-baryonic dark matter,
 which relates the mass of the dark matter ($M'$) with the mass of the baryonic matter ($M$) via an acceleration parameter $a_0$;
\begin{eqnarray}
\label{clustermasspintro}
\dfrac{M'}{M}
& = & \dfrac{\alpha}{\left[\,1+ 
\left(r/r_\textrm{MDM}\right)\,\right]}\left[\dfrac{a_0^2}{(a_\mathrm{obs}+a_0)^2 - a_0^2}\right]
\;,
\end{eqnarray}
where $a_{obs}$ is the observed acceleration, $r$ the radial distance, $r_{MDM}$ is a dark matter distance scale, and $\alpha$ is constant factor that is of order 1 for galaxies and 100 for galaxy clusters.\footnote{The value of $\alpha$ for galaxy clusters is currently not well-constrained. Values between $\sim50 - 100$ fit the data in our sample well. In this paper, we use $\alpha=50$ for our galaxy cluster fits.}
Note that for the case of galaxies $r/r_{MDM} \to 0$, and then (see Eq.~\ref{Mprime} in \S \ref{mdmentropic}) the mass profile reduces simply to
\begin{eqnarray}
\dfrac{M'}{M}
& = & 
\dfrac{1}{2}\left[\dfrac{a_0^2}{(a_\mathrm{obs}+a_0)^2 - a_0^2}\right]
\;.
\end{eqnarray}
Thus this profile works both on galactic and cluster scales, and how well it works can be pictorially represented in Fig. 1 and Fig. 2.

Fig. 1 shows a very tight correlation between baryonic matter and dark matter in galaxies. This figure is similar to the one presented in Ref.~\citen{McGaugh:2016leg}, but for a different data set. We use galactic rotation curve data from the sample of Ursa Major galaxies represented in Fig. 3. The data is fit with a modified version of Eq.~\ref{McGaughRelation}:
\begin{equation}
a_\mathrm{obs} \;=\; 
\dfrac{a_\mathrm{bar}}{1-e^{-\sqrt{a_\mathrm{bar}/z\ac}}},
\label{modified_McGaughRelation}
\end{equation}
where
\begin{equation}
z= \dfrac{\alpha}{\left[\,1+ \left(r/r_\textrm{MDM}\right)\,\right]}
\end{equation}
is the prefactor in Eq.~\ref{clustermasspintro} appropriate for galaxies. Of course, numerically, this is the same formula used in Ref.~\citen{McGaugh:2016leg}. However, inclusion of $z$ allows for consistency when we go to the galaxy cluster scale, where Eq.~\ref{McGaughRelation} does not fit the data well.

Fig. 2 shows a correlation beween baryonic matter and dark matter in the sample of thirteen galaxy clusters presented in \S\ref{clusters_subsection}. The black squares represent the data fitting functions developed in Ref.~\citen{Vikhlinin:2005mp}. Our fitting function for galaxy clusters has the same form as the function used for galaxies, Eq.~\ref{modified_McGaughRelation}, with $z$ appropriate for the galaxy cluster scale. We plot the function for two values of the acceleration scale: For the dashed red line, we use $\ac$, and for the solid red line, we use $a_0$. Note that we use the same scale distance $r_{MDM}$ for all galaxy clusters in this plot, while in the fits presented later, this scale is allowed to vary for different clusters. Using a single value increases the scatter in the data.

\begin{figure*}
  \includegraphics[angle=0,width=1.0\textwidth]{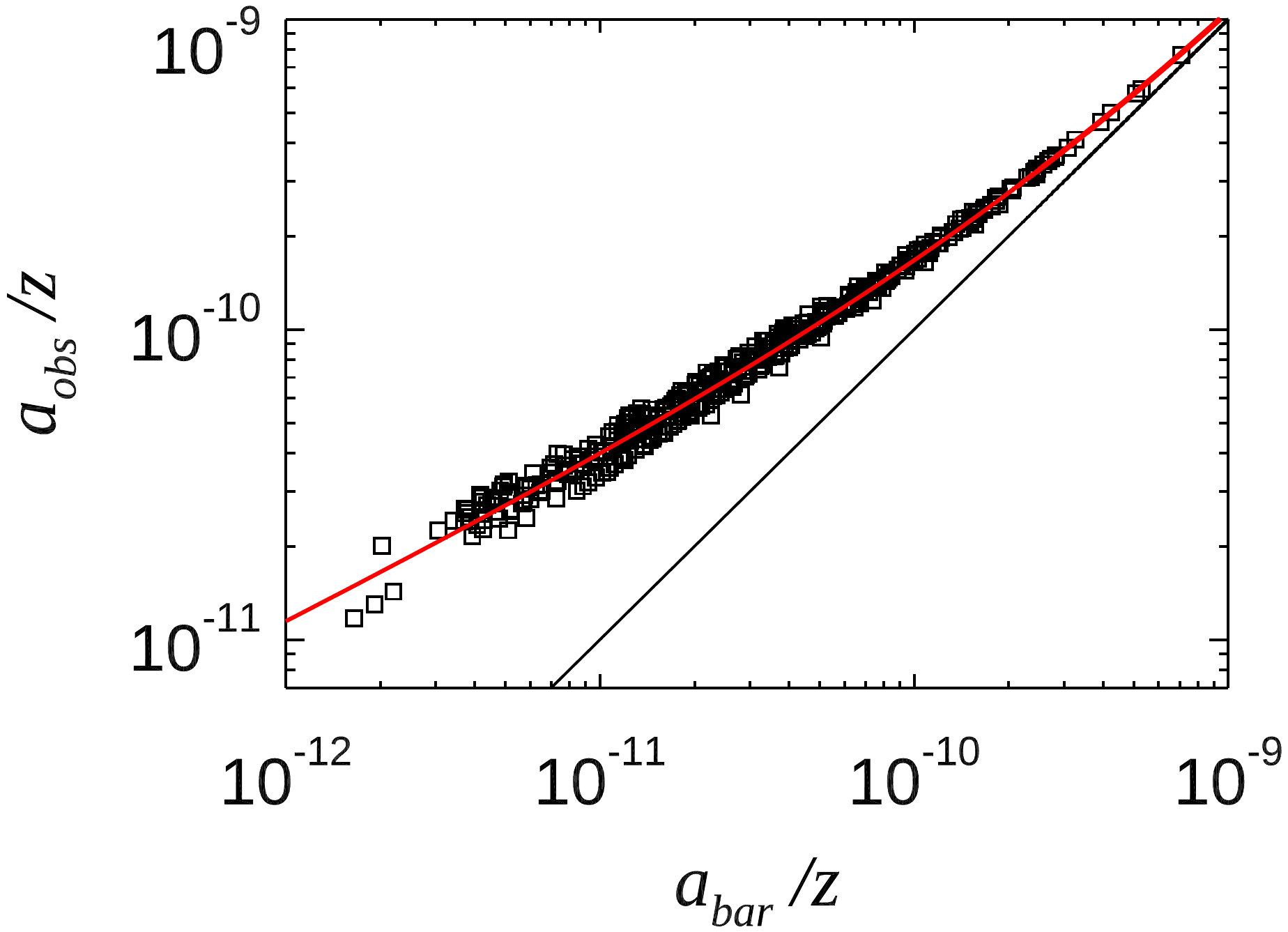}
  \caption{Comparison of observed accelerations and accelerations expected from baryons in galaxies. The black squares are 386 data points from a sample of 30 galaxies presented in this paper. The black line is what we expect from Newtonian physics and no dark matter. The red line is the prediction of Eq.~\ref{modified_McGaughRelation}.}
  \label{fig_compare_g_galaxies}
\end{figure*}

\begin{figure*}
  \includegraphics[angle=0,width=1.0\textwidth]{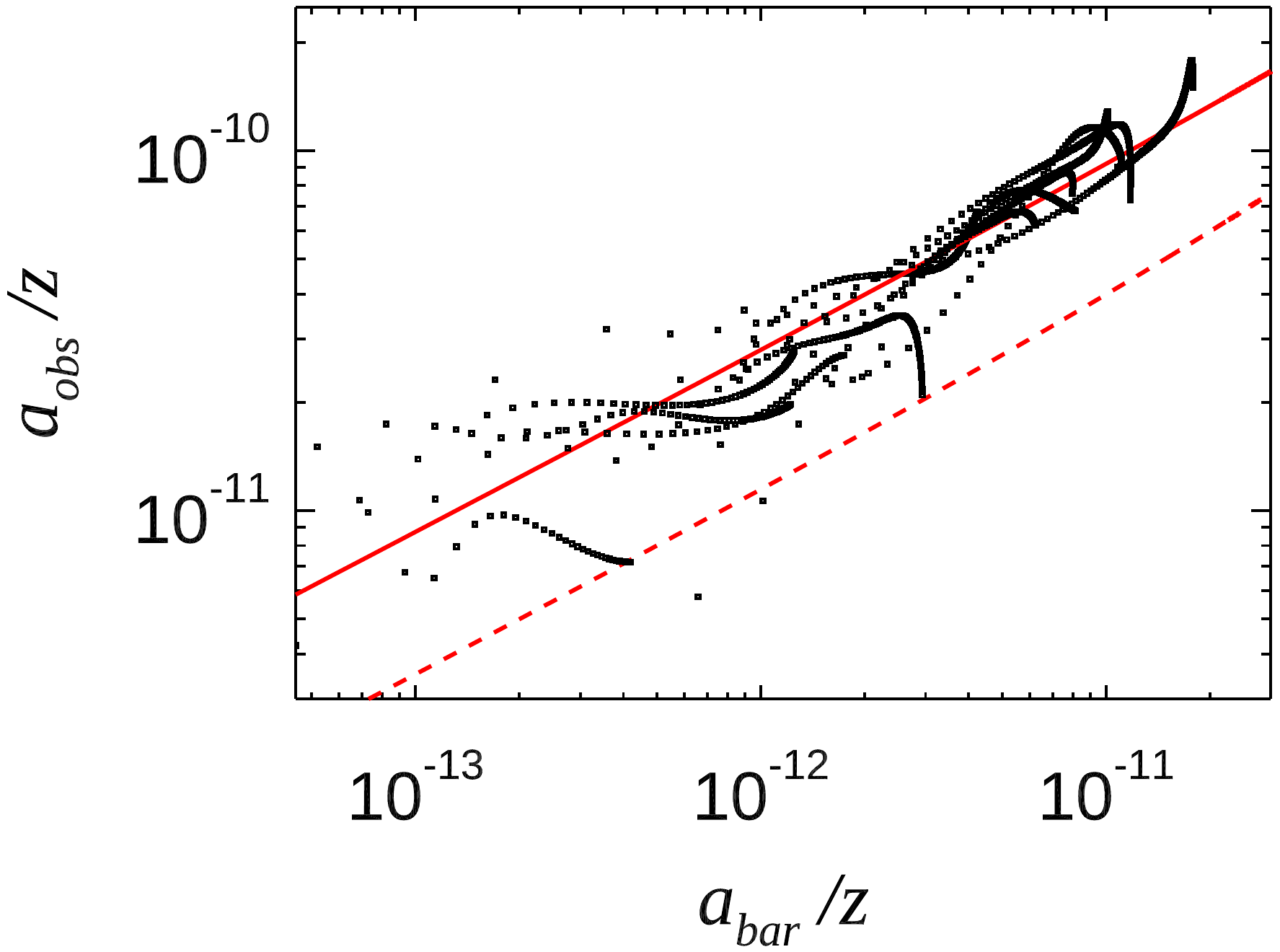}
  \caption{Comparison of observed accelerations and accelerations expected from baryons in galaxy clusters. The black squares represent fits to data from a sample of 13 galaxy clusters\cite{Vikhlinin:2005mp}. The solid and dashed red lines are the predictions of Eq.~\ref{modified_McGaughRelation} using $a_0$ and $\ac$, respectively.}
  \label{fig_compare_g_clusters}
\end{figure*}

The paper is organized as follows: First we give a heuristic argument for the modified dark matter (MDM) profile
based on gravitational thermodynamics. 
In particular, we review and then generalize the
entropic gravity/gravitational thermodynamics arguments of 
Jacobson \cite{Jacobson:1995ab} and Verlinde
\cite{Verlinde:2010hp,Verlinde:2016toy} to de-Sitter space with positive
cosmological constant to construct a dark matter model which 
is sensitive to $\Lambda$ and thus to the fundamental acceleration \cite{Ho:2010ca,Ho:2011xc,Ho:2012ar}.
In this context we explain the relation between MDM and the entropic gravity
proposal of Verlinde. 
Then we discuss how MDM captures the observed data on galactic, cluster, and even cosmological scales. 
We emphasize that modified dark matter
captures the successful features of 
CDM at cluster and cosmological scales, but it effectively behaves like MOND at 
the galactic scales.  (The may
explain the apparent failure of MOND at large scales.)
We subject MDM to observational tests
with galactic rotation curves for 30 galaxies and observed mass profiles 
for 13 galactic clusters. \cite{Edmonds:2013hba,Edmonds:2016tio}
We show that MDM is in some sense more economical than CDM in fitting
data at the galactic scale, and it is superior to MOND at the cluster 
scale. Also, based on very general arguments, MDM also works on cosmological scales where it is
consistent with the $\Lambda CDM$ paradigm.
We conclude the review with a few comments about some non-local non-particle properties of MDM as well as about its possible fundamental origin
in quantum gravity.

\section{Constructing Modified Dark Matter (MDM)}

In this section we aim to outline the basic logic of our proposal.
We want to construct a mass profile for non-baryonic dark matter, that is sensitive to the fundamental acceleration whose value is set by the observed cosmological constant. We also want to tie this dark matter mass profile to the baryonic mass profile via the fundamental acceleration parameter on galactic scales, as indicated by data. However, we also aim to have enough flexibility not to correlate the dark matter mass profile to the baryonic mass profile on cluster and cosmological scales.
So the question is: how can this be achieved? One might think that this is impossible, given the fact that the canonical mass profiles for CDM are arrived at after laborious numerical simulations of structure formation.

The idea here is that the acceleration can be re-interpreted in terms of temperature of the Unruh-Hawking kind \cite{Davies:1974th,Unruh:1976db}, and that in turn, such temperature can also be corrected by the presence of the cosmological constant, due to the fact that maximally symmetric spaces with positive cosmological constant, that is, asymptotically de Sitter spaces, also have a characteristic temperature associated with their cosmological horizons. This temperature can be rephrased as the fundamental acceleration. Furthermore, any excess temperature can be interpreted as excess energy, and thus as extra matter source. 
Thus, the fundamental origin of dark matter is tied to the thermal properties associated with gravity in the context of 
effective quantum field theory in curved spacetime.
Note that according to this proposal any excess source and the usual visible matter sources could be related via the corresponding temperature.
Therefore, in principle, the dark matter and visible, baryonic matter, could be related, and this relation can, in principle, involve the fundamental acceleration parameter.
Finally, given the fact that temperature can be red-shifted by using the well-known Tolman-Ehrenfest formula, we can
in principle argue for different mass profiles on different scales, while still maintaining the explicit dependence of the dark matter profile on the fundamental acceleration. As we will, see this flexibility allows us to account for the observed data both on galactic and cluster scales. 

Technically, our approach can be traced to the work of Jacobson \cite{Jacobson:1995ab} regarding the thermodynamics of Einstein's gravitational equations.
We are interested in a slight modification of his argument that utilizes the thermodynamics of de Sitter space.
Thus we consider a local observer with acceleration $a$ in a spatially flat 
de Sitter space. 
Jacobson's idea was to start with the thermodynamic relation
\begin{equation} \label{firstlaw}
dE \;=\; T dS.
\end{equation}
in Rindler spacetime.
By examining the ways in which temperature $T$, entropy $S$ and 
energy $E$ are
defined, and by utilizing the Raychaudhuri focusing equation, Jacobson
deduced the integral form of the Einstein
equation. For $T$, it is natural to use the Unruh temperature
associated with the local accelerating 
(Rindler) observer \cite{Davies:1974th,Unruh:1976db}
\begin{equation}
T \;=\; \frac{\hbar a}{2\pi c k_B}\;.
\end{equation}
For $S$, the holographic principle 
\cite{tHooft:1993dmi,Susskind:1994vu} can be invoked to give 
\begin{equation}
S \;=\; \frac{c^3 A}{4\GN\hbar}\;,
\label{entropy}
\end{equation}
where $A$ is the area of the Rindler horizon.  Here $E$ denotes the
integral of the energy momentum tensor of matter
\begin{equation}
E \;=\; \int T_{\alpha\beta} k^{\alpha} k^{\beta},
\label{energy}
\end{equation}
where $k^{\alpha}$ are appropriate unit vectors.

The variation of the area $A$ in the holographic principle expression
is given in terms of the expansion of the horizon
generators the focusing of which is associated with the 
energy flux flowing across the horizon.
For an instantaneously stationary local Rindler horizon (required for
equilibrium thermodynamics), the shear and vorticity terms can be neglected
in the integration of the Raychaudhuri equation 
for the expansion of the horizon generators
to yield
\begin{equation}
\dfrac{\delta A}{\delta\lambda} 
\;=\; R_{\alpha \beta} k^{\alpha} k^{\beta} + \cdots\;,
\end{equation}
where $\lambda$
is the appropriate affine parameter. Using the above relations, Jacobson
obtained
\begin{equation}
8\pi\GN \int T_{\alpha \beta} k^{\alpha} k^{\beta} \;=\;
\int R_{\alpha \beta} k^{\alpha} k^{\beta} 
\;.
\end{equation}
The cosmological constant appears as an integration constant in this approach, and application of local
conservation of energy and momentum finally yields the Einstein equation.
\footnote{
Note that, in the derivation, the Ricci tensor $R_{\alpha \beta}$ enters
the discussion via the expression for $dS$ in Eq.~(\ref{entropy}).  
Thus, to
remain consistent with general relativity, we should preserve the 
holographic
scaling of the area.  This fact will be useful in the construction of MDM 
mass profile.}
In the 
following two sub-sections, we construct MDM by two related 
arguments both inspired by Jacobson's insight.

\subsection{MDM and Gravitational Thermodynamics}

We can argue for the MDM profile by using a heuristic argument as follows. Consider
modifying Jacobson's argument \cite{Jacobson:1995ab} by 
introducing a fundamental 
acceleration that is related to the cosmological constant.
We assume that Einstein's theory of gravity is valid and we want a standard 
energy-momentum
tensor. The first condition requires that we preserve the holographic 
scaling of the area.
Consequently the second condition, in conjunction with the form of the 
thermodynamic relation,
demands that we change the temperature while preserving the entropy. Our 
model is therefore
given by the thermodynamic relation
\begin{equation}\label{deftilde}
d\tilde {E} \;=\; \widetilde{T} dS\;.
\end{equation}
We note that, since the Unruh
temperature knows the inertial properties and
is fixed by the background, 
the
additional part of the energy-momentum tensor (coming from a modified 
temperature) will
also know the inertial properties and the background.

Here we want to point out that in these arguments one always has to absorb the acceleration factor in the Einstein tensor, in order to end up with the correct normalization of
Einstein's equations. This acceleration-dependent factor will be crucial in our argument for the emergence of the dark matter mass profile that depends on the observed acceleration as well as the fundamental acceleration parameter, and that is still
subject to the canonical force law, as implied by the canonical Einstein equations.

Now consider a local observer with local acceleration $a$ in de Sitter 
space, where $a_0 = c^2 
\sqrt{\Lambda/3} = c H_0$.
The Unruh temperature experienced by this observer is\footnote{If the four-dimensional de Sitter space is envisioned as a hyperboloid embedded in a flat, five-dimensional space, the effective five-dimensional acceleration ($a_5$) is related to the four-dimensional acceleration ($a_4$) as $a_5 = \sqrt{a_4^2 + a_0^2}$.} 
\cite{Deser:1997ri,Jacobson:1997ux}
\begin{equation}
T_{a_0 +a} \;=\; \dfrac{\hbar}{2\pi c k_B} \sqrt{\,a^2+a_0^2\,}\;.
\label{Temp}
\end{equation}
However, since de Sitter space has a cosmological horizon, it has a horizon 
temperature associated with $a_0$. 
We thus define the following {\it effective} 
temperature (so that for zero acceleration we get zero temperature)
\begin{eqnarray}\label{net}
\widetilde{T} \;\equiv\; T_{a_0+a} - T_{a_0} \;=\; \frac{\hbar}{2\pi c k_B} 
\left(\,\sqrt{a^2+a_0^2} - a_0 \,\right) \;\equiv\; 
\frac{\hbar \tilde{a}}{2\pi c k_B}
\;.
\end{eqnarray}

In analogy with the normalized temperature $\widetilde{T}$,
the normalized energy is given by 
$
\tilde{E} \;=\; E_{a_0 +a} - E_{a_0}\;.
$
Writing the temperature as
$T_{a_0 +a} = T + T'$ where 
the $T$ part corresponds to the Unruh temperature of the 
observer moving with the 
Newtonian acceleration $a_N$ (in the correspondence 
limit $a_0 \ll a=a_N$), and using
$
d E_{a_0+a} \;=\; T_{a_0+a} d S\;,
$ 
we can also write
$
d E +d E' \;=\; T d S + T' d S\;.
$
If we interpret the original $dE = T dS$ as corresponding to 
baryonic (unprime) matter, then
$
d E' \;=\; T'  dS \;=\; \frac{T'}{T} dE\;
$
corresponds to dark (prime) matter.
By expanding the formula for the de Sitter temperature 
Eq.~(\ref{Temp}) (relating $T'$ to $T$), using Eq.~(\ref{energy})
(relating $E$ to $T_{\alpha \beta}$) and the fact that the
energy density is the $00$ component of the energy-momentum
tensor, we have a relation between 
the dark and visible matter
\begin{equation}
M' \;=\;  \dfrac{a_0^2}{2a^2} \,M\;.
\label{fMDM2}
\end{equation}
This dark matter profile is qualitatively the same as the one 
(Eq.~(\ref{Mprime})) in our original papers on MDM 
\cite{Ho:2010ca,Ho:2011xc,Ho:2012ar}. 

\subsection{MDM and Entropic Gravity}
\label{mdmentropic}

The argument given in Ref.~\citen{Ho:2010ca} is 
a simple generalization of Verlinde's recent proposal of entropic 
gravity \cite{Verlinde:2010hp}, inspired by Jacobson's work 
\cite{Jacobson:1995ab},
for $\Lambda = 0$ to the case of 
de-Sitter space with a
positive $\Lambda$.  Let us first review Verlinde's view 
of Newton's second law $\vec{F} = m \vec{a}$. Consider a
particle with mass $m$ approaching a holographic screen
at temperature $T$.  Using the first law of thermodynamics to introduce 
the concept of entropic force
\begin{equation}
F \;=\; T \dfrac{\Delta S}{\Delta x}\;,
\end{equation}
and invoking Bekenstein's original arguments \cite{Bekenstein:1973ur}
concerning the entropy $S$ of black holes,
\begin{equation}
\Delta S \;=\; 2\pi k_B \frac{mc}{\hbar} \Delta x \;,
\end{equation}
Verlinde found
\begin{equation}
F \;=\; 2\pi k_B \frac{mc}{\hbar} T\;. 
\end{equation}
With the aid of
the formula for the Unruh temperature
associated with a uniformly accelerating (Rindler) observer, 
\begin{equation}
k_B T \;=\; \dfrac{\hbar a}{2 \pi c}\;,
\label{UnruhTemp}
\end{equation}
Verlinde then obtained $\vec{F} = m \vec{a}$.

The already cited formula for the effective temperature, \cite{Unruh:1976db,
Davies:1974th,Hawking:1974sw} as measured by a
non-inertial observer with
acceleration $a$ relative to an inertial observer,
is 
\begin{equation}
\widetilde{T} \;=\; \frac{\hbar
\tilde{a}}{2\pi k_B c}\;,
\end{equation}
with $\tilde{a} = \sqrt{a^2+a_0^2} - a_0$. \cite{Deser:1997ri}
This reduces to the usual temperature for Rindler observers 
by neglecting $a_0$. 
Note that
\begin{equation}
\sqrt{a^2+a_0^2} - a_0
\;\approx\; \begin{cases}
a & (a\gg a_0)\;, \\
\dfrac{a^2}{2a_0^2} & (a\ll a_0)\;.
\end{cases} 
\end{equation}
The entropic force (in de-Sitter space) 
is hence given by the replacement 
of $T$ and $a$ by
$\widetilde{T}$ and $\tilde{a}$ respectively, leading to
\begin{equation}
F \;=\;  m \left[\sqrt{a^2+a_0^2}-a_0\right] \;\equiv\; m a_\mathrm{obs} \;,
\label{modforce}
\end{equation}
%
where we have changed the notation from $\tilde{a}$ defined in Eq.~(\ref{net}) to
$a_\mathrm{obs}$ to emphasize its meaning as the observed acceleration.
This means that the above mass profile needs to be rewritten as
\begin{equation}
M' = \frac{1}{2} 
\left(\,\frac{a_0}{a}\,\right)^2\, M
\equiv \frac{1}{2} 
\left(\,\frac{a_0^2}{(a_\mathrm{obs}+a_0)^2 -a_0^2}\,\right)\, M\;,
\label{Mprime}
\end{equation}

Here we emphasize one very important point. Given the fact that we have not changed Einstein's equations, we cannot change the Newton law of motion. However, the Newtonian, observed acceleration $a_\mathrm{obs}$, does depend on the
fundamental acceleration parameter $a_0$, as well as the auxiliary acceleration $a$ that appears in the Unruh formula for the de Sitter space. This hidden dependence, originating in gravitational thermodynamics, finds its explicit realization in the dark matter profile, which now depends on $a_0$ and $a_\mathrm{obs}$. Thus, in some sense, we are managing to bootstrap 
the observed acceleration, the dark matter mass $M'$, and the visible matter mass $M$.
Thus the relevant Newtonian dynamics is described via
\begin{equation}  
a_\mathrm{obs} \;=\; \frac{\GN(M + M')}{r^2}\;\;\equiv\; \frac{\GN M}{r^2}[1 + \frac{1}{2} 
\left(\,\frac{a_0^2}{(a_\mathrm{obs}+a_0)^2 -a_0^2}\,\right)\,]\;.
\label{MDMeom}
\end{equation}

To summarize: dark matter exists in our scheme, but its mass profile is tied to the baryonic mass profile (at least on galactic scales) via the observed acceleration and the fundamental acceleration parameter, set by the cosmological constant.
Such dark matter is most probably of a non-local, non-particle kind, because of the fact that its mass profile is
tied to the baryonic mass profiles as well as to the inertial properties, encoded in the observed acceleration.
Furthermore, as we will show next, the MOND-like scaling, found to work beautifully at the galactic scale, 
is simply a
manifestation of such non-local dark matter, answering the question raised in Ref.~\citen{Kaplinghat:2001me}.
Dark matter of this kind can behave as if there is no dark
matter but MOND. (That was why we used to call it ``MONDian dark matter".
However, we find the Modified Dark Matter more appropriate, given the new, modified, form of
the dark matter profile.)
Intriguingly the dark matter profile we have
obtained relates, at the galactic scale,
dark matter ($M'$), dark energy ($\Lambda$) and ordinary matter
($M$) to one another.  In the concluding section, we will speculate on a microscopic basis for 
the dark matter's dependence on $\Lambda$.\\

Finally, let us comment on the origin of an effective MOND in our scheme.
In order to compare to the usual MOND reasoning, we revert to the notation that involves the auxiliary acceleration
$a$. First we note that for $ a \gg a_0$, we have $F/m \approx a$ which gives 
$a = a_N \equiv \GN M/r^2$, 
the familiar Newtonian value for the acceleration due to 
the source $M$. But for $a \ll a_0$, 
$F \approx m \frac{a^2}{2\,a_0},$ so
the terminal velocity $v$ of the test mass $m$ in a 
circular motion with radius $r$
should be determined from
\,$ m a^2/(2a_0) = m v^2/r$.  In this small acceleration regime,
the observed flat galactic rotation curves ($v$ being independent 
of $r$) now require
$ a \approx \left( 2 a_N \,a_0^3 \, / \pi \right)^{\frac14}$.
But that means $F \approx m \sqrt{a_N \ac}\,$.
This is the modified Newtonian dynamics (MOND)
scaling \cite{Milgrom:1983ca,Milgrom:1983pn,Milgrom:1983zz}, 
discovered by Milgrom who introduced the
critical acceleration parameter $\ac = a_0/ (2 \pi) = c H_0 / (2 \pi)$ 
(with $H_0$ being
the Hubble parameter) by hand to phenomenologically explain the flat galactic
rotation curves.
\footnote{As far as the knowledge of $\ac \sim c H_0$ is concerned, 
Kaplinghat and Turner
\cite{Kaplinghat:2001me} have argued that the acceleration scale $c H_0$ may 
arise naturally within CDM models. The CDM obtains information on $c H_0$
from the simple fact that it is evolving in a universe expanding 
at the rate of $H_0$,
whereas the coincidence $\ac \sim O(1) c H_0$ is more of a numerical accident.
But even with this knowledge, CDM has problems at the galactic scale as 
mentioned above.} 
Thus, as advertised, we have recovered MOND with the correct
magnitude for the critical galactic acceleration parameter $\ac \sim 
10^{-8}
cm/s^2$.\footnote{The emergence of this scale in non-standard dark matter models has
been also addressed in Ref.~\cite{Khoury:2014tka, Berezhiani:2015bqa,Berezhiani:2015pia}.}
From our perspective, MOND is a {\it classical} phenomenological
consequence of the MDM mass profile, obtained from
gravitational thermodynamics in de Sitter space (with the $\hbar$ dependence
in $T \propto \hbar$ and $S \propto 1/\hbar$ canceled out). 
\cite{Ho:2010ca}  As a bonus,
we have also recovered the observed Tully-Fisher relation ($v^4 \propto M$).

\section{Observational Tests of MDM}

As a preparation for our discussion on testing MDM (and CDM and MOND),
let us first collect all the relevant formulas for the 
various (effective) mass profiles.
For MDM, the equation of motion (Eq.~(\ref{MDMeom})) reads
\begin{equation}
a_N \Big[\,1+f_\mathrm{MDM}(a/a_0)\,\Bigr]
\;=\; \sqrt{a^2 + a_0^2} - a_0^{\phantom{1}} 
\;\equiv\; a_\mathrm{obs}\;,
\label{MDM-EQM}
\end{equation}
where 
\begin{equation}
\dfrac{M'}{M} \;=\; f_\mathrm{MDM}(a/a_0) \;\equiv\;  
f_\mathrm{MDM}\left(\sqrt{(a_\mathrm{obs}+a_0)^2 - a_0^2}\Big/a_0\right)\;.
\end{equation}
%
%
For comparison, let us include the dynamical masses predicted 
by CDM and MOND. For CDM, 
we use the \cite{Navarro:1995iw,Navarro:1996gj} (NFW) density profile,
\begin{equation}
\rho_\mathrm{NFW}(r) \;=\; 
\dfrac{\rho_0}{\dfrac{r}{r_\textrm{CDM}}\left(1+\dfrac{r}{r_\textrm{CDM}}
\right)^2}
\;,
\label{NFWrho}
\end{equation}
to determine the mass predicted by CDM, where
$r_\textrm{CDM} = r_{200}/c$,
and $r_{200}$ designates the edge of the halo, within which 
objects are assumed to be virialized, 
usually taken to be the boundary at which the halo density 
exceeds 200 times that of the background. The parameter $c$ 
(not to be confused with the speed of light) 
is a dimensionless number that indicates how centrally 
concentrated the halo is.

For an effective MOND, one can write the left-hand-side of Newton's equation as
\begin{equation}
\dfrac{1}{\mu(a/\ac)}\dfrac{G M}{r^2}
\;=\; \dfrac{\GN(M+M')}{r^2}
\;,
\end{equation} 
and interpret $M$ and
\begin{equation}
M'
\;=\; M\left[\dfrac{1}{\mu(a/\ac)}-1\,\right]
\;\equiv\; M f_\mathrm{MOND}(a/\ac)
\;,
\label{fdef}
\end{equation}
as the mass of baryonic matter and the {\it effective} 
mass of non-baryonic dark matter respectively enclosed 
within the same sphere.
Then we have that
\begin{equation}
a_N\Bigl[\,1 + f_\mathrm{MOND}(a/\ac) \,\Bigr] \;=\; a\;.
\label{MOND-EQM3}
\end{equation}
Solving this equation for the acceleration $a$ will also determine 
$M'=M f_\mathrm{MOND}(a/\ac)$.
The dark matter distribution determined in this fashion
would precisely reproduce the results of MOND without modifying inertia 
or the law of gravity.
Therefore for MOND, 
we have, from Eq.~(\ref{fdef}),
\begin{equation}
\dfrac{M'(r)}{M(r)} \;=\; f_\mathrm{MOND}(a(r)/\ac) \;=\; 
\dfrac{1}{\mu(a(r)/\ac)}-1\;.
\label{MpMOND}
\end{equation}
Assuming a spherically symmetric distribution, 
the effective dark matter density profile for MOND is given by
\begin{equation}
\rho_\mathrm{MOND}(r) \;=\; \dfrac{1}{4\pi r^2}\dfrac{d}{dr}M'(r)\;.
\end{equation}

\subsection{Galactic Rotation Curves}

In order to test MDM with galactic rotation curves, we fit computed rotation
curves to a selected sample of Ursa Major galaxies given in Ref.~\citen{Sanders:1998gr}. 
The sample contains both high surface brightness (HSB) and
low surface brightness (LSB) galaxies. The rotation curves, predicted by MDM
as given above by 
%
\begin{equation}
F 
\;=\;  m \left[\sqrt{a^2+a_0^2}-a_0\right]
\;=\; m a_N \left[ 1 + \dfrac{1}{2} \left( \dfrac{a_0}{a} \right)^2 \right]
\;,
\end{equation} 
or equivalently, in terms of the observed acceleration $a_\mathrm{obs}$,
\begin{equation}
F 
\;=\; m a_\mathrm{obs}
\;=\; m a_N 
\left[ 1 + \dfrac{1}{2} \left\{ \dfrac{a_0^2}{(a_\mathrm{obs}+a_0)^2 - a_0^2} \right\} \right]
\;,
\end{equation}
along with $ F = m v^2 / r$ for circular orbits, can be
solved for $a_\mathrm{obs}(r)$ (or $a(r)$) and $v(r)$. In Ref.~\citen{Edmonds:2013hba}, we
fit these to the observed rotation curves as determined in Ref.~\citen{Sanders:1998gr},
using a least-squares fitting routine. As in Ref.~\citen{Sanders:1998gr}, the
mass-to-light ratio $M/L$, which is our {\it only} fitting parameter for
MDM, is assumed constant for a given galaxy but allowed to vary between
galaxies. Once we have, for example, the auxiliary acceleration $a(r)$, we can find the MDM density profile by using
%
$M' \;\approx\; \dfrac{1}{2} \left(\,\dfrac{a_0}{a}\,\right)^2\, M\;,$
%
to give
\begin{equation}
\rho'(r) \;=\; \left( \frac{\ac}{r} \right)^2 \frac{d}{dr} \left(
\frac{M}{a^2} \right)\;.
\end{equation}

\begin{figure*}
  \includegraphics[angle=180,width=0.5\textwidth]{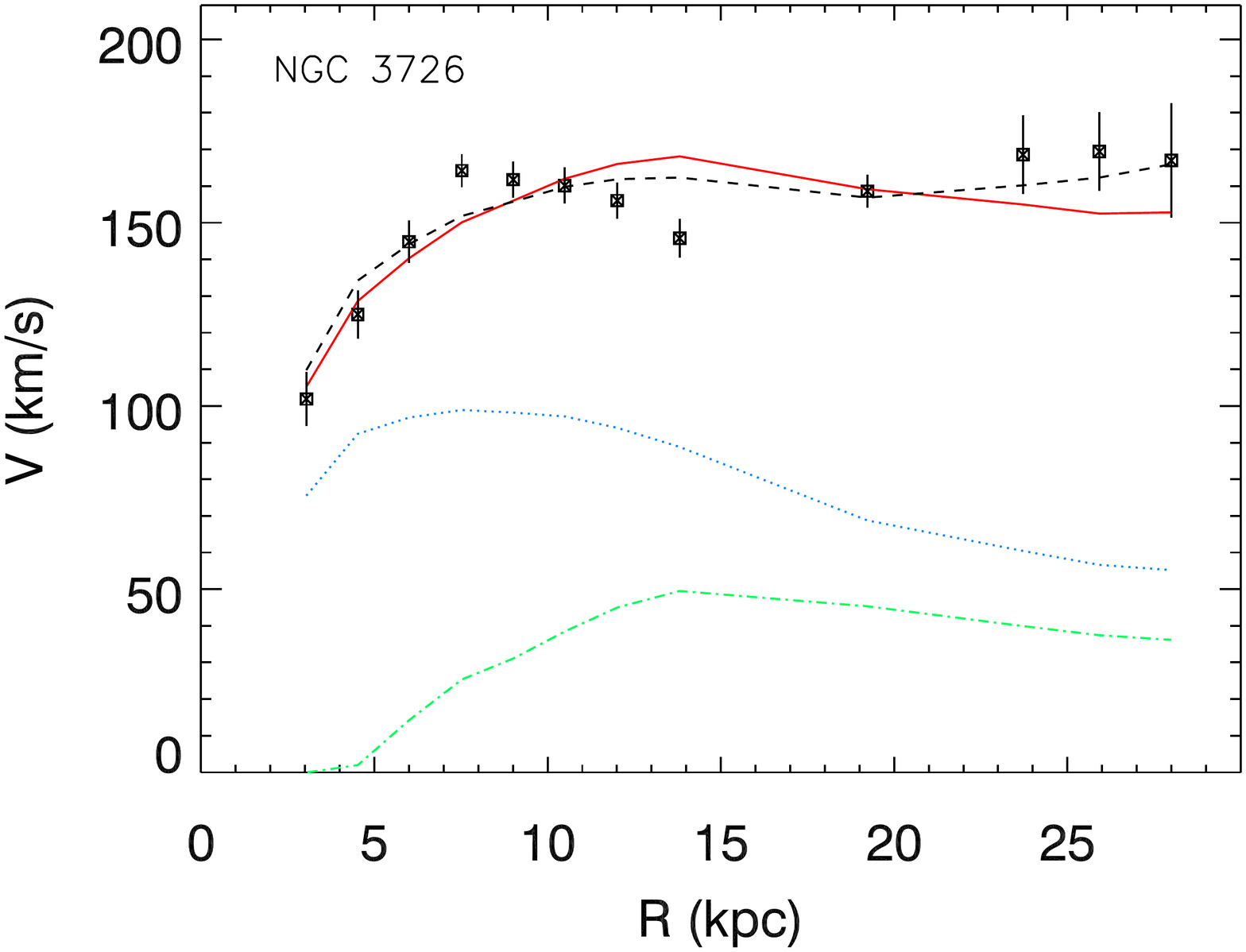}
  \includegraphics[angle=180,width=0.5\textwidth]{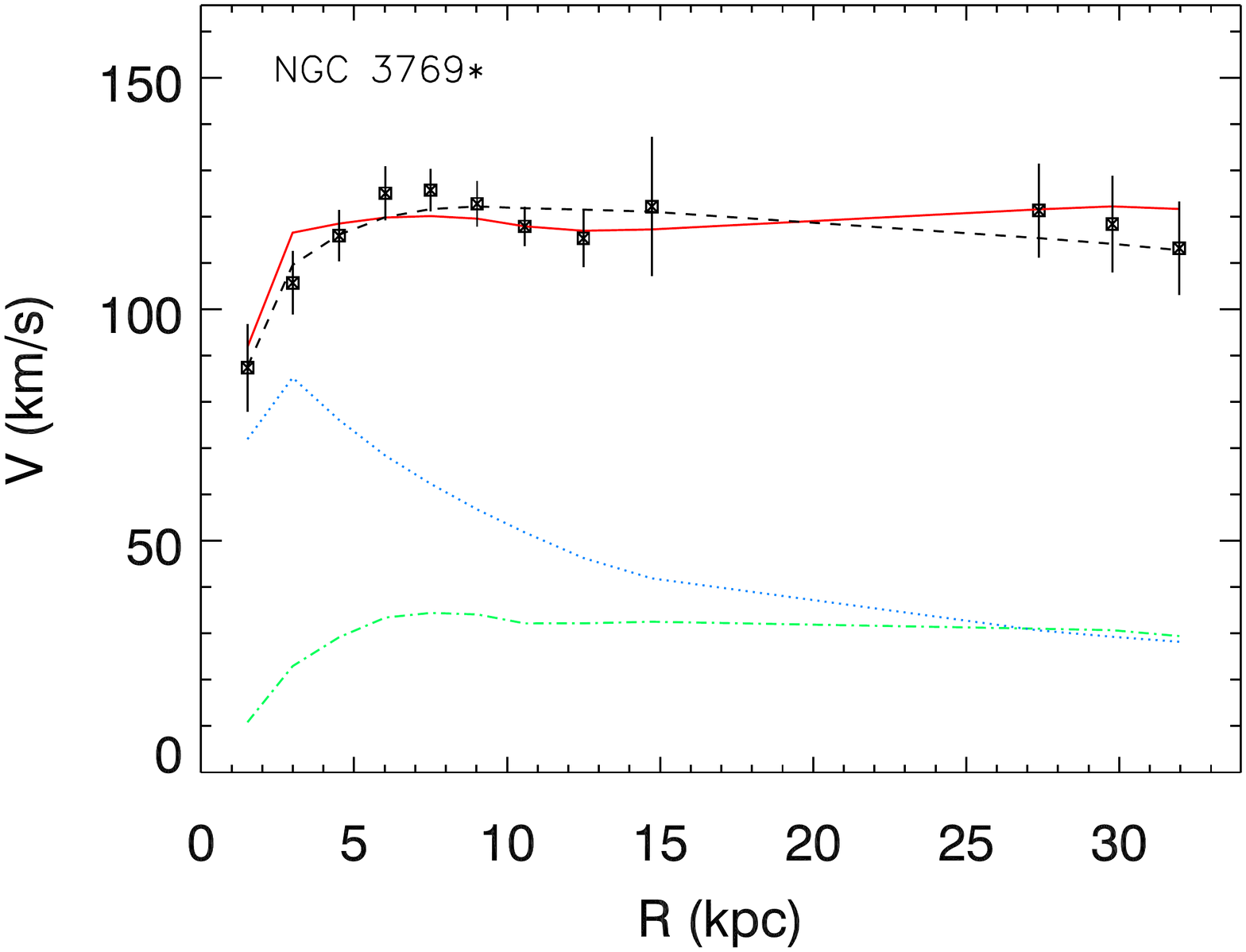}
  \\
  \includegraphics[angle=180,width=0.5\textwidth]{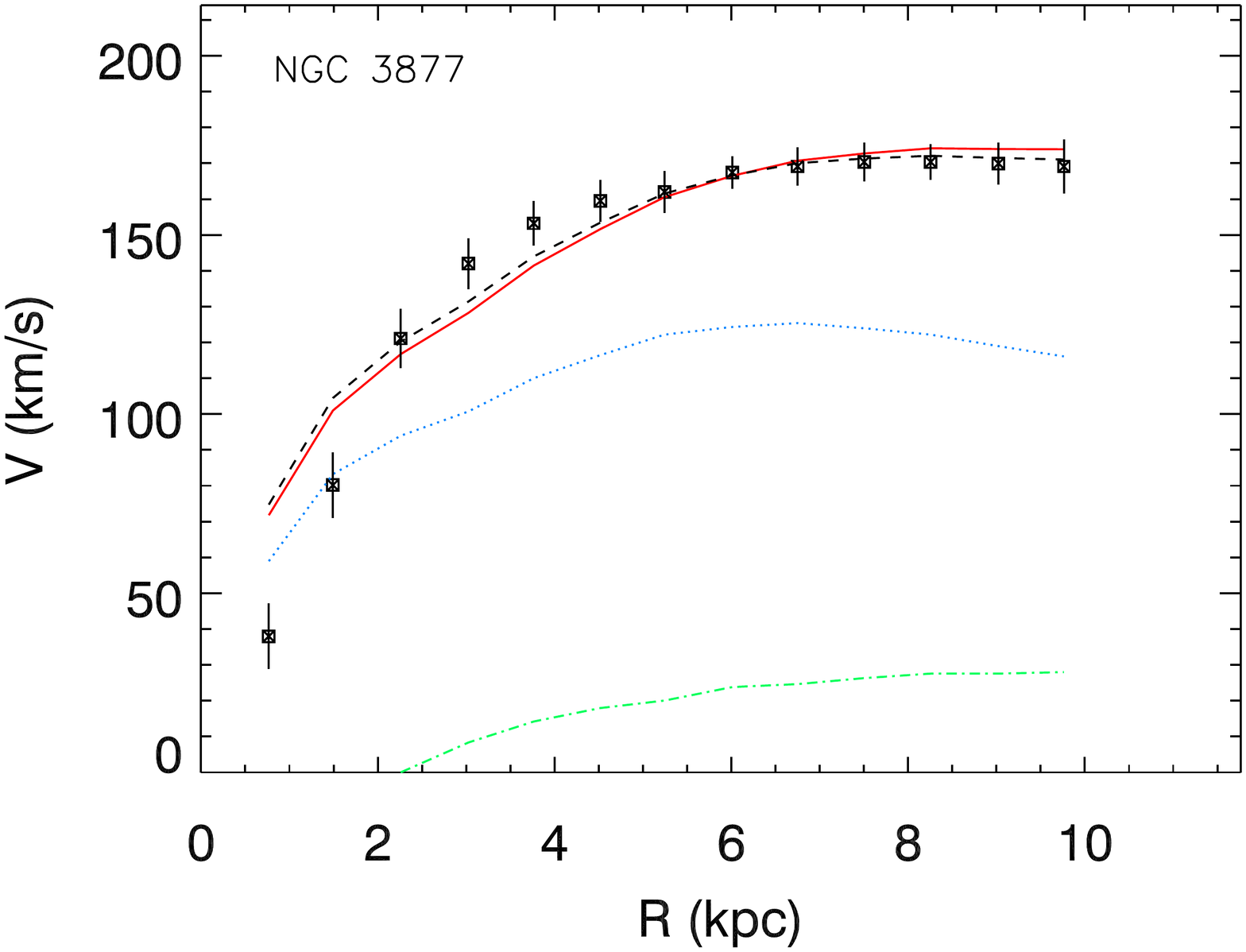}
  \includegraphics[angle=180,width=0.5\textwidth]{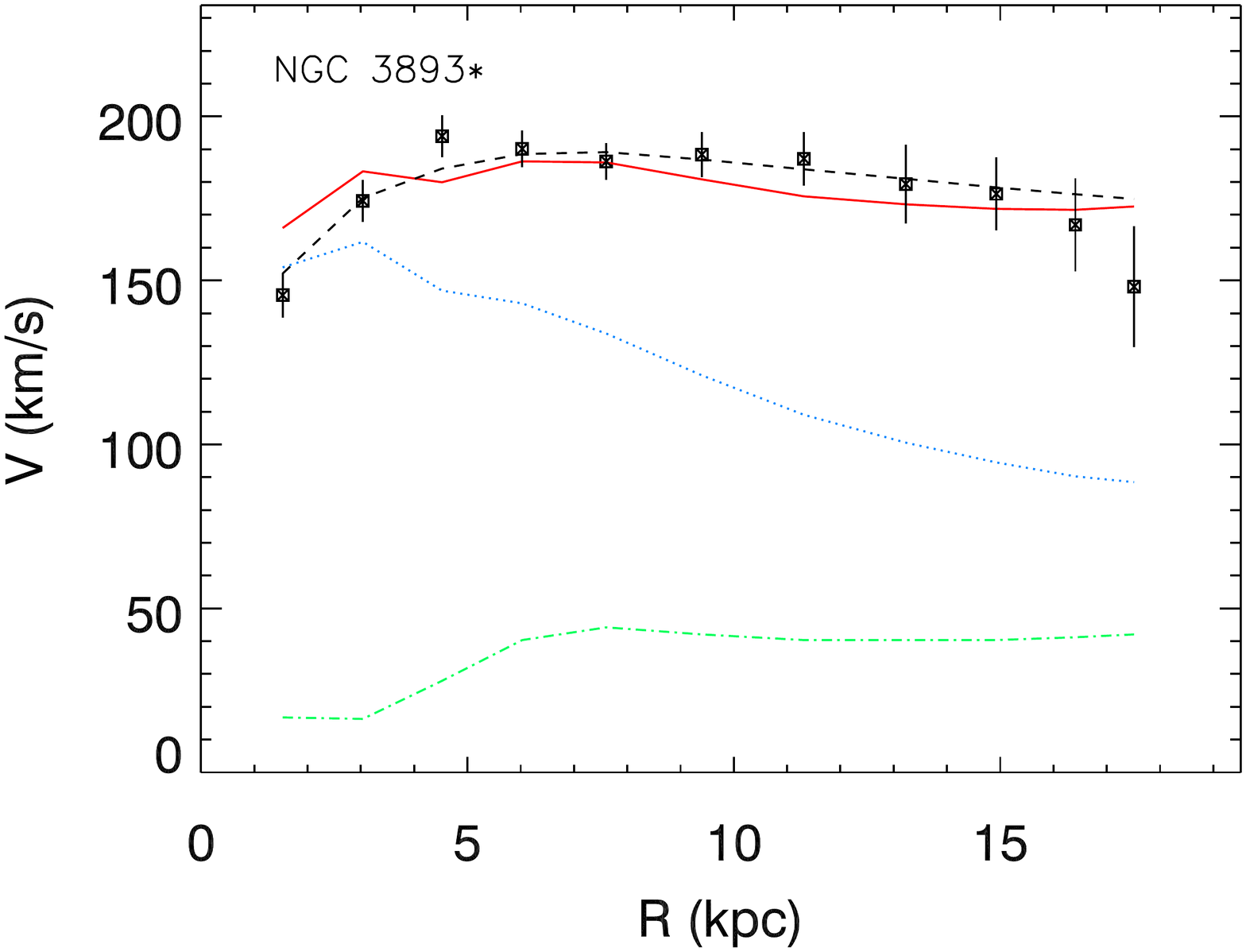}
  \\
  \includegraphics[angle=180,width=0.5\textwidth]{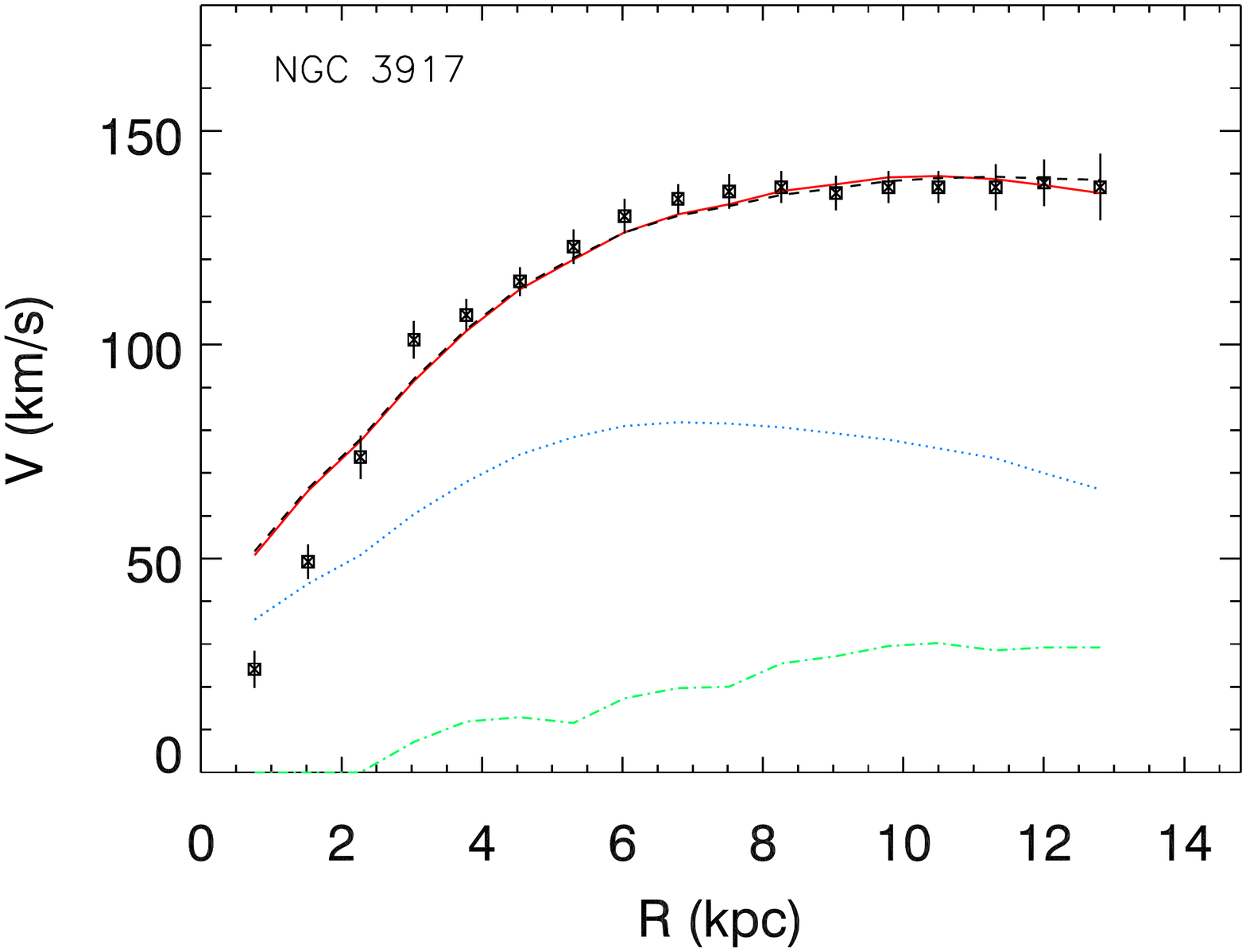}
  \includegraphics[angle=180,width=0.5\textwidth]{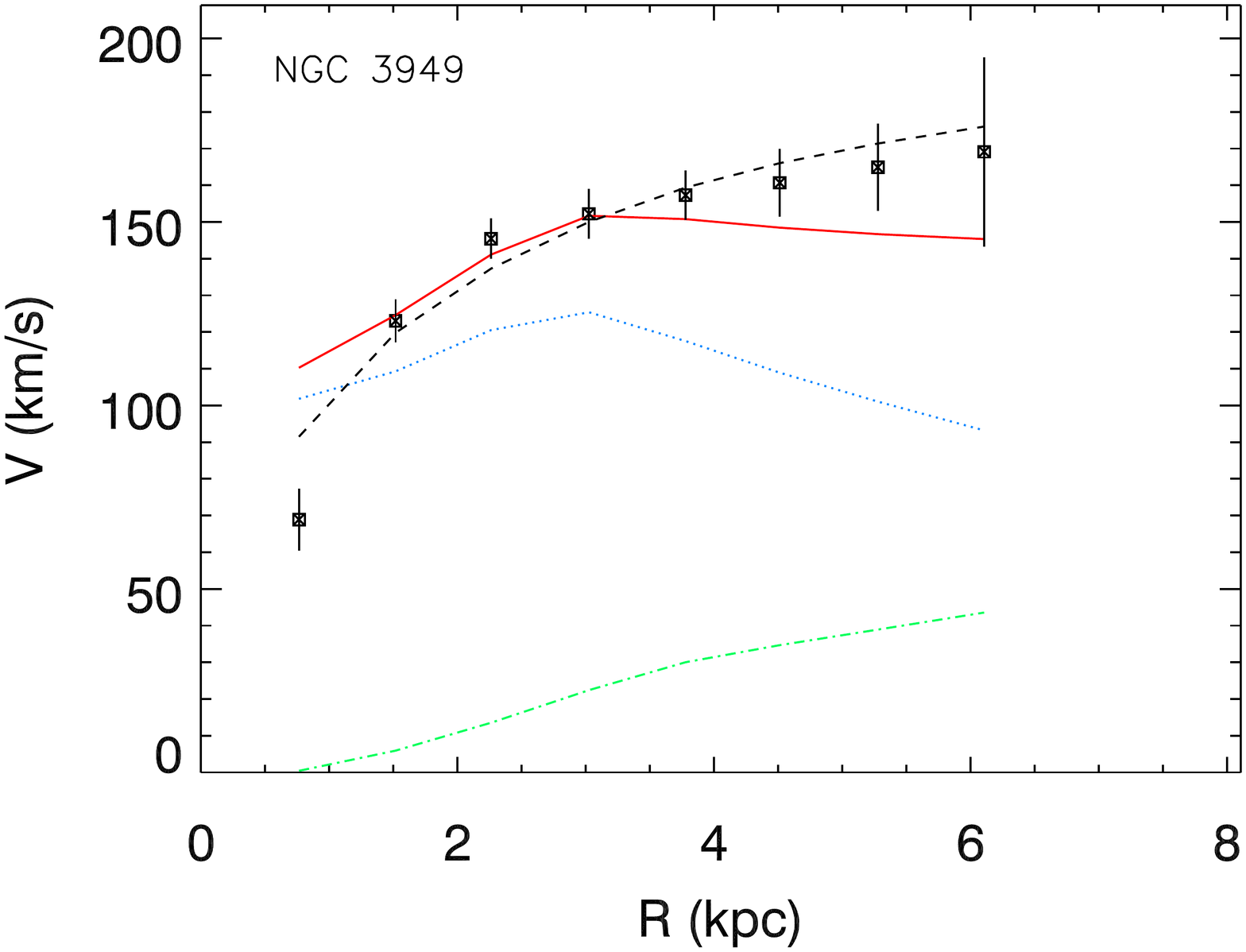}
  \caption{Galactic rotation curves. The observed rotation curve is depicted by points with error bars. The solid red and dashed black lines are the MDM and CDM rotation curves, respectively. Newtonian curves for the stellar and gas components of the baryonic matter are depicted by dotted blue and dot-dashed green lines, respectively. The mass of the stellar component is derived from the 
$M/L$ ratio determined from MDM
fits to the rotation curve. These figures are based on
the generalized mass profile}
  \label{fig:GRC}
\end{figure*}

Rotation curves predicted by MDM are shown in Fig.~\ref{fig:GRC}. Both the MDM and CDM models fit the data well; 
but note that while the MDM fits use only 1 free parameter,
for the CDM fits one needs to use 3 free parameters.  Thus the MDM model is
a more
economical model than CDM in fitting data at the galactic scale.

Shown in Fig.~\ref{fig:hsb_dens} are the dark matter density
profiles predicted by MDM (solid lines) and CDM (dashed lines).

Finally we should point out that
the rotation curves predicted by MDM and MOND have been found 
\cite{Edmonds:2013hba} to be virtually indistinguishable
over the range of observed radii and both employ only 1 free parameter.

%

\renewcommand{\thefigure}{\arabic{figure} (Cont.)}
\addtocounter{figure}{-1}

\begin{figure*}
  \includegraphics[angle=180,width=0.5\textwidth]{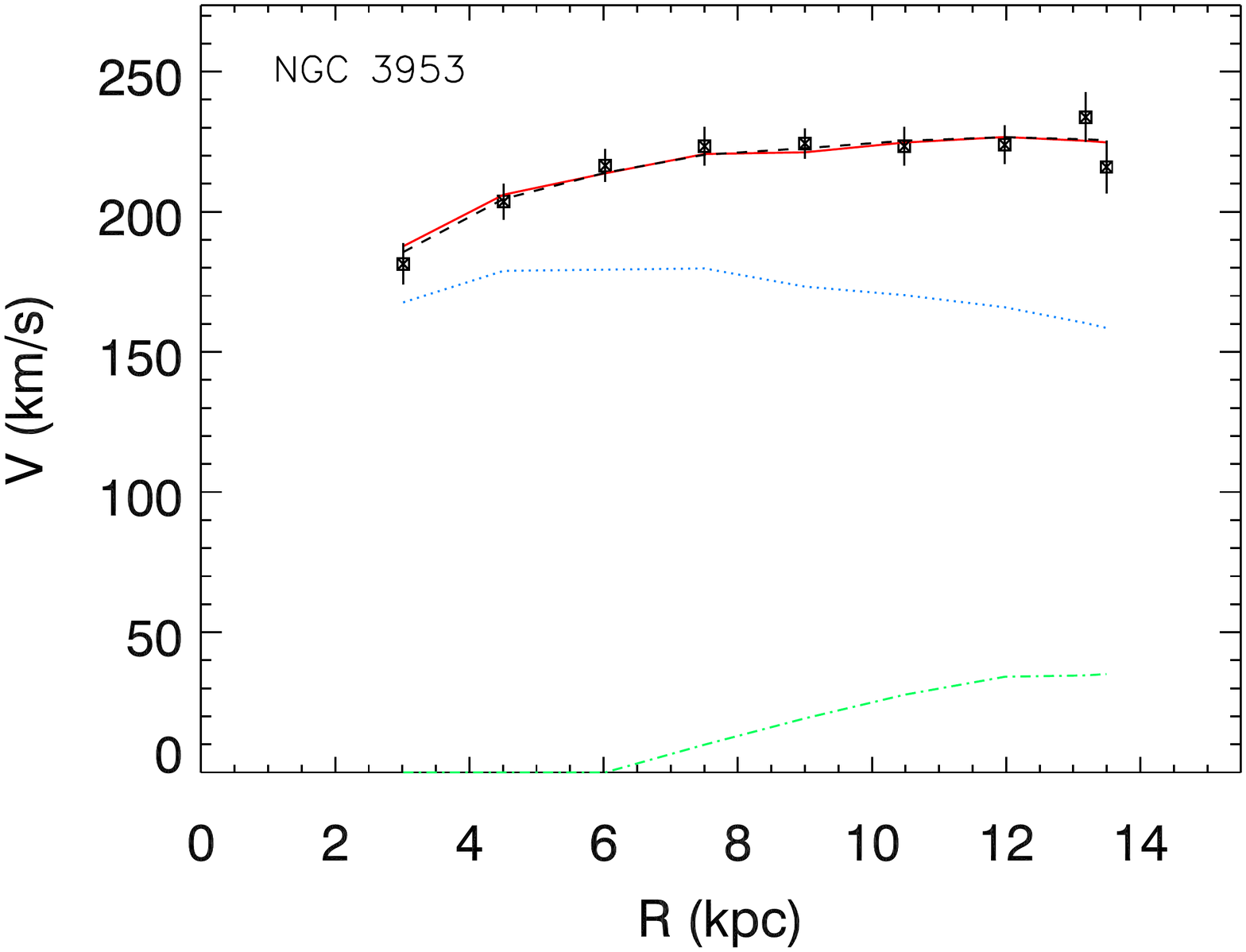}
  \includegraphics[angle=180,width=0.5\textwidth]{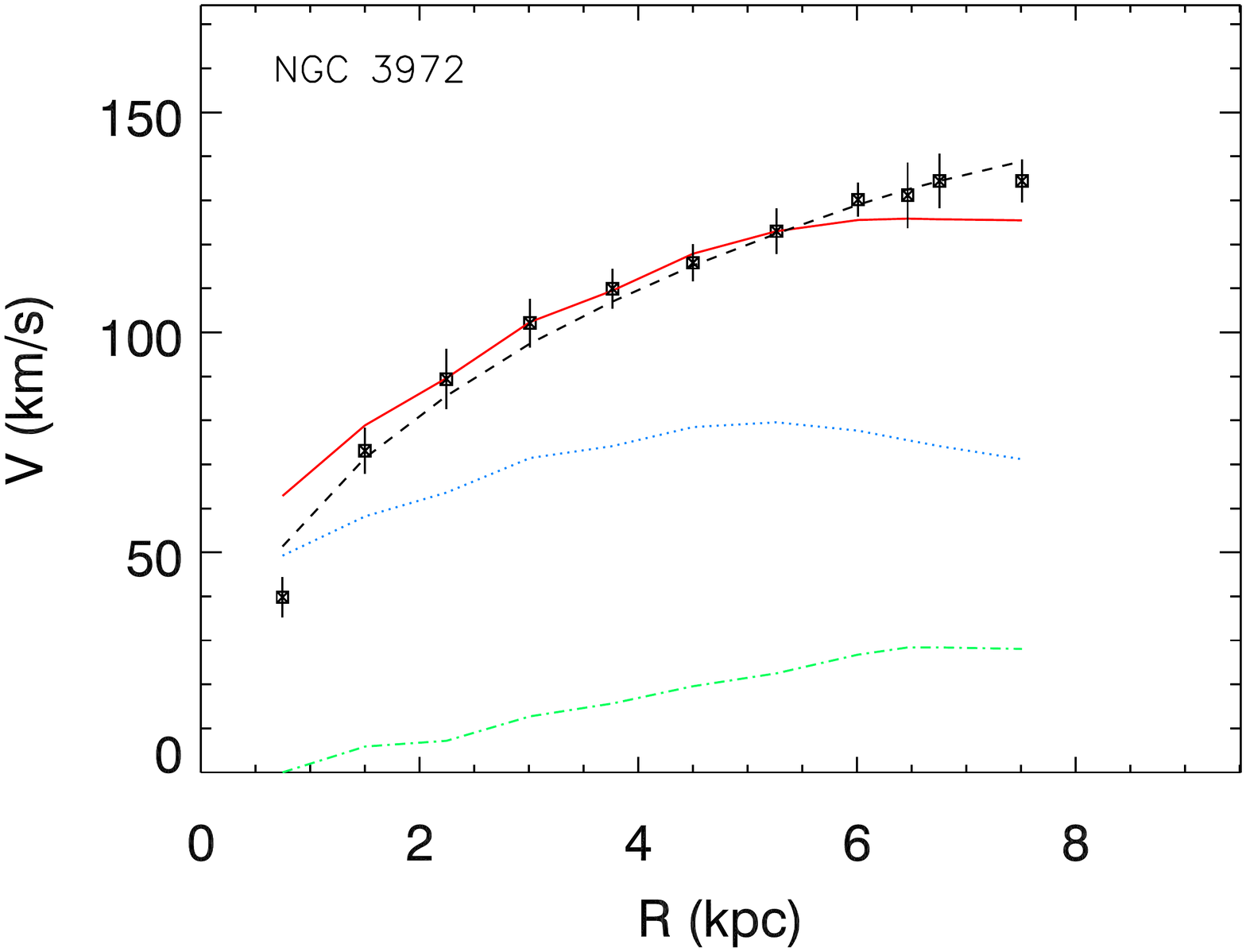}
  \\
  \includegraphics[angle=180,width=0.5\textwidth]{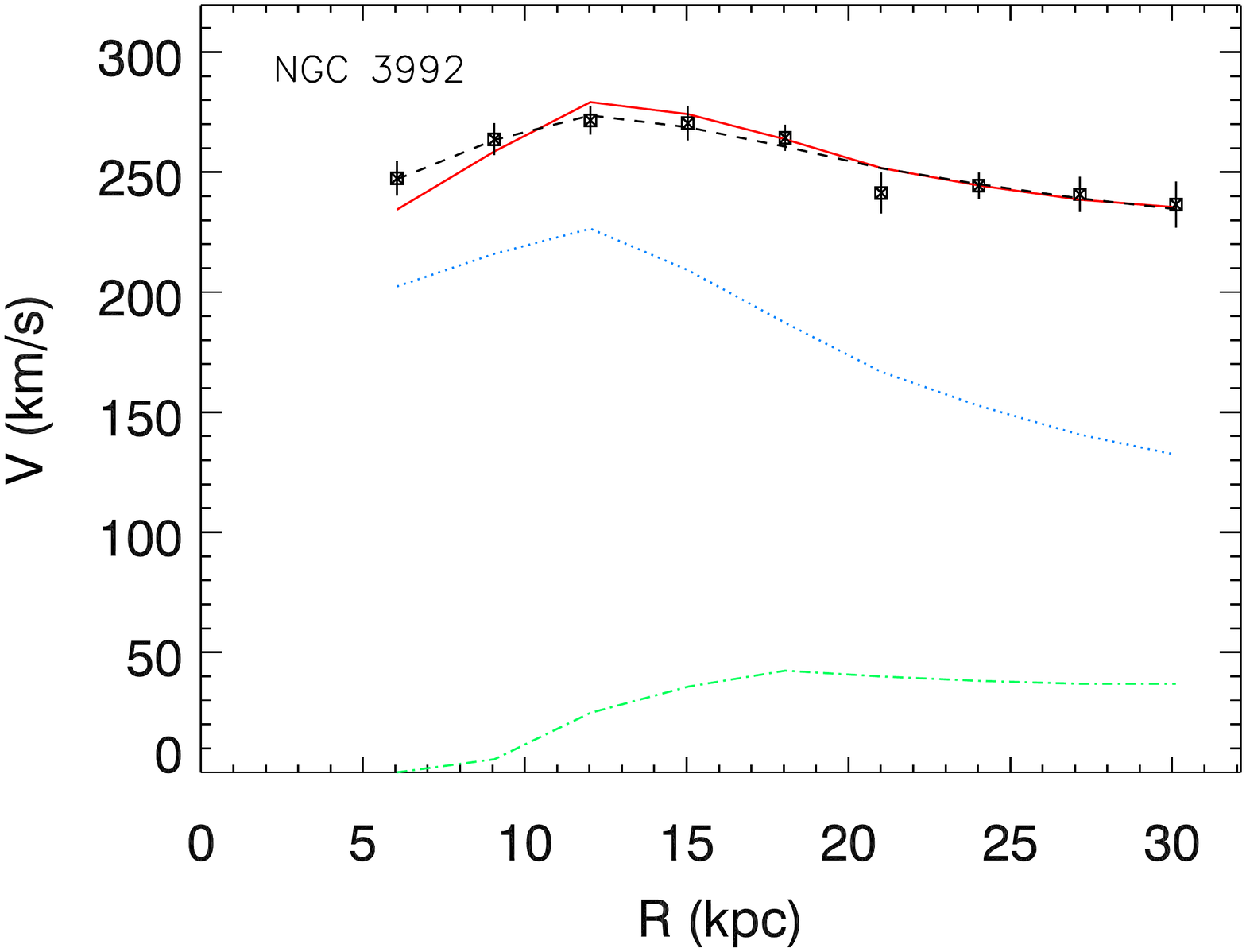}
  \includegraphics[angle=180,width=0.5\textwidth]{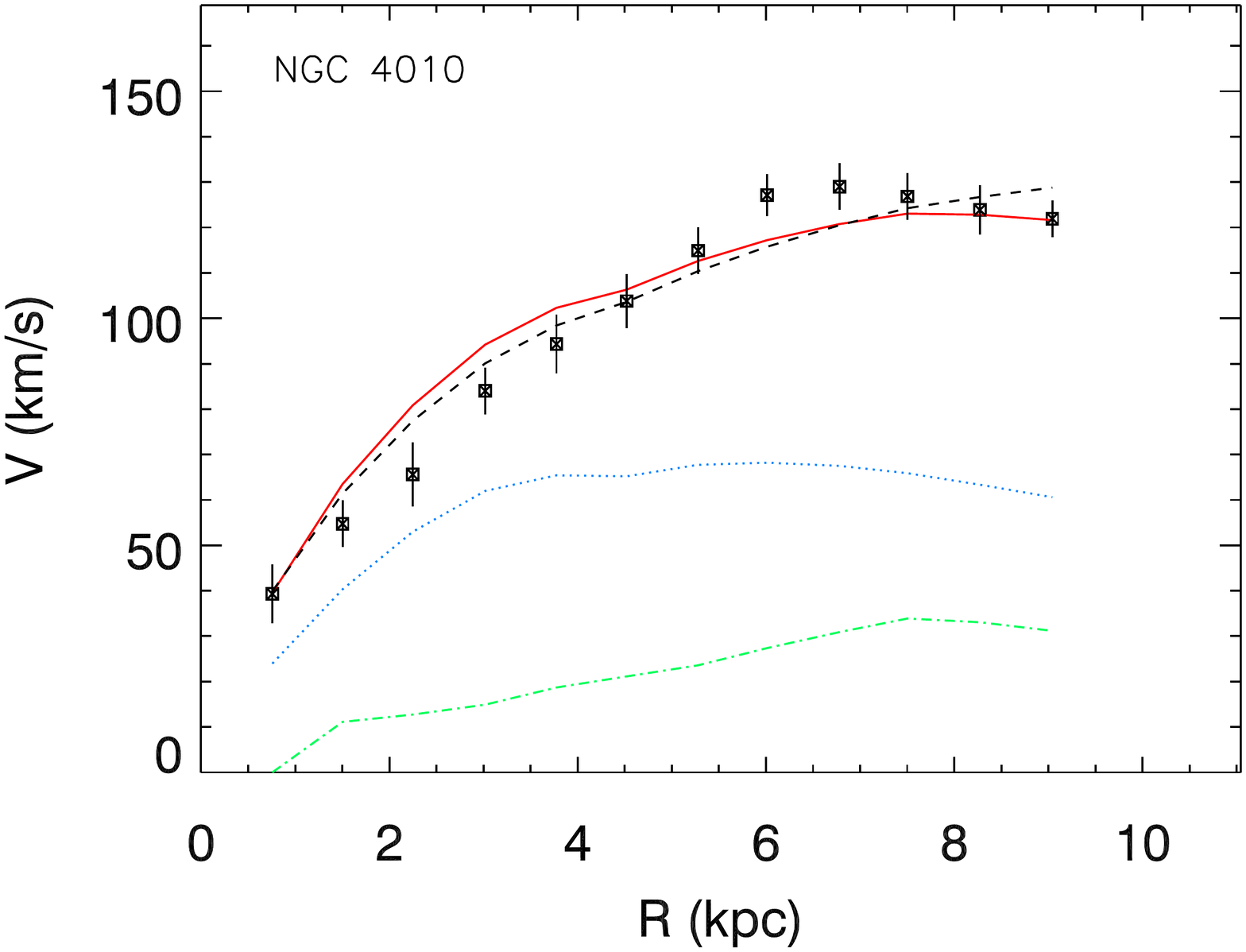}
  \\
  \includegraphics[angle=180,width=0.5\textwidth]{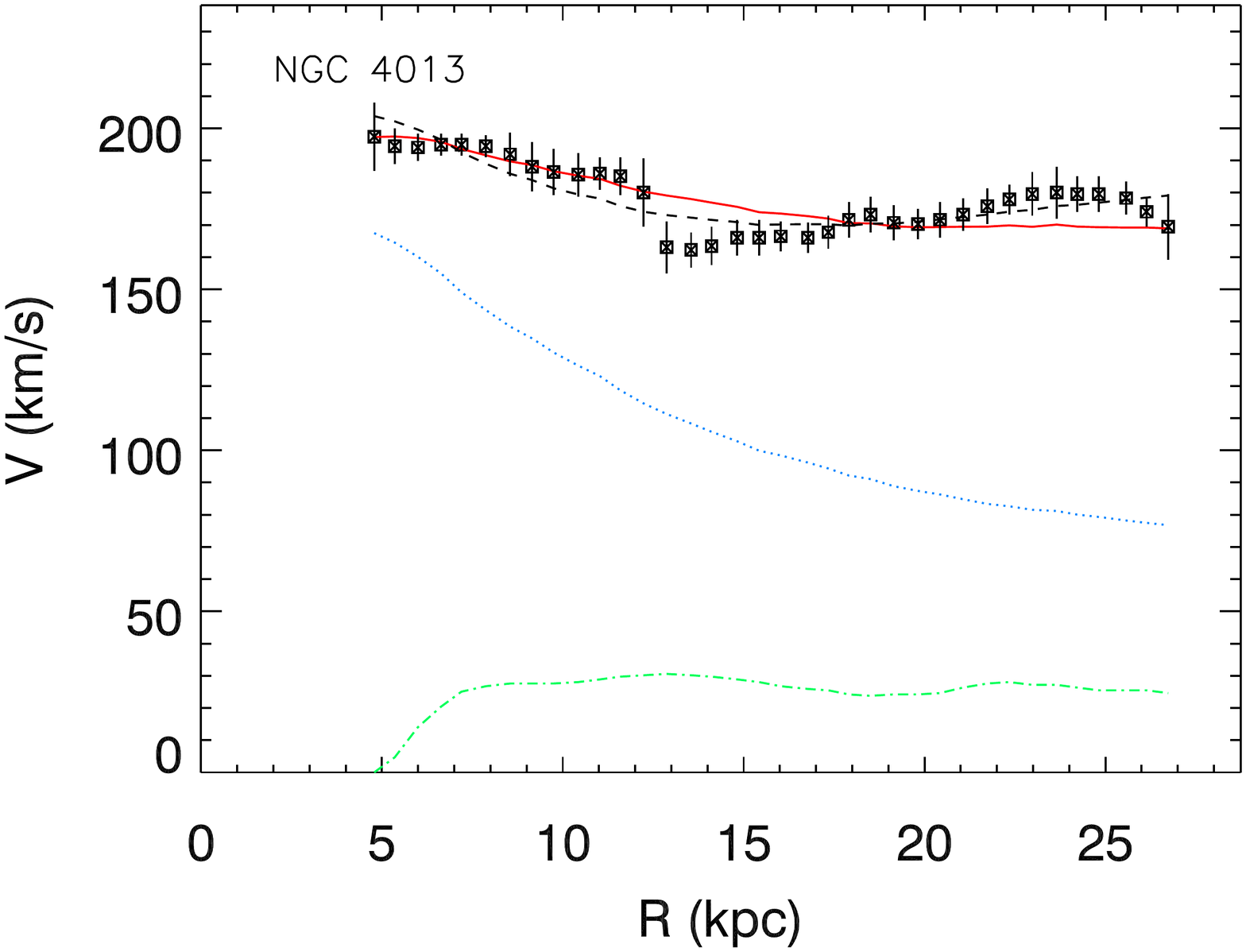}
  \includegraphics[angle=180,width=0.5\textwidth]{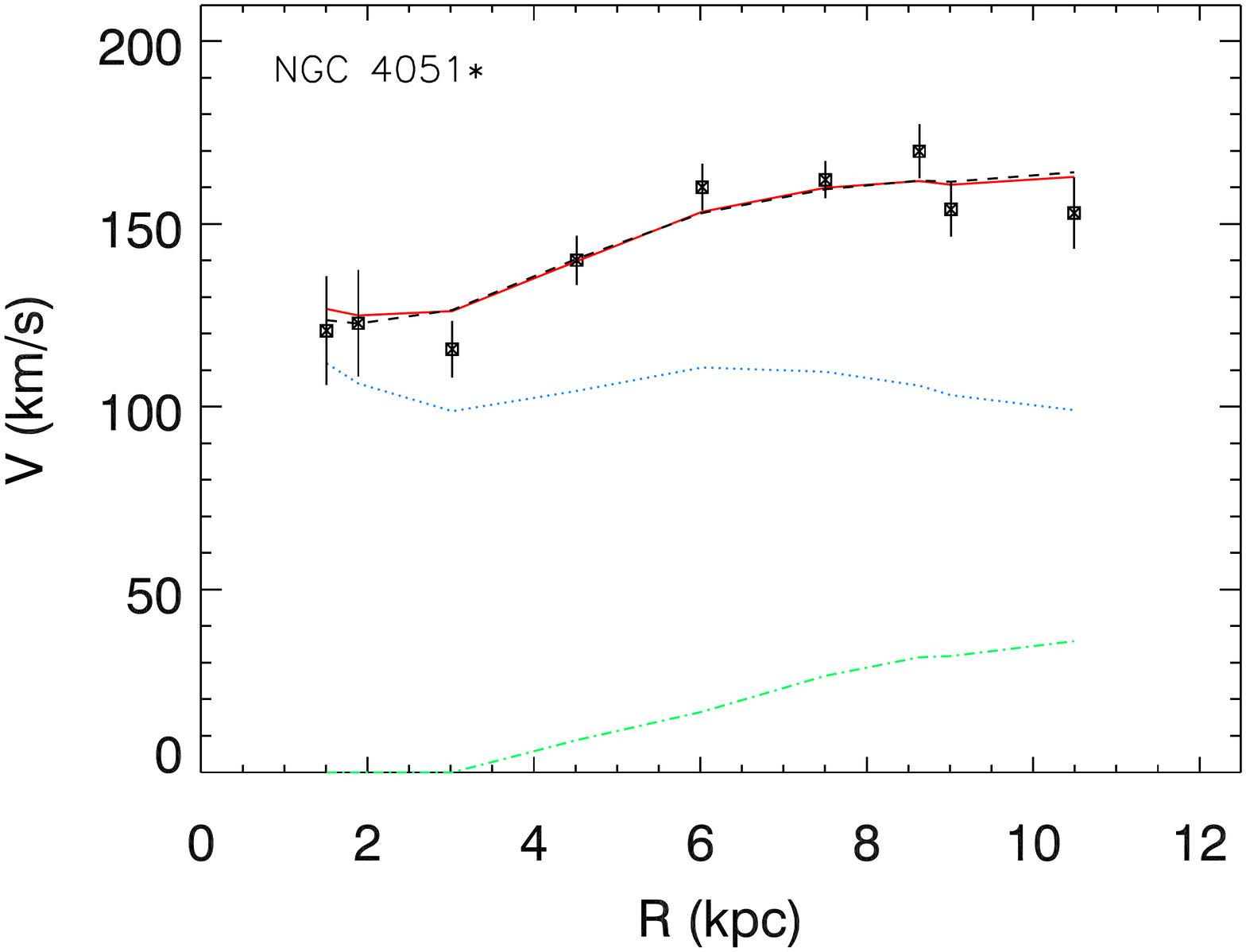}
  \caption{}
\end{figure*}

\addtocounter{figure}{-1}

\begin{figure*}
  \includegraphics[angle=180,width=0.5\textwidth]{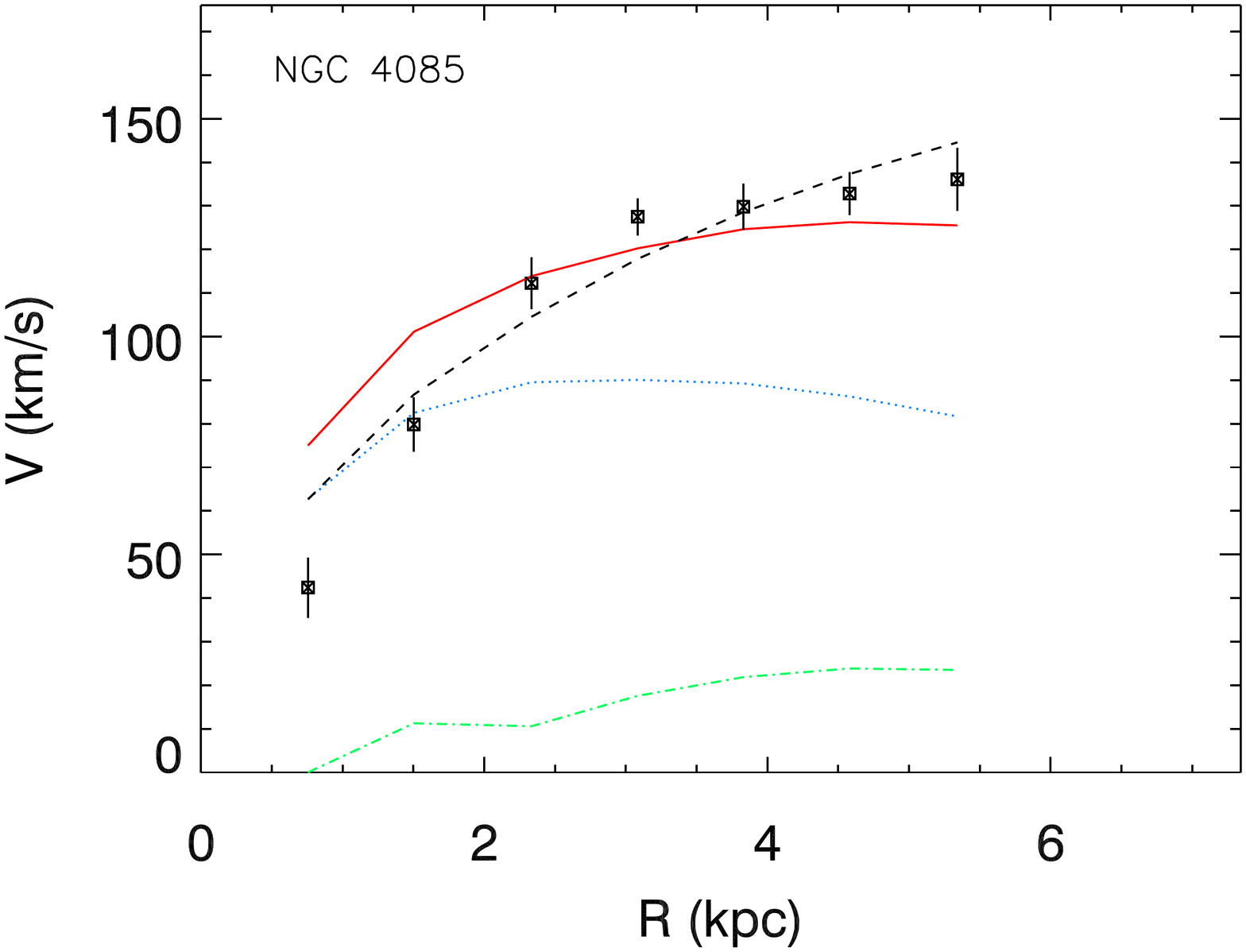}
  \includegraphics[angle=180,width=0.5\textwidth]{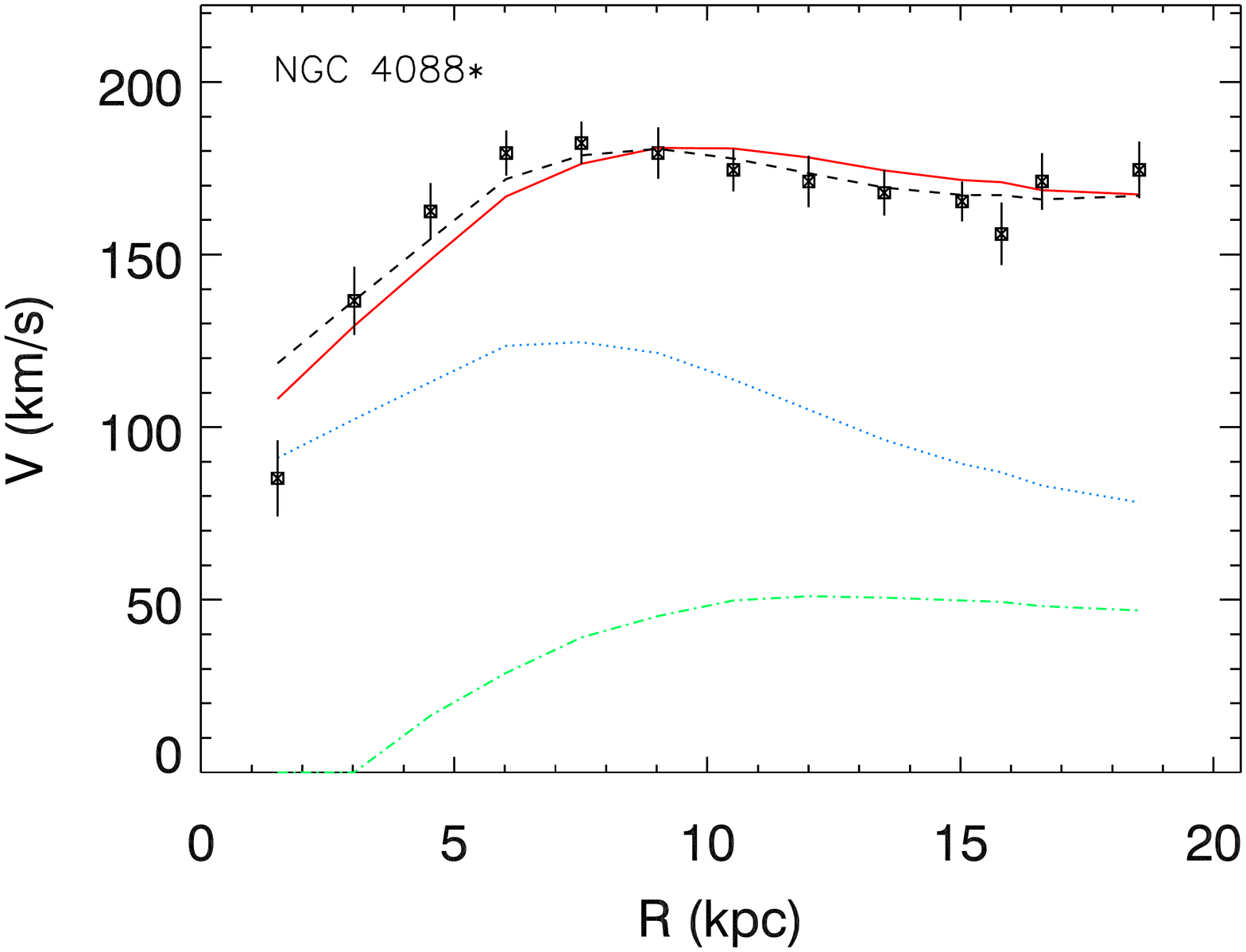}
  \\
  \includegraphics[angle=180,width=0.5\textwidth]{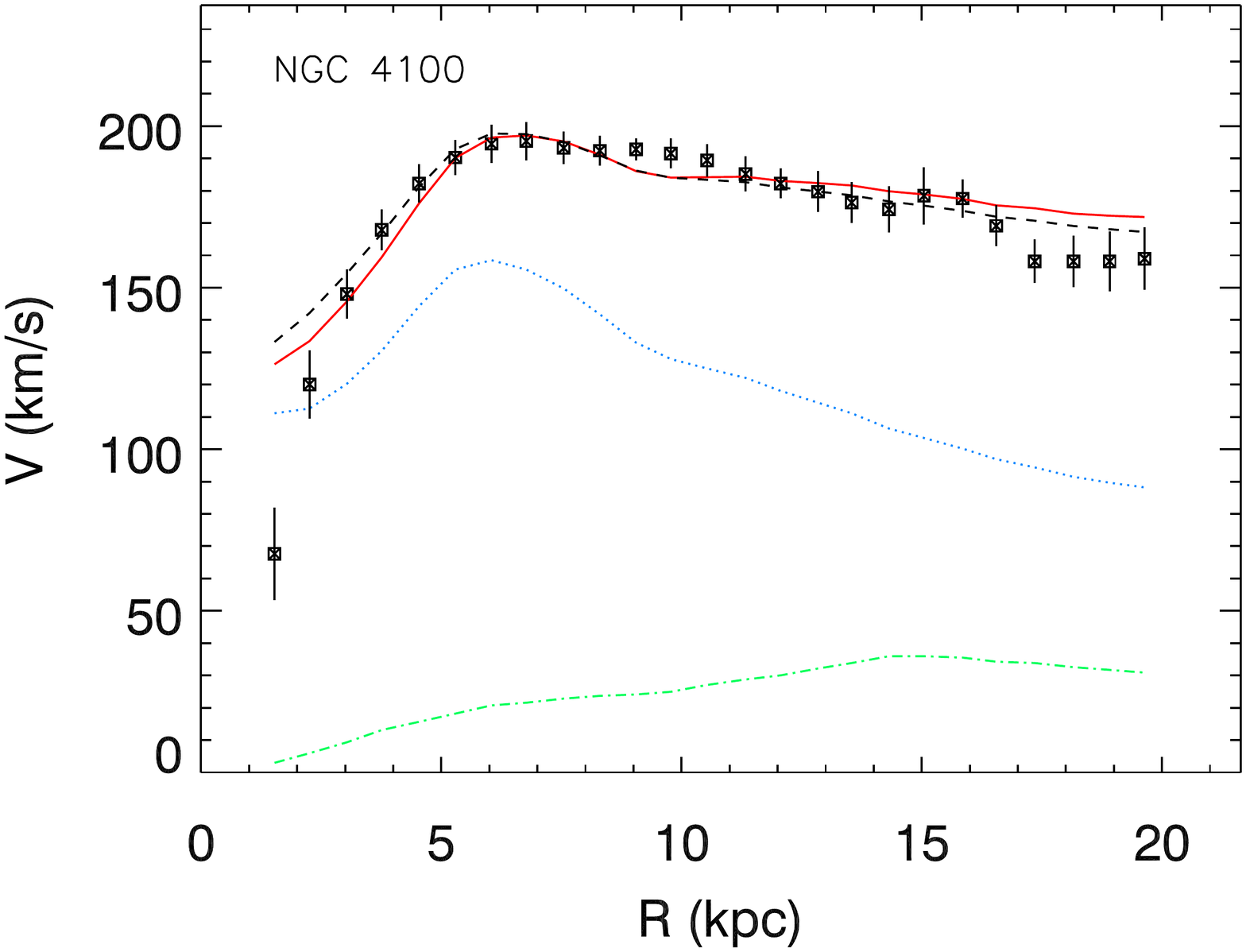}
  \includegraphics[angle=180,width=0.5\textwidth]{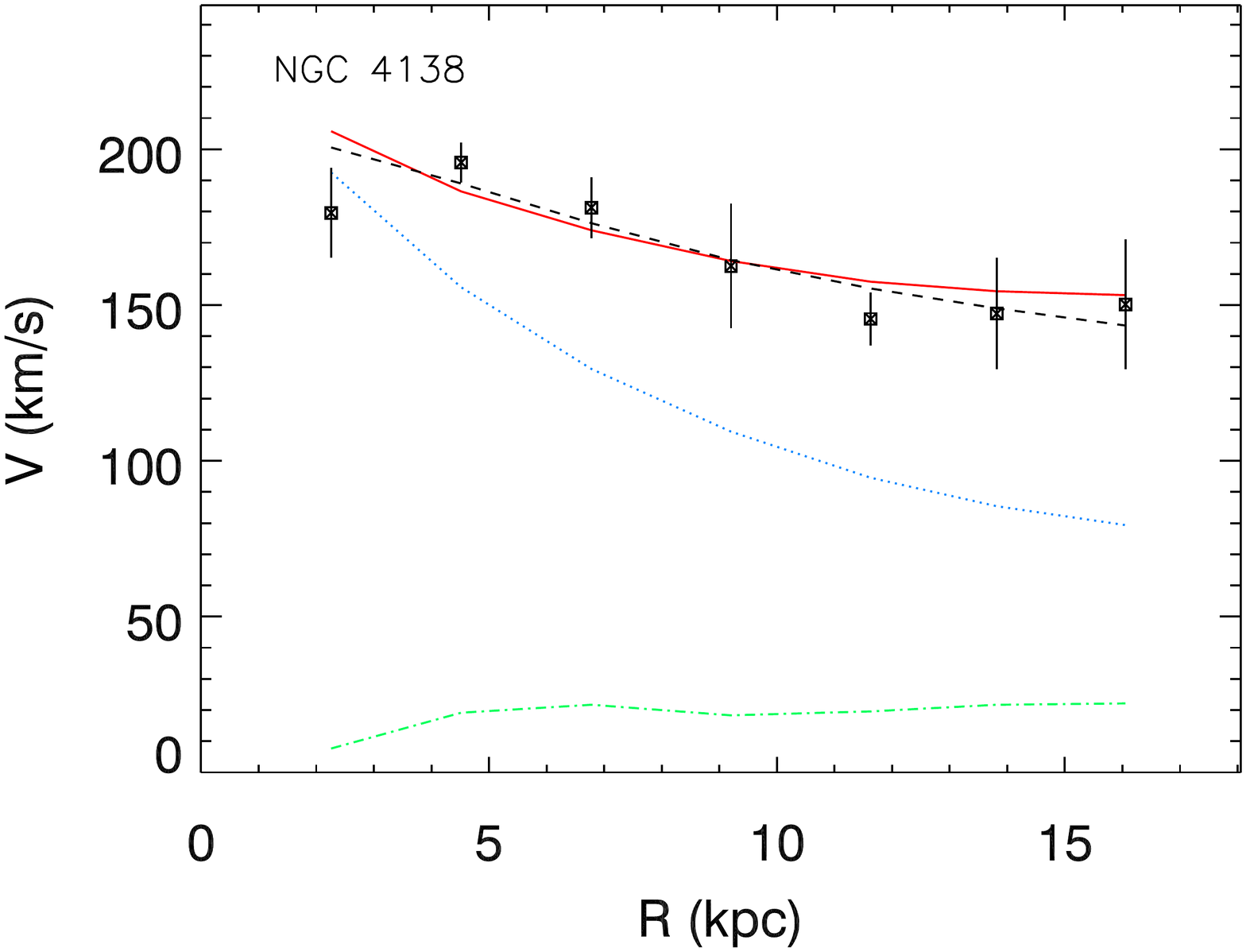}
  \\
  \includegraphics[angle=180,width=0.5\textwidth]{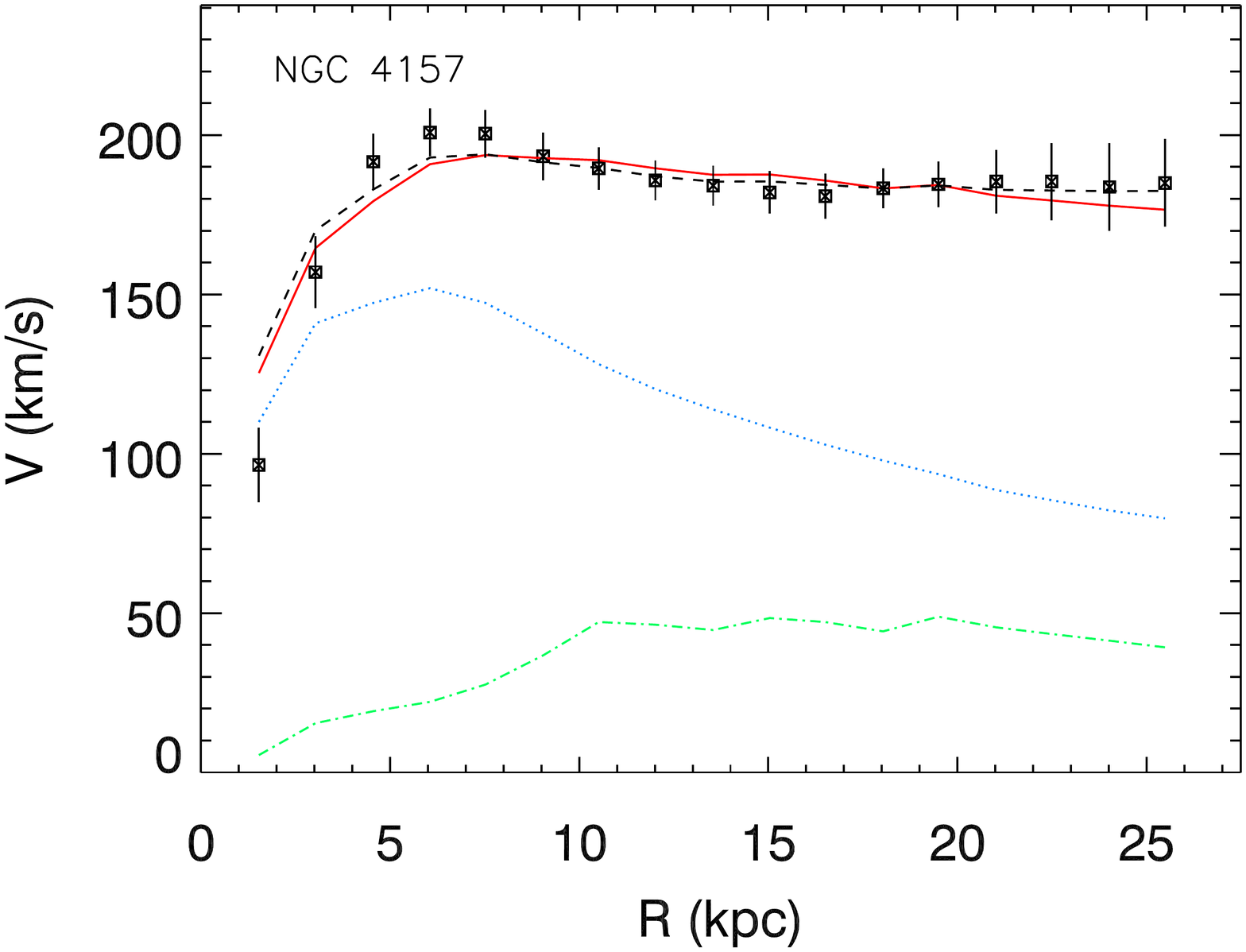}
  \includegraphics[angle=180,width=0.5\textwidth]{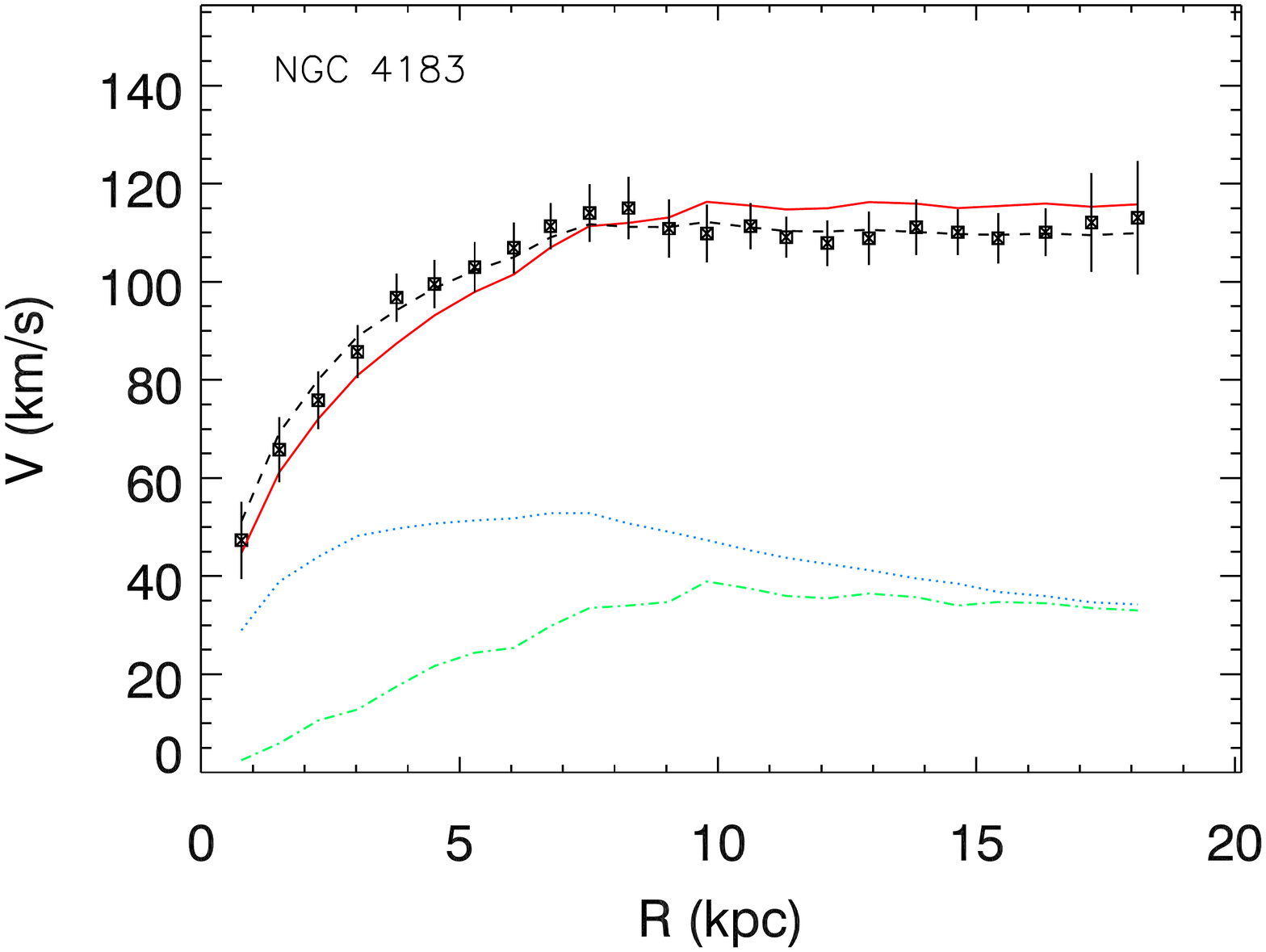}
  \caption{}
\end{figure*}

\addtocounter{figure}{-1}

\begin{figure*}
  \includegraphics[angle=180,width=0.5\textwidth]{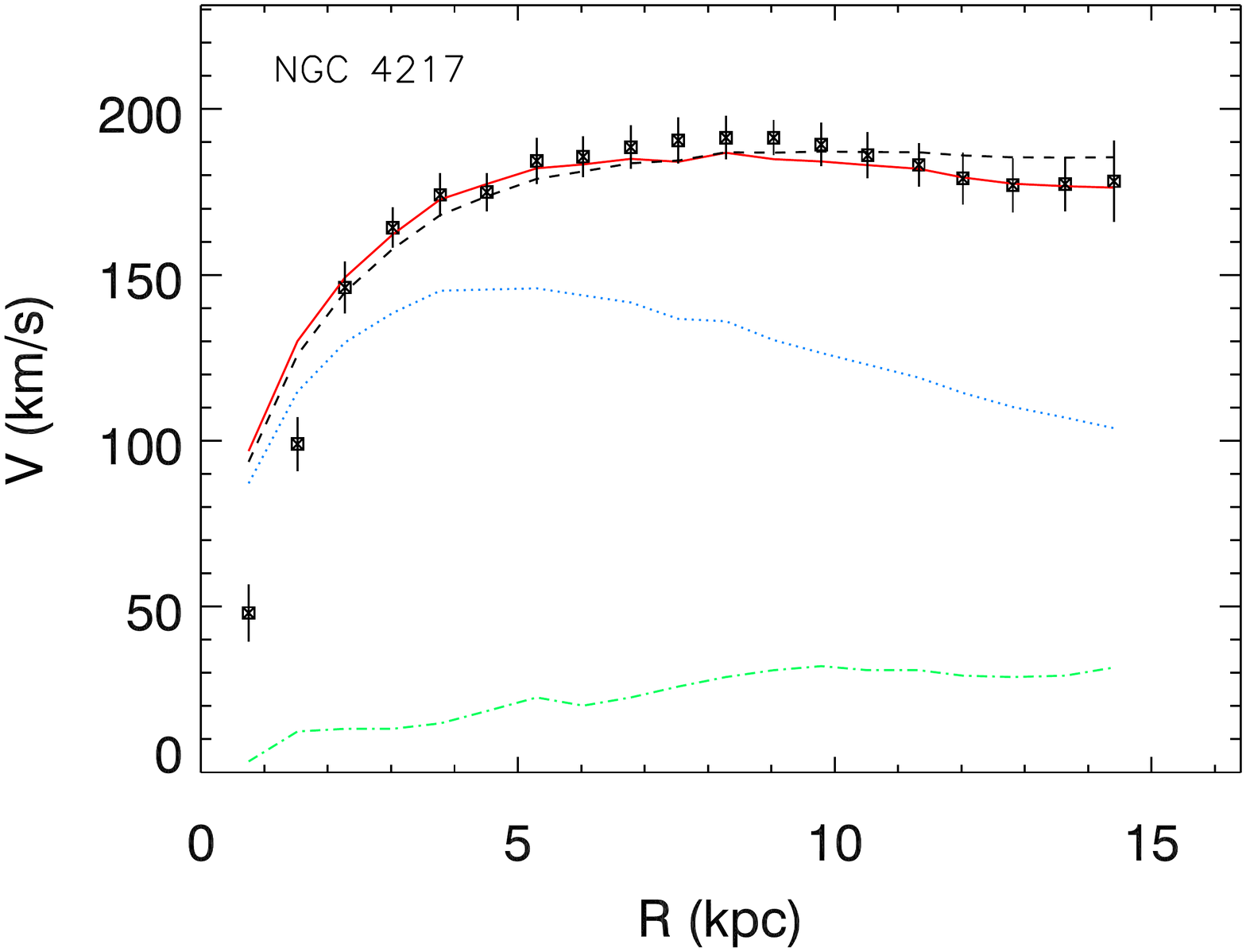}
  \includegraphics[angle=180,width=0.5\textwidth]{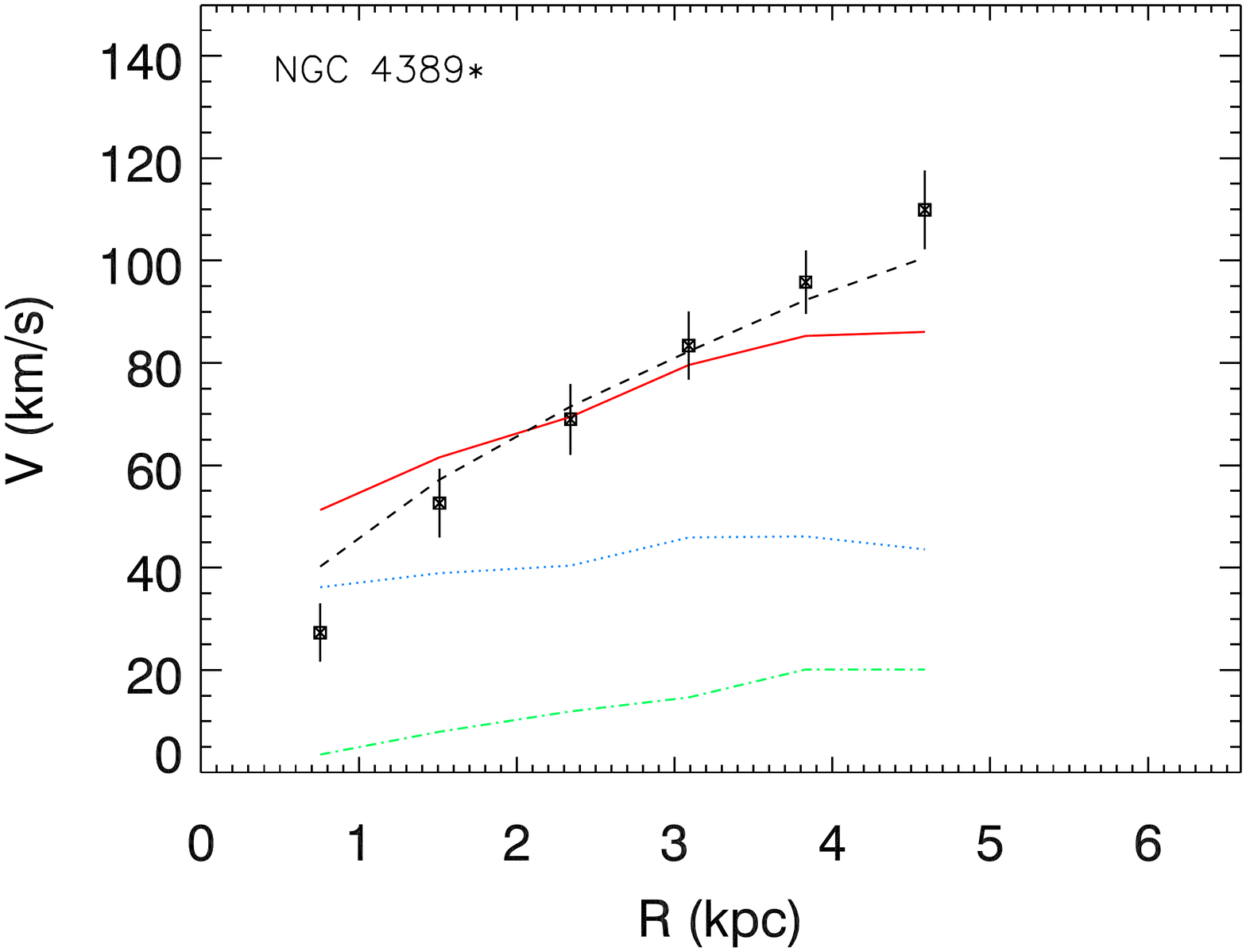}
  \\
  \includegraphics[angle=180,width=0.5\textwidth]{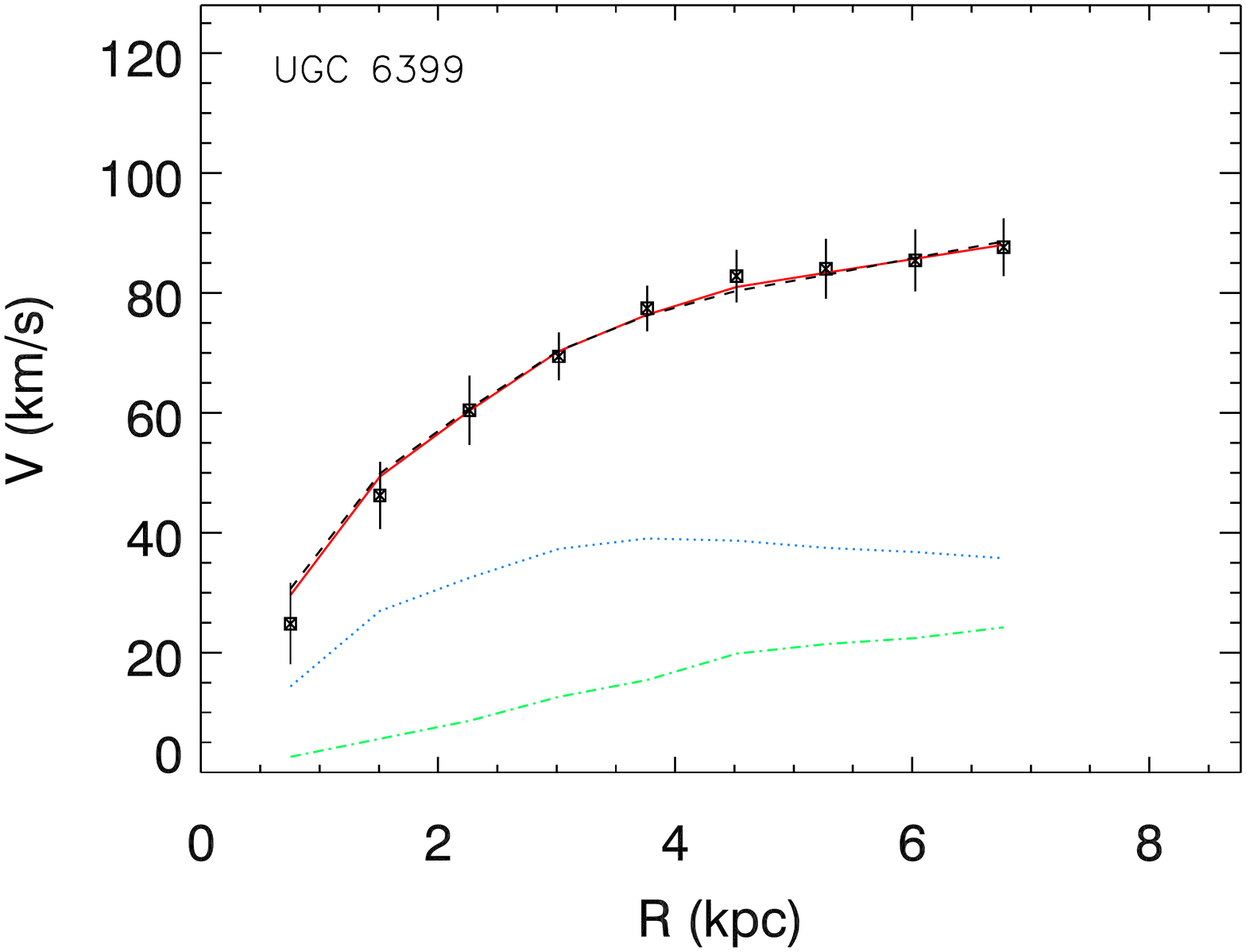}
  \includegraphics[angle=180,width=0.5\textwidth]{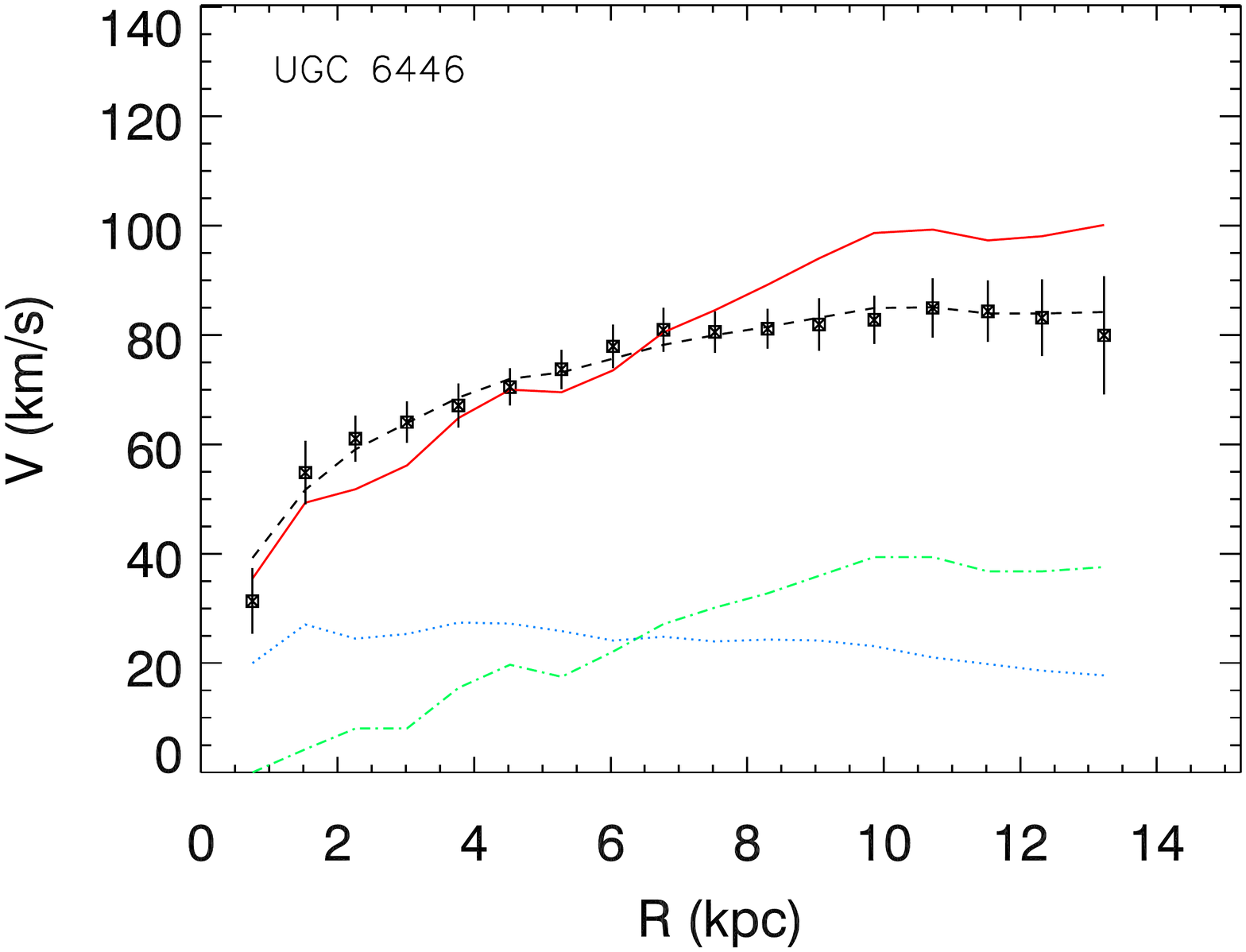}
  \\
  \includegraphics[angle=180,width=0.5\textwidth]{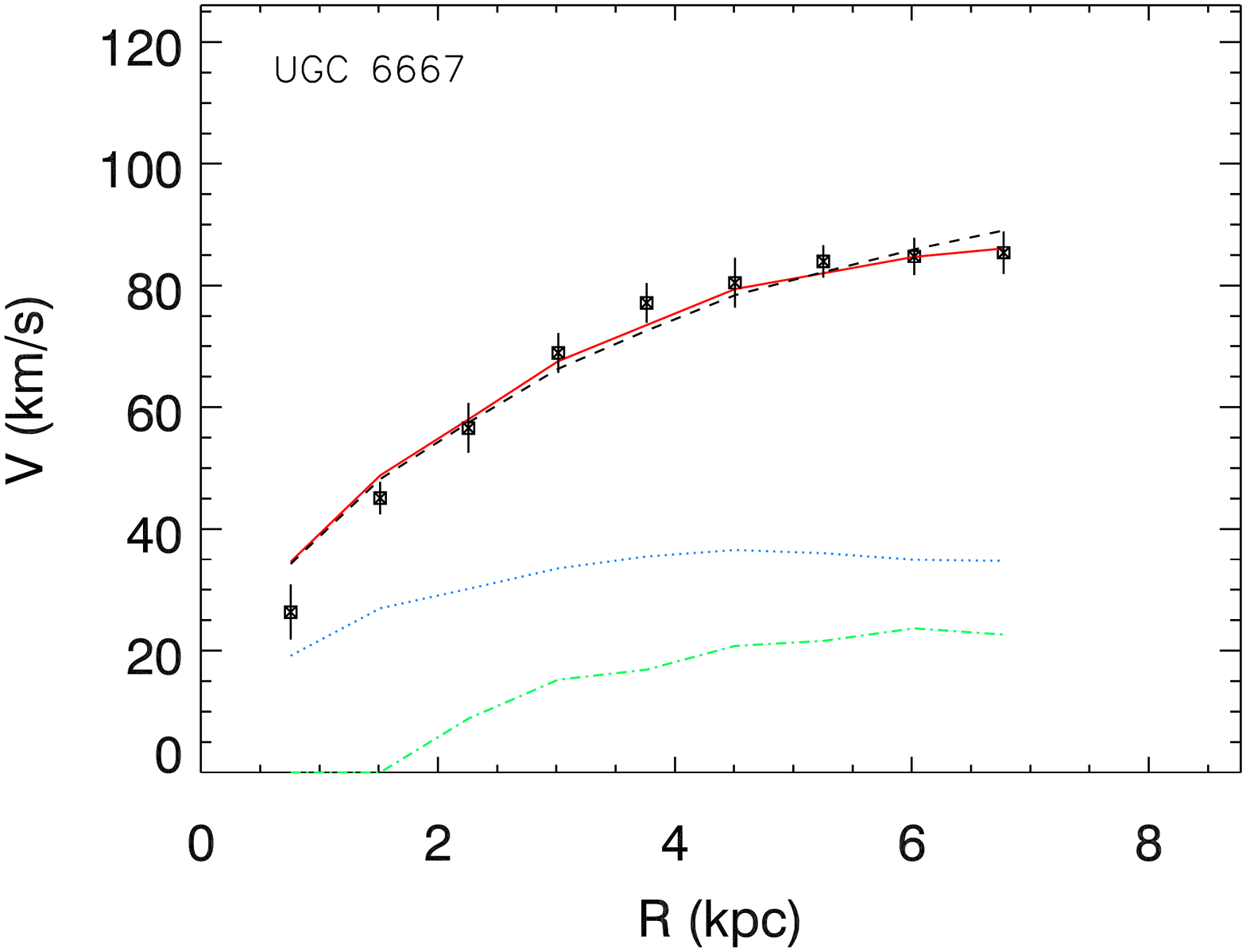}
  \includegraphics[angle=180,width=0.5\textwidth]{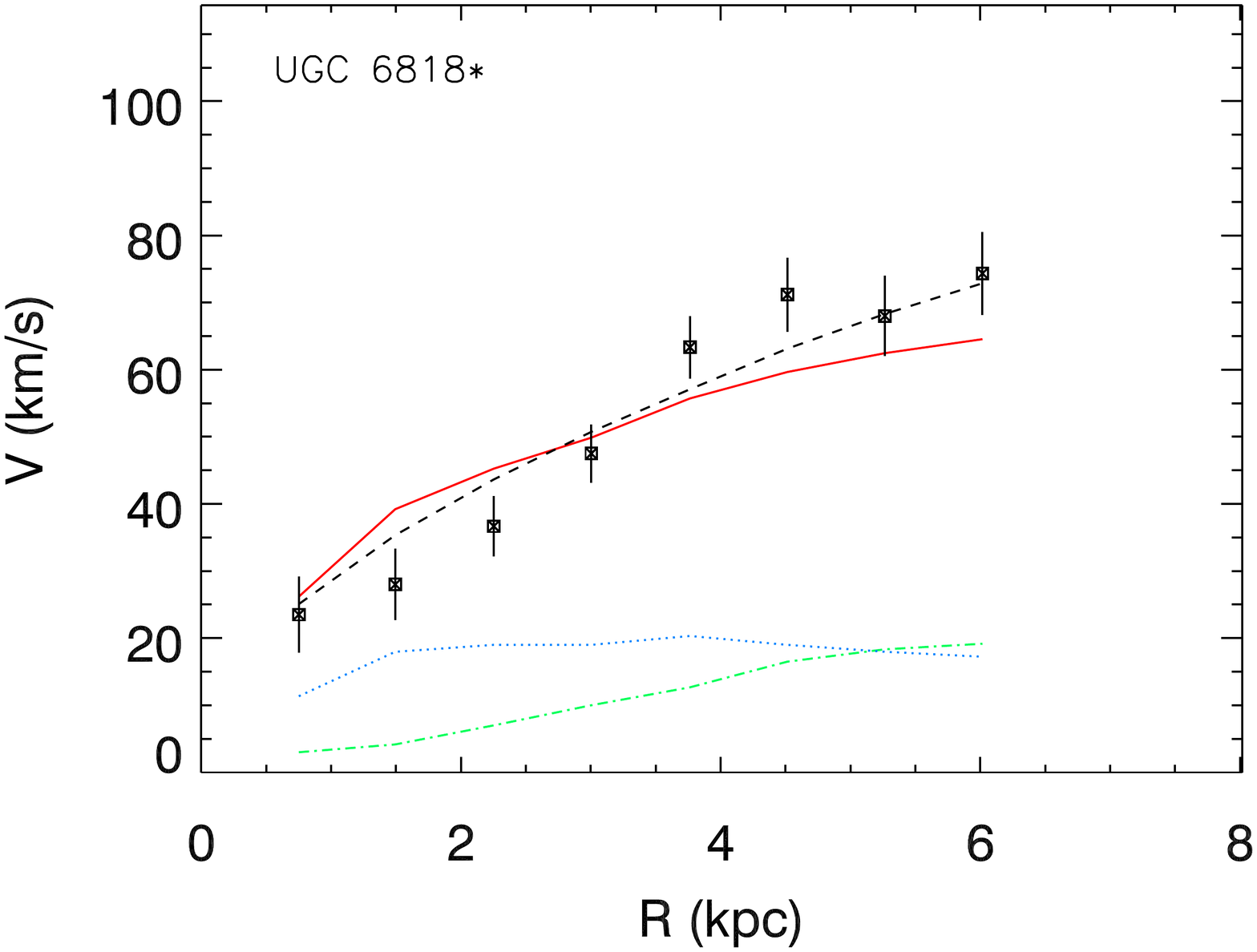}
  \caption{}
\end{figure*}

\addtocounter{figure}{-1}

\begin{figure*}
  \includegraphics[angle=180,width=0.5\textwidth]{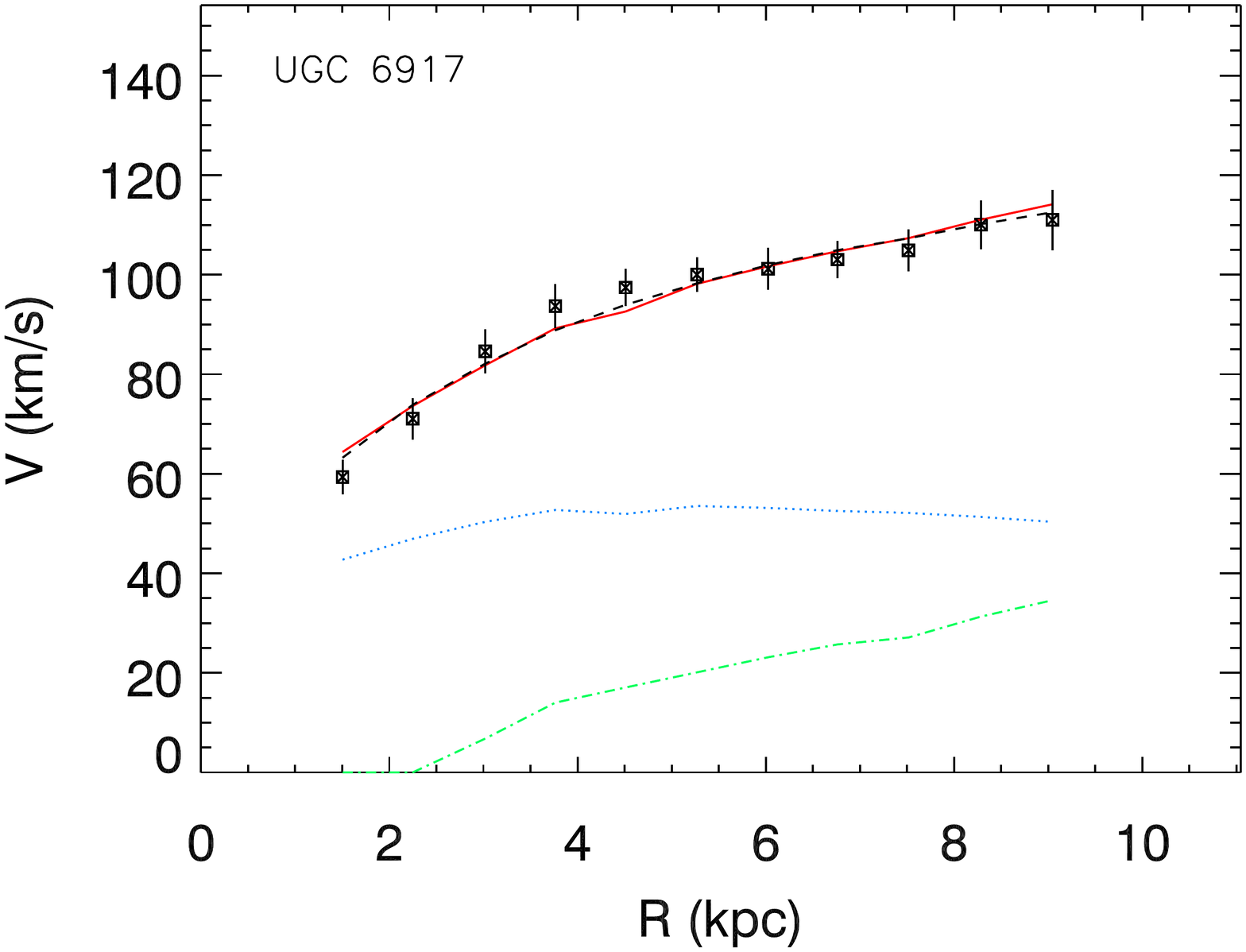}
  \includegraphics[angle=180,width=0.5\textwidth]{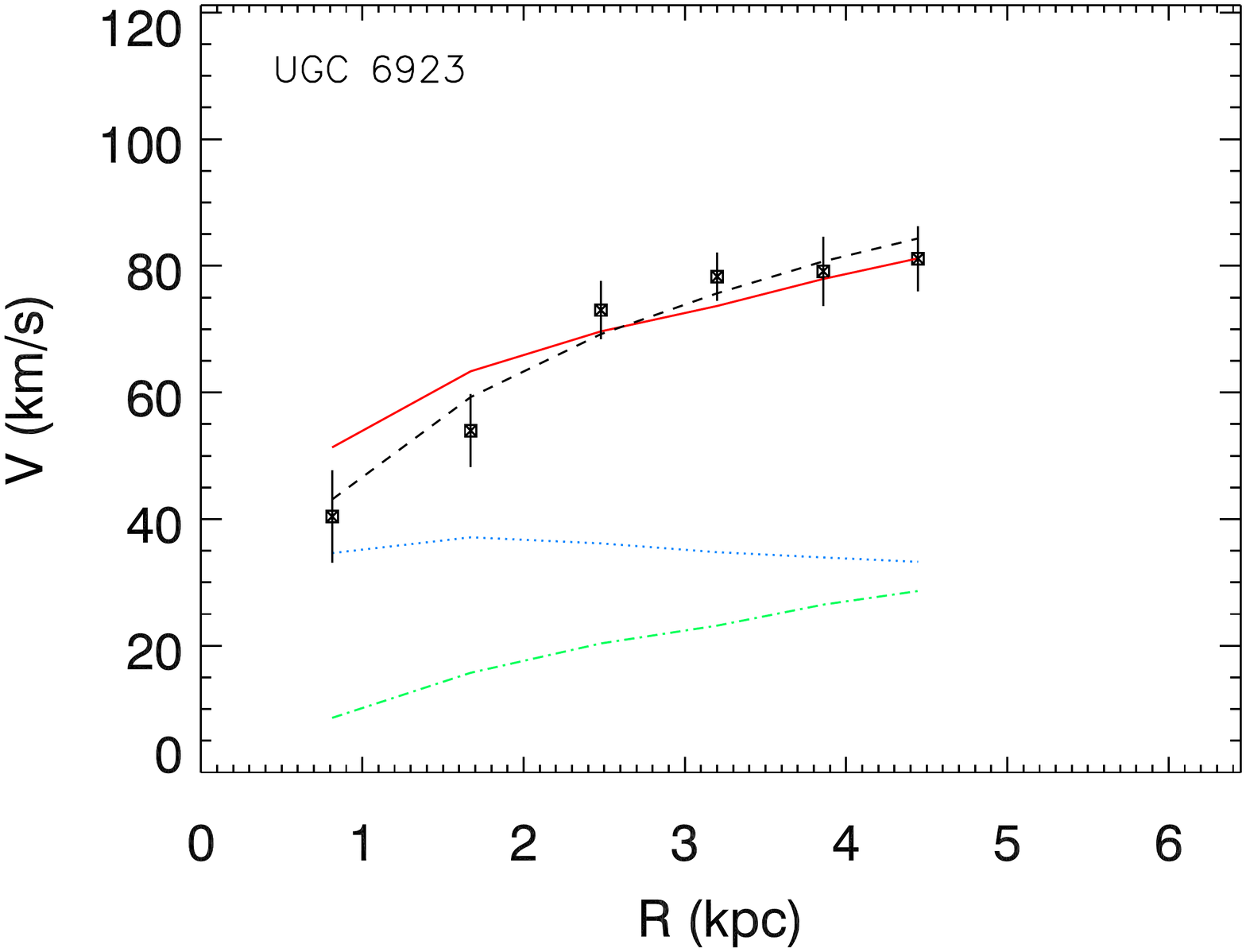}
  \\
  \includegraphics[angle=180,width=0.5\textwidth]{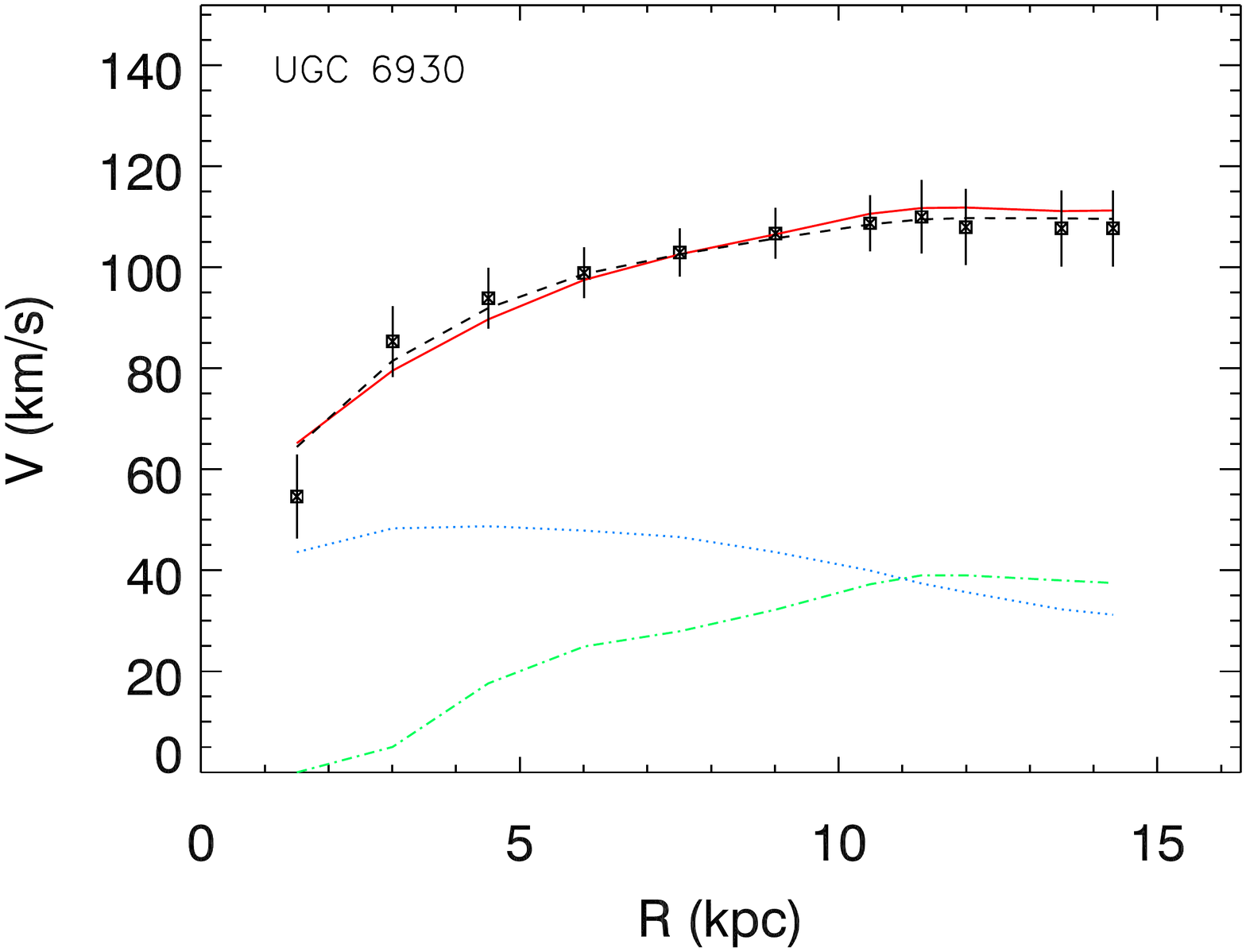}
  \includegraphics[angle=180,width=0.5\textwidth]{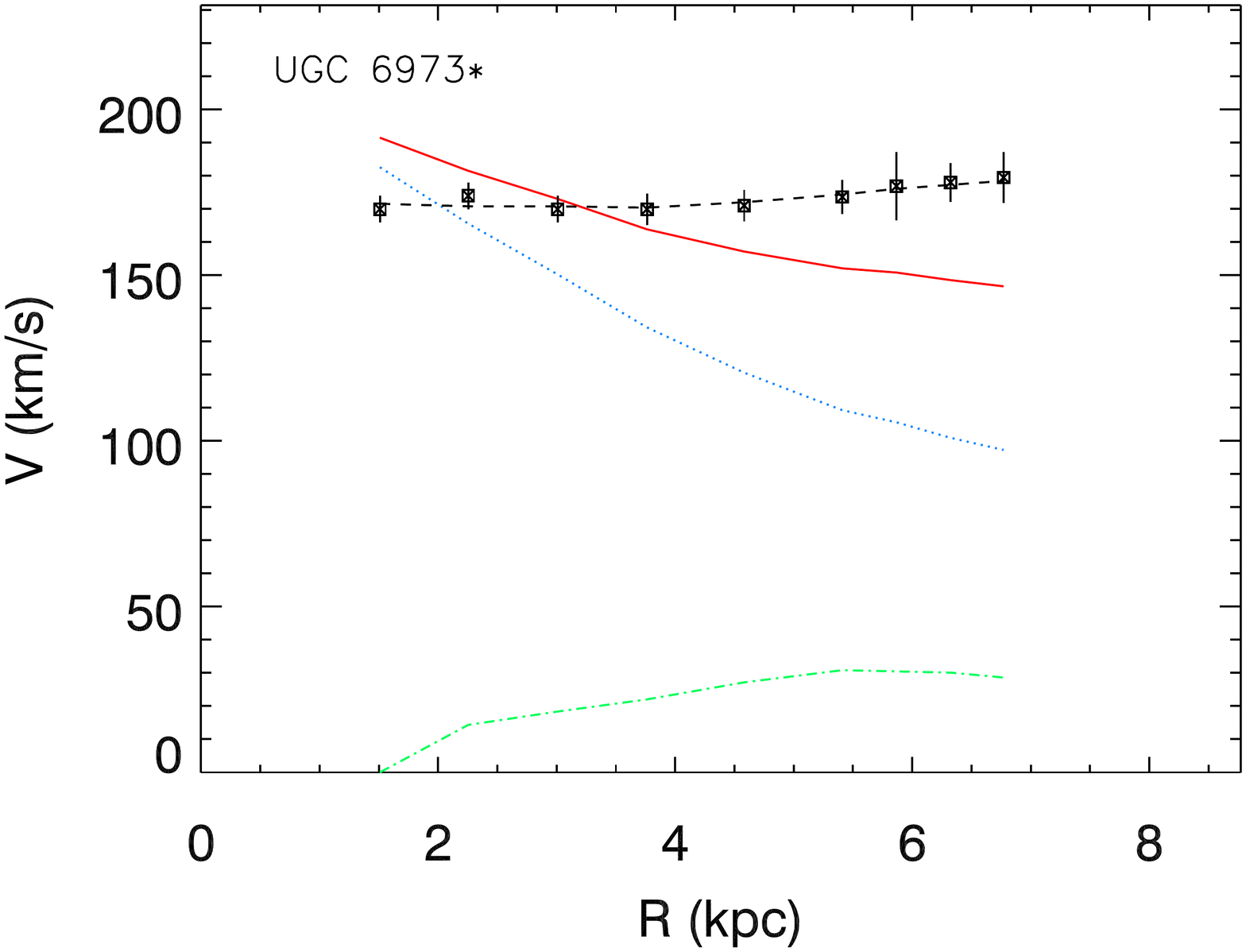}
  \\
  \includegraphics[angle=180,width=0.5\textwidth]{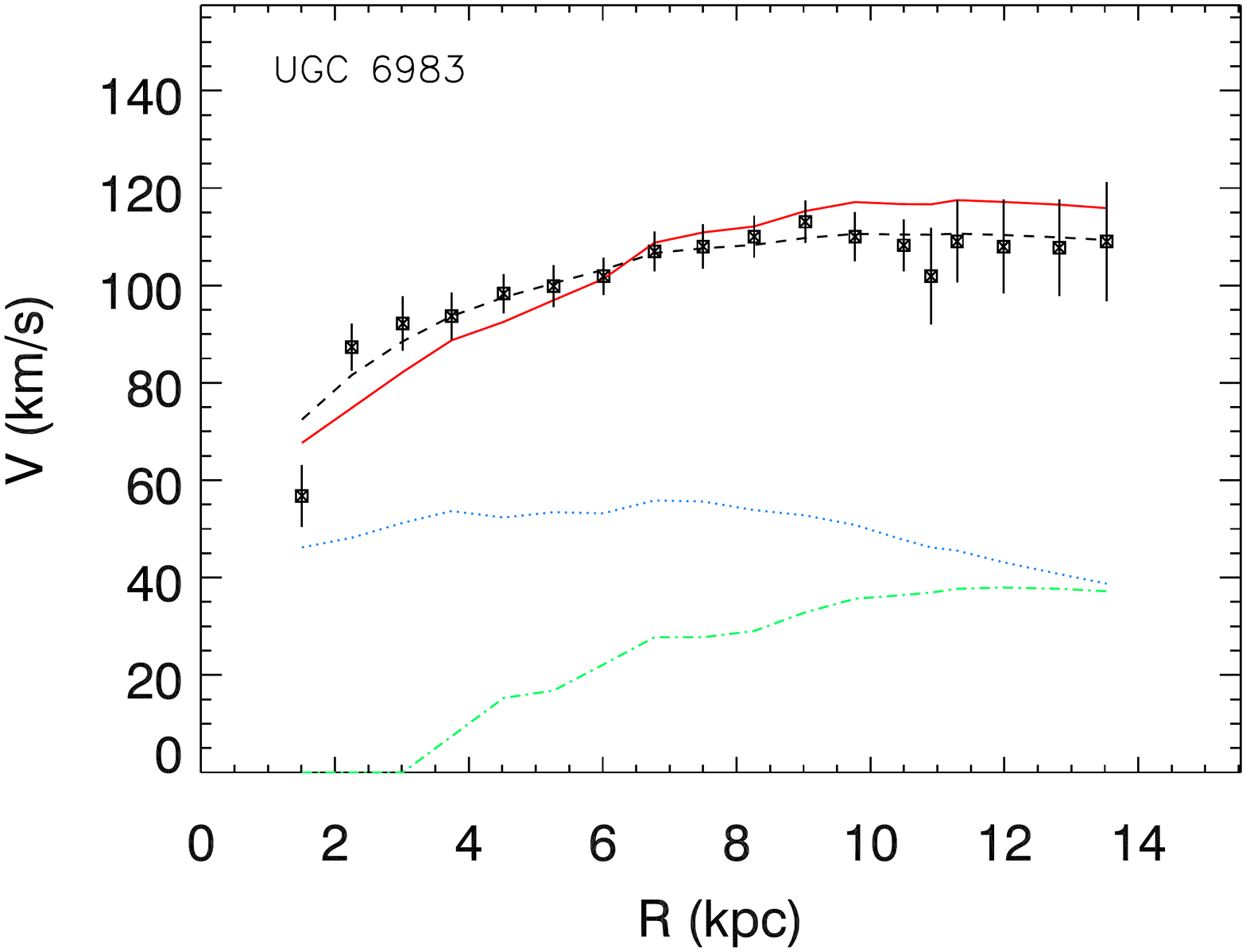}
  \includegraphics[angle=180,width=0.5\textwidth]{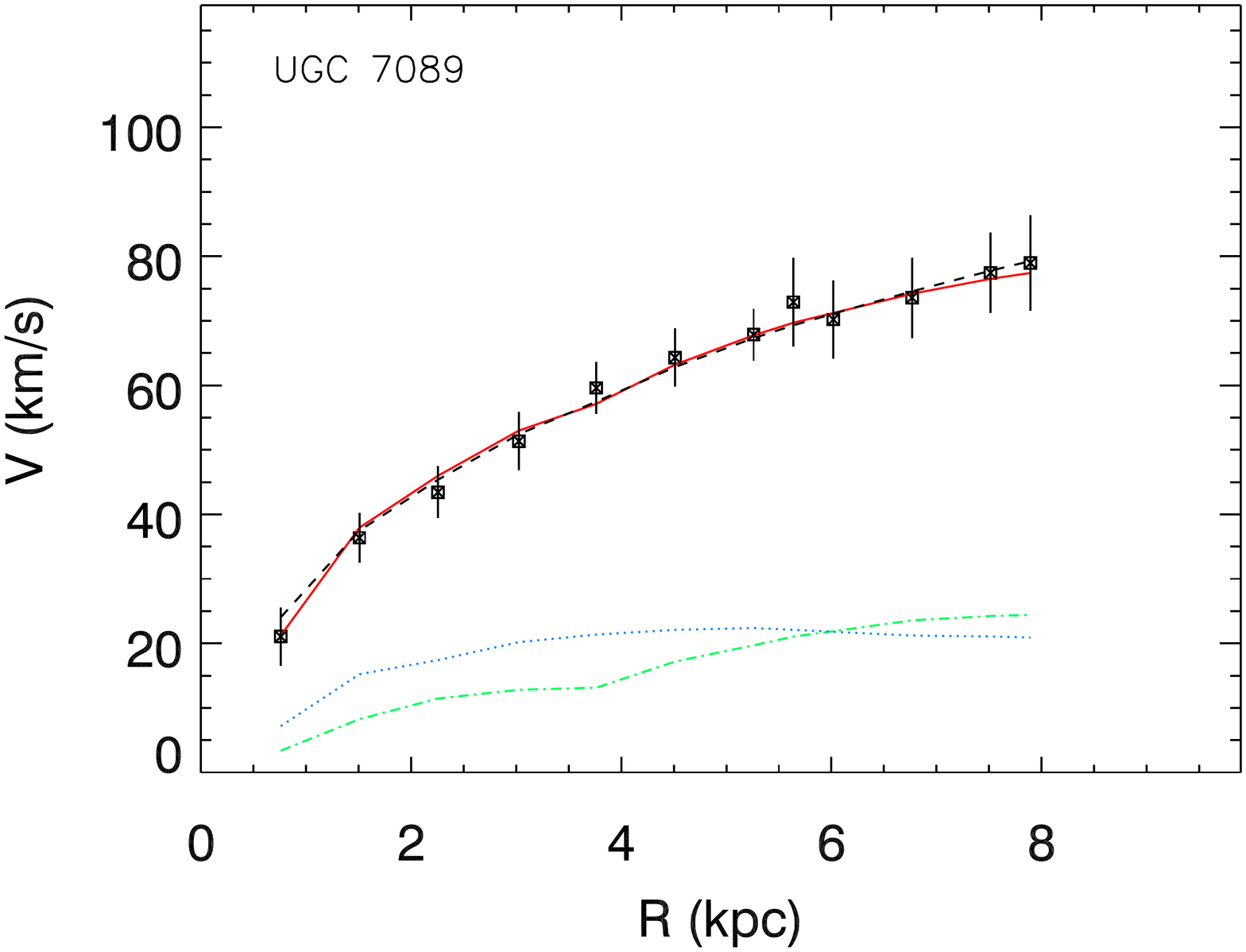}
  \\
  \caption{}
\end{figure*}

\renewcommand{\thefigure}{\arabic{figure}}


\begin{figure}[!ht]
  \includegraphics[angle=90,width=0.48\textwidth]{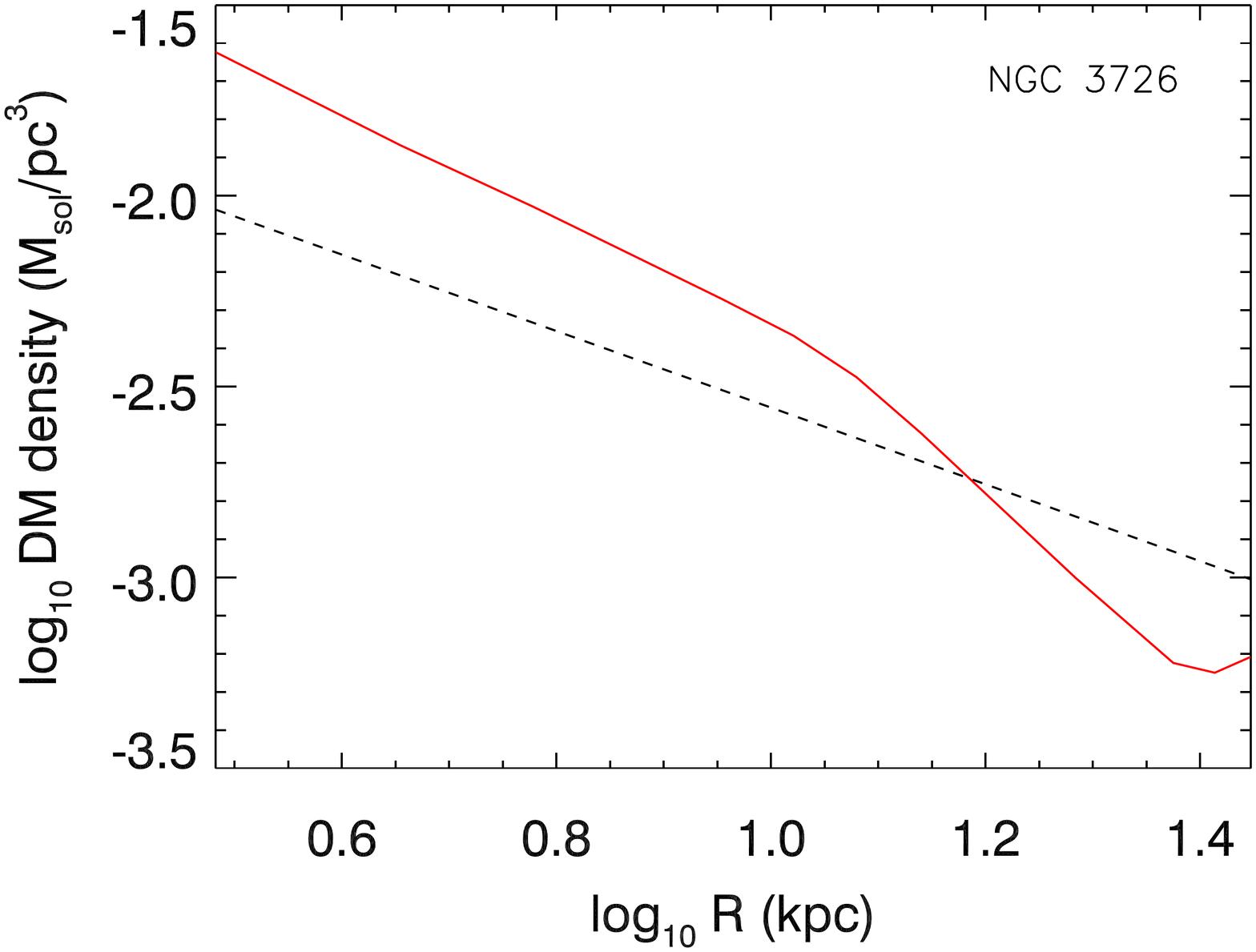}
  \includegraphics[angle=90,width=0.48\textwidth]{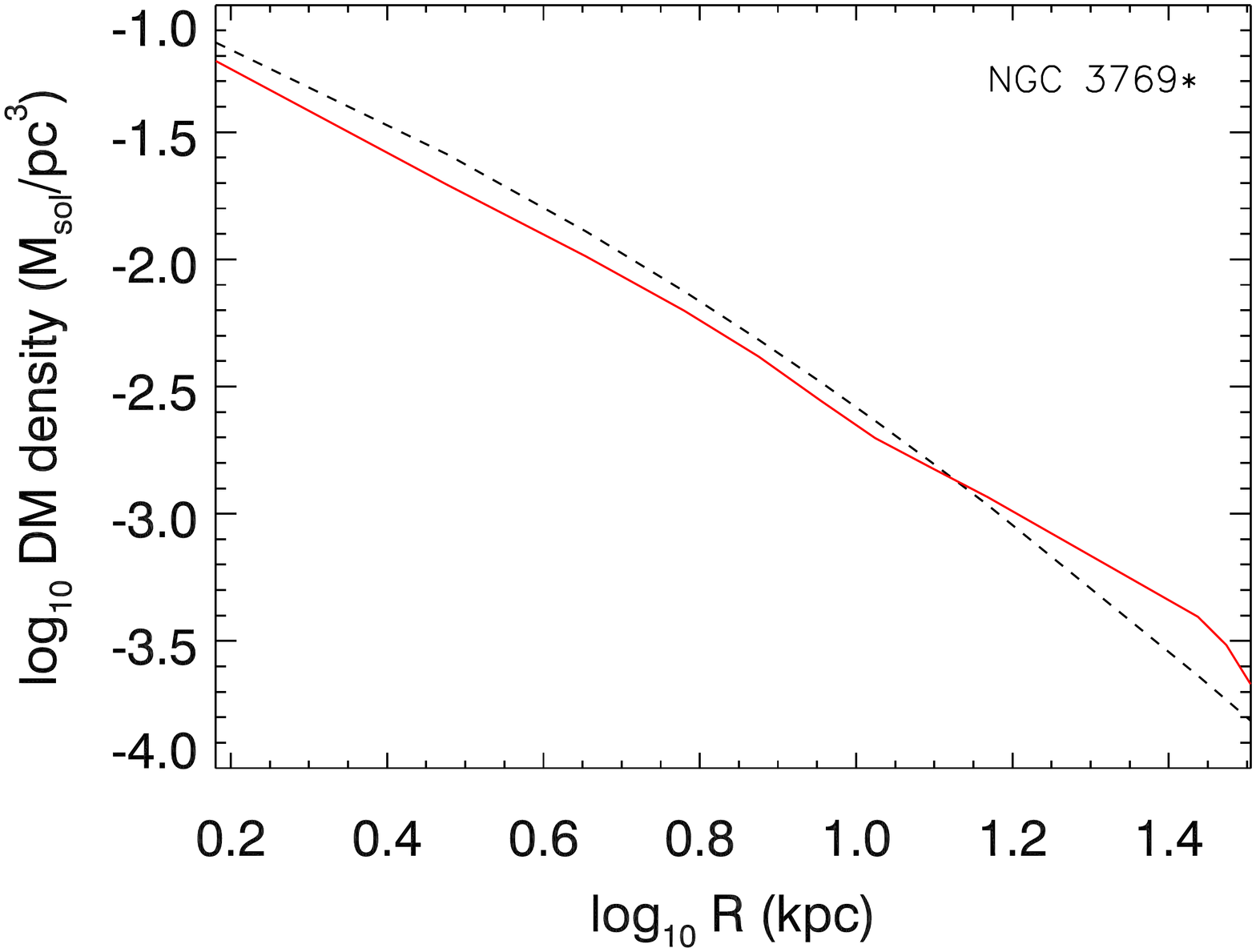}
  \\
  \includegraphics[angle=90,width=0.48\textwidth]{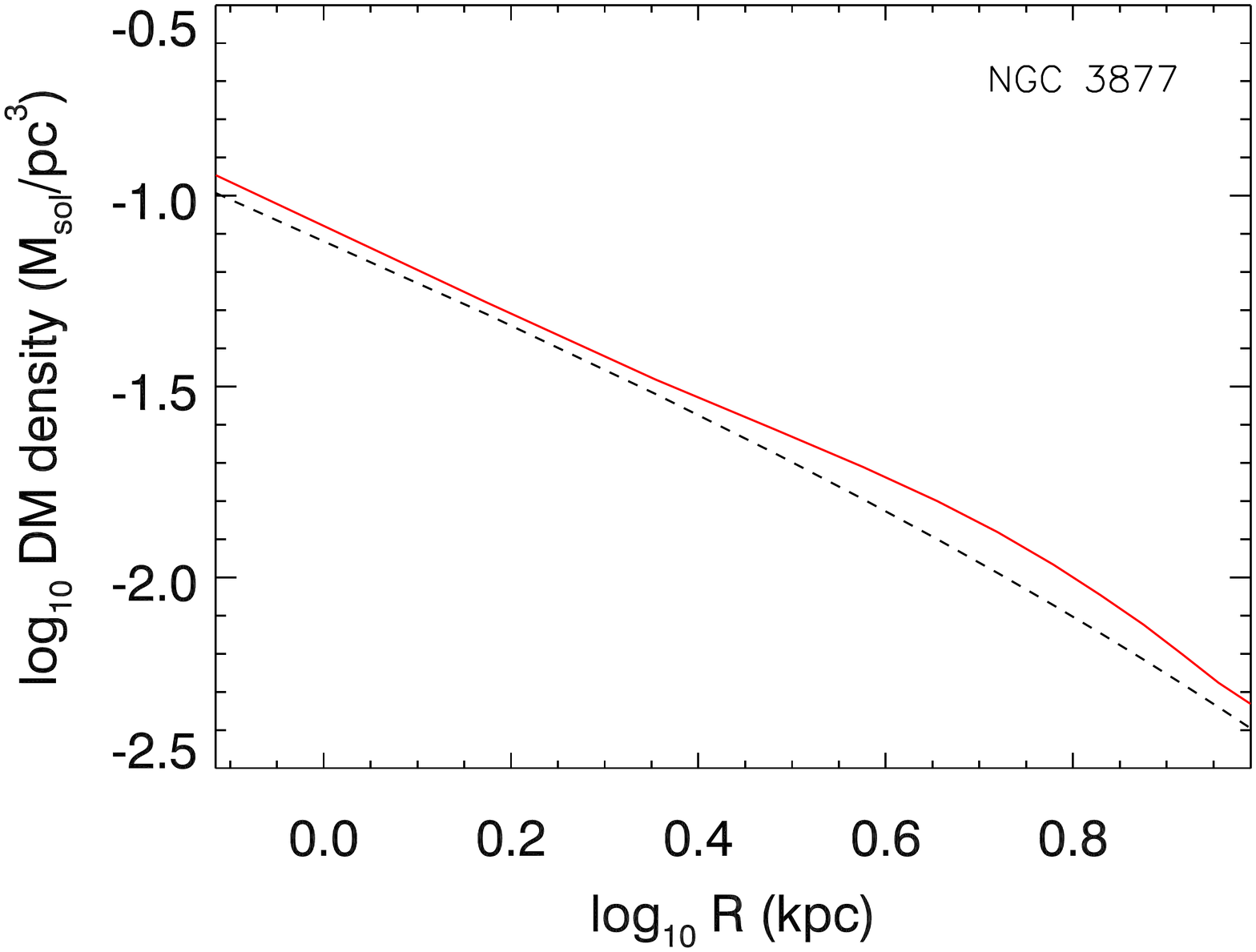}
  \includegraphics[angle=90,width=0.48\textwidth]{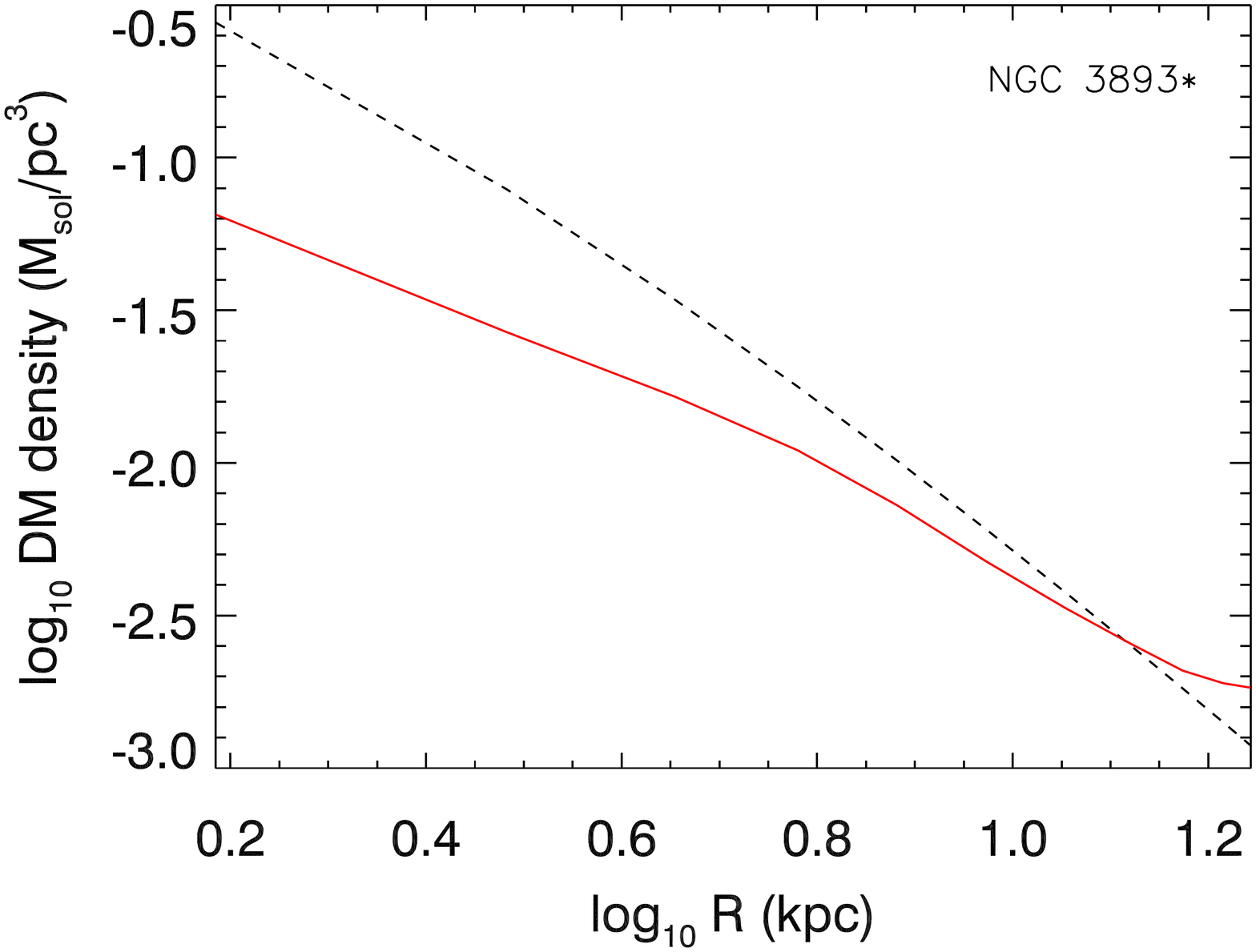}
  \\
  \includegraphics[angle=90,width=0.48\textwidth]{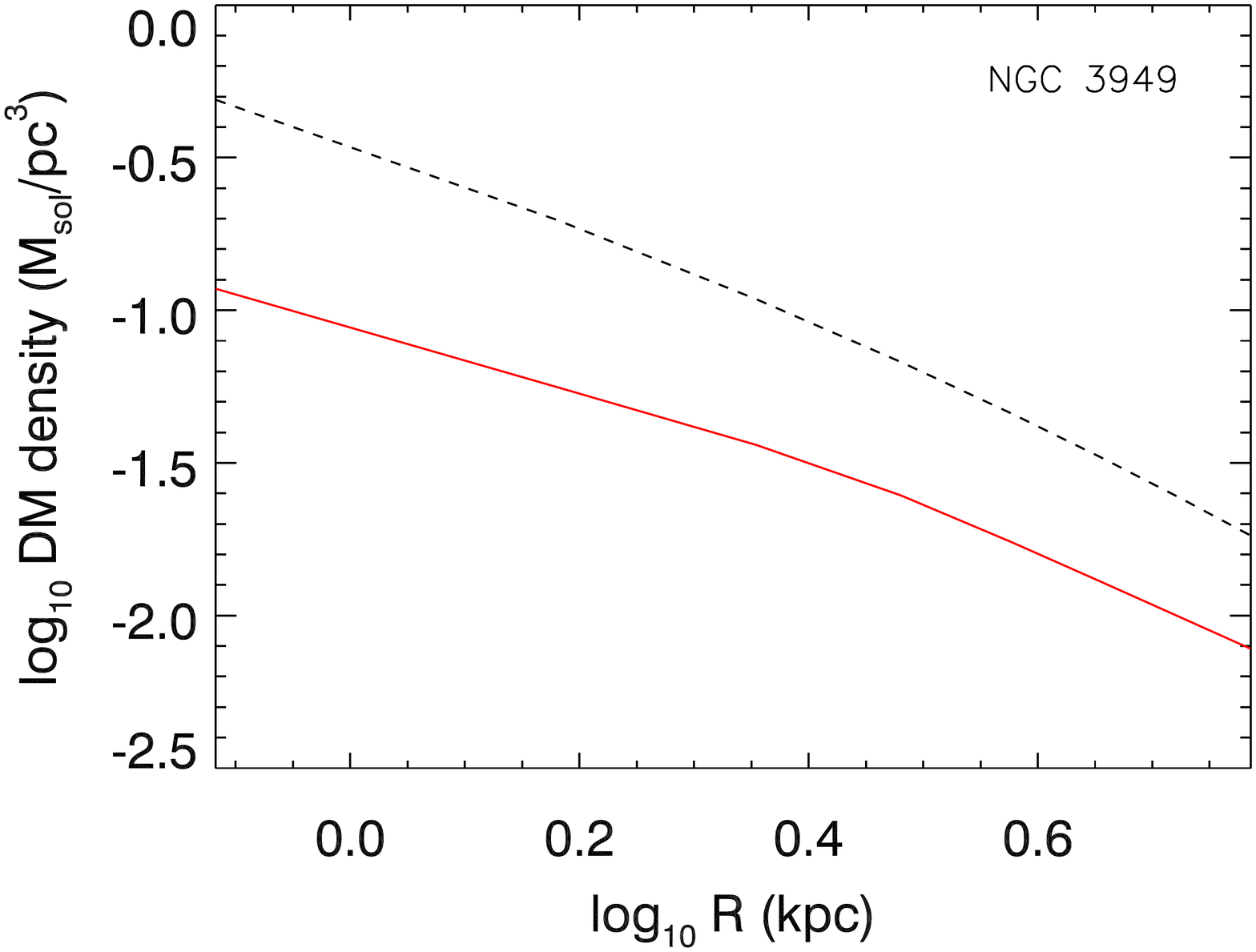}
  \includegraphics[angle=90,width=0.48\textwidth]{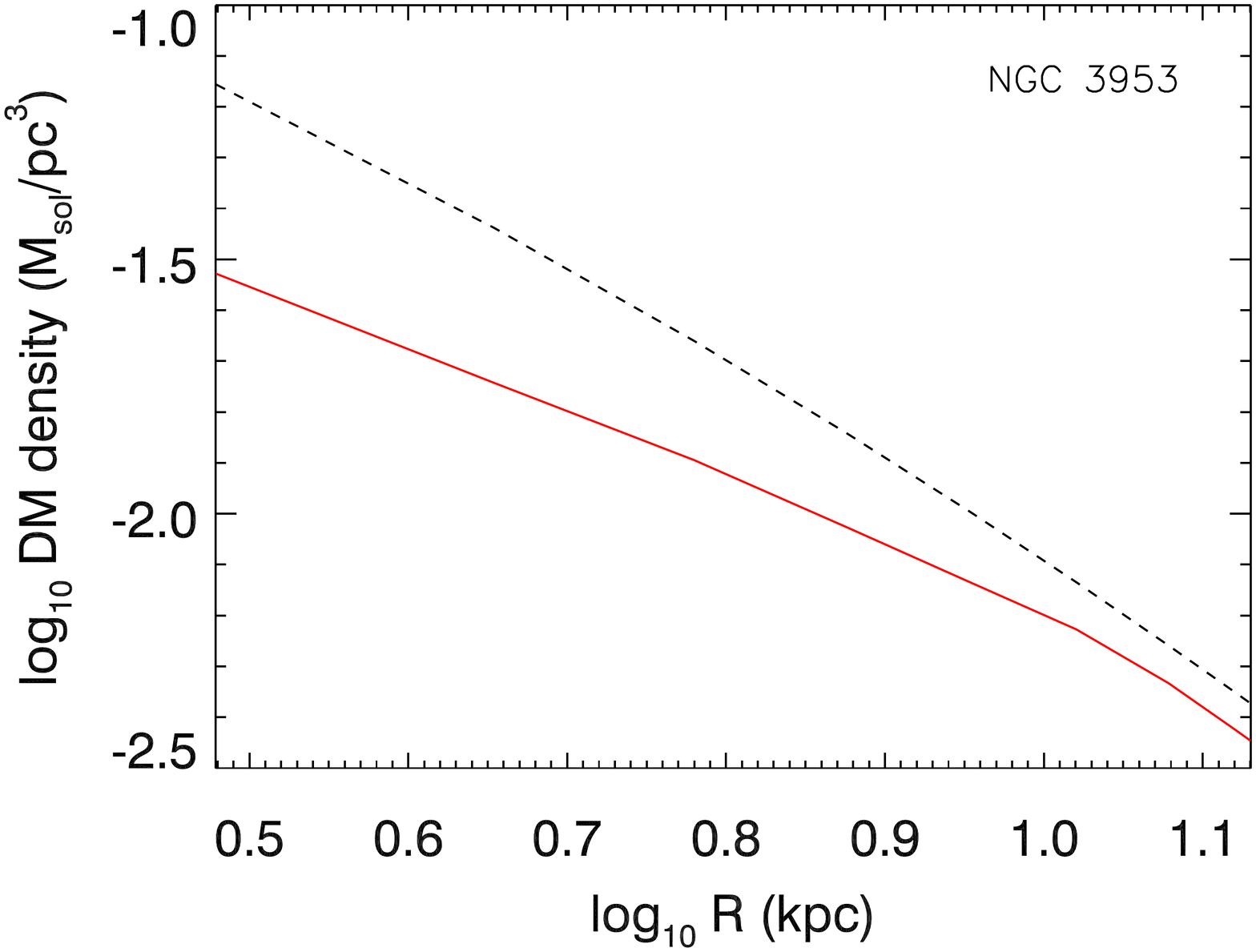}
  \caption{Dark matter density profiles for our sample of galaxies. Density profiles for MDM and CDM are depicted as solid and dashed lines, respectively.}
  \label{fig:hsb_dens}
\end{figure}

\renewcommand{\thefigure}{\arabic{figure} (Cont.)}
\addtocounter{figure}{-1}

\begin{figure}[!ht]
  \includegraphics[angle=90,width=0.48\textwidth]{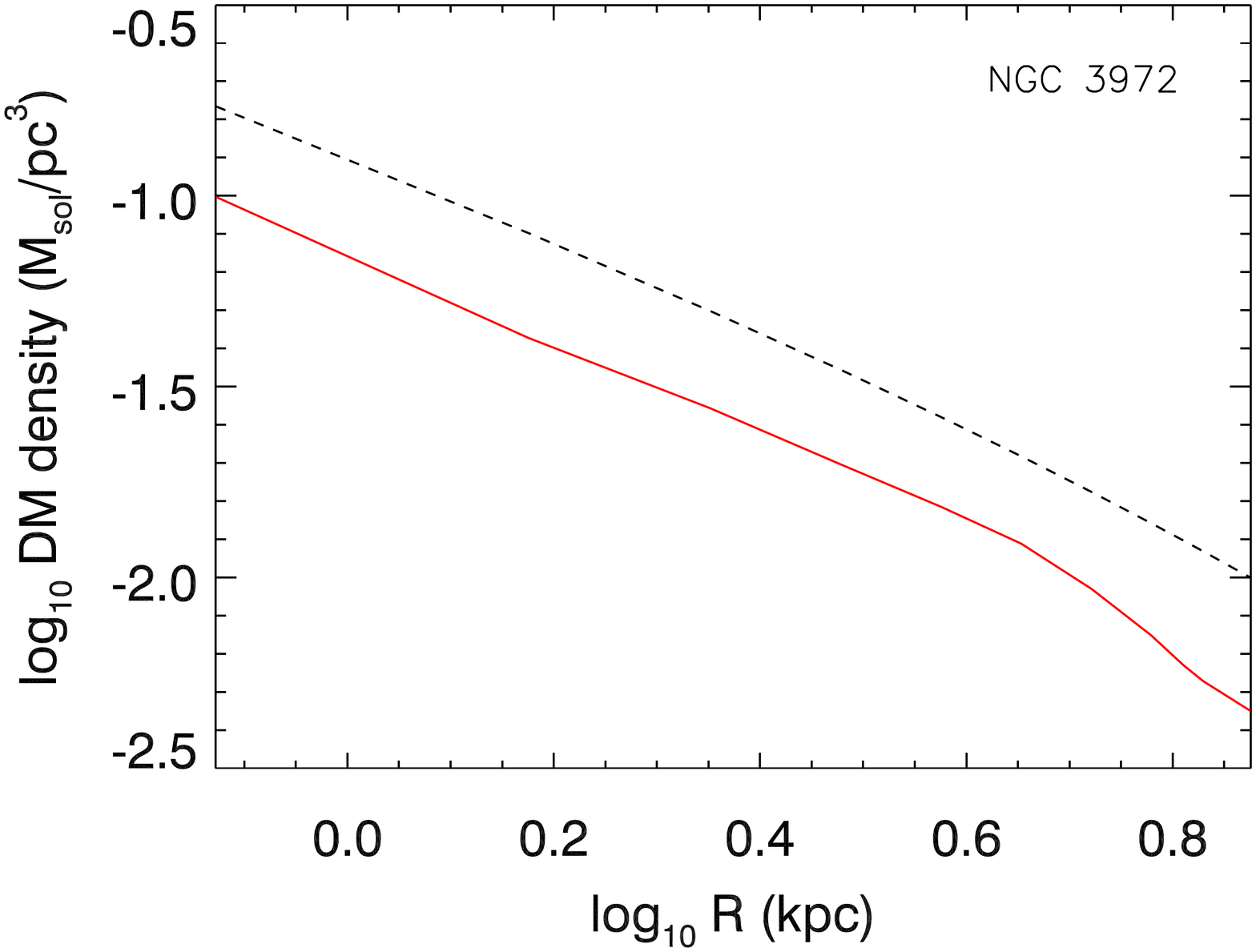}
  \includegraphics[angle=90,width=0.48\textwidth]{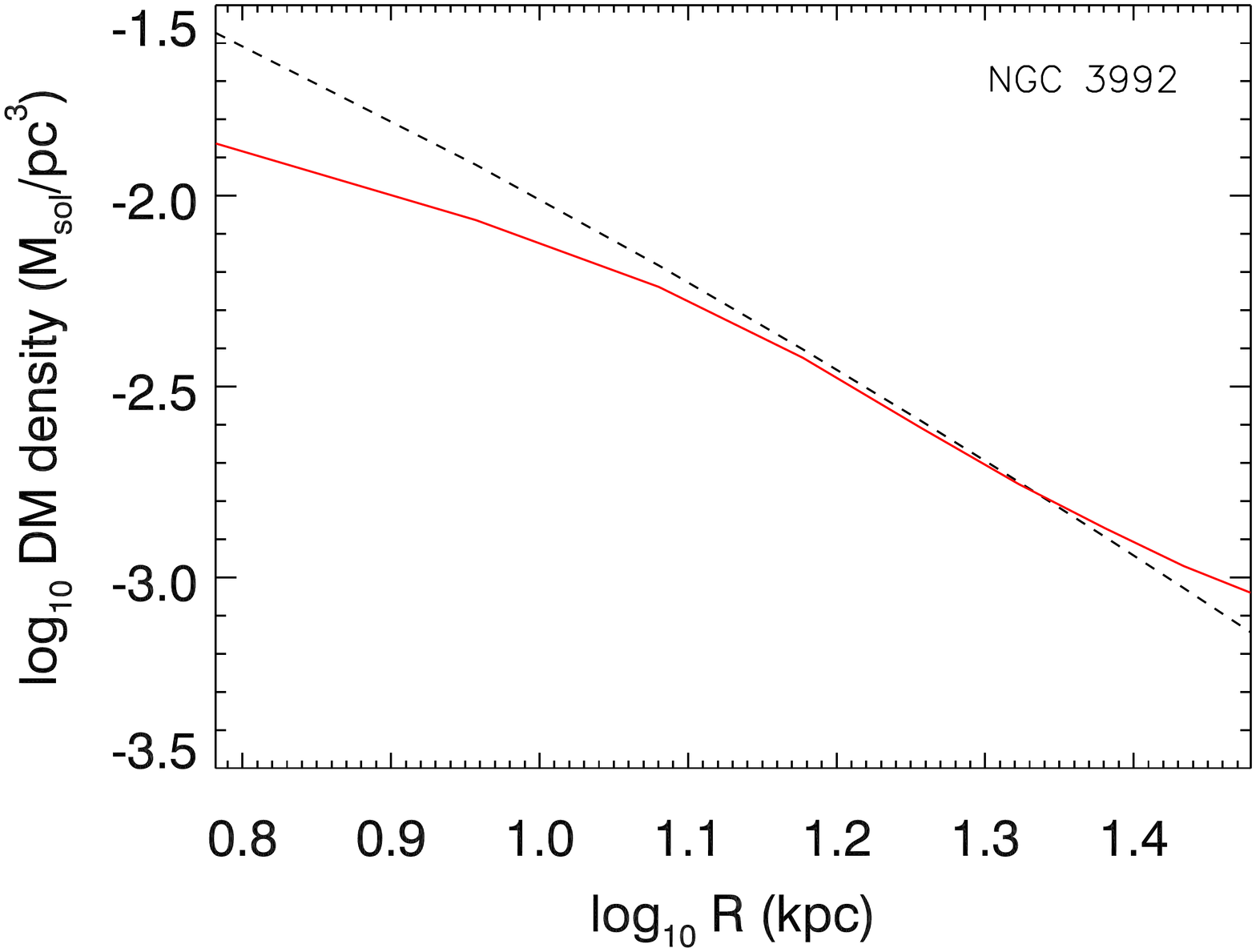}
  \\
  \includegraphics[angle=90,width=0.48\textwidth]{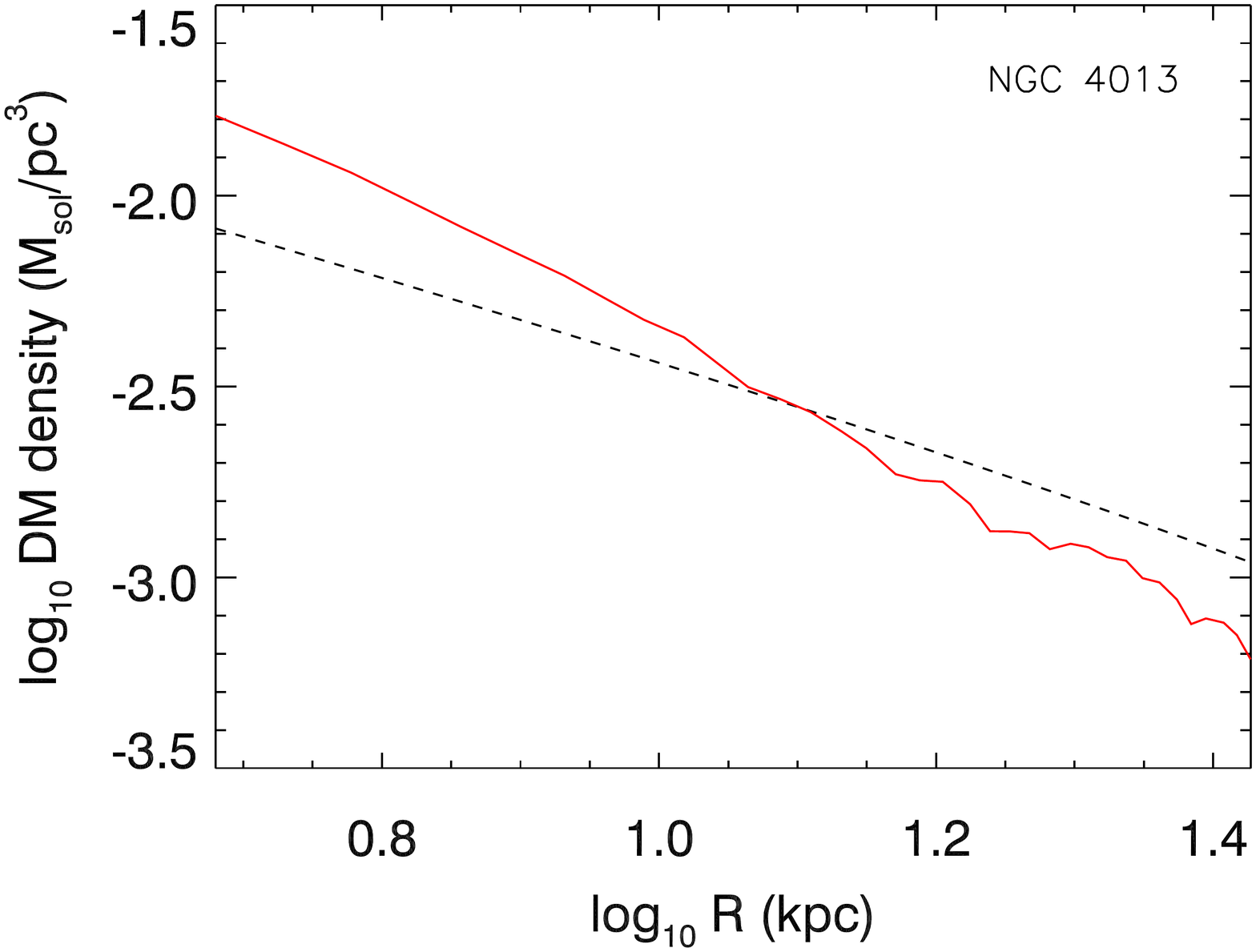}
  \includegraphics[angle=90,width=0.48\textwidth]{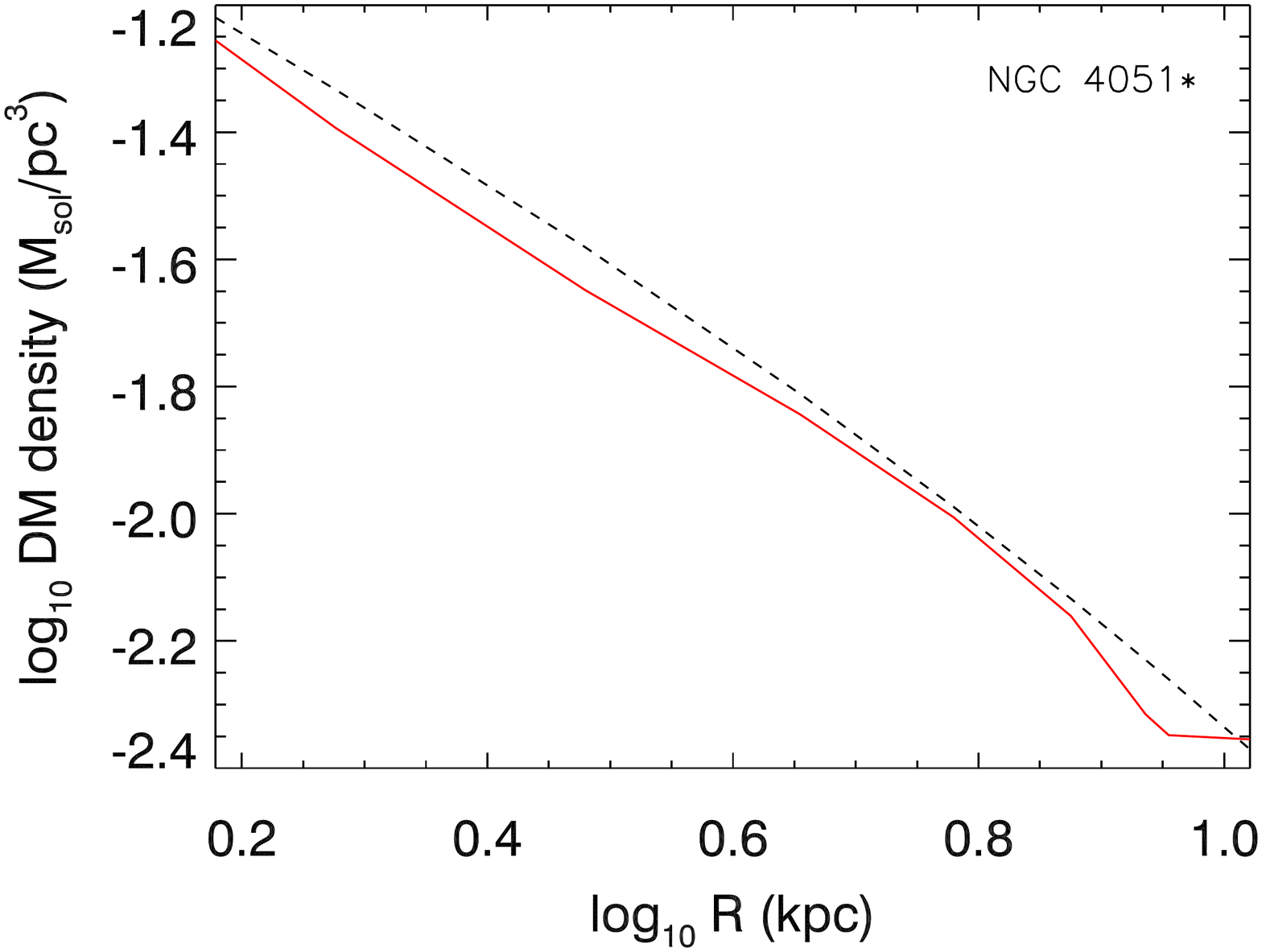}
  \\
  \includegraphics[angle=90,width=0.48\textwidth]{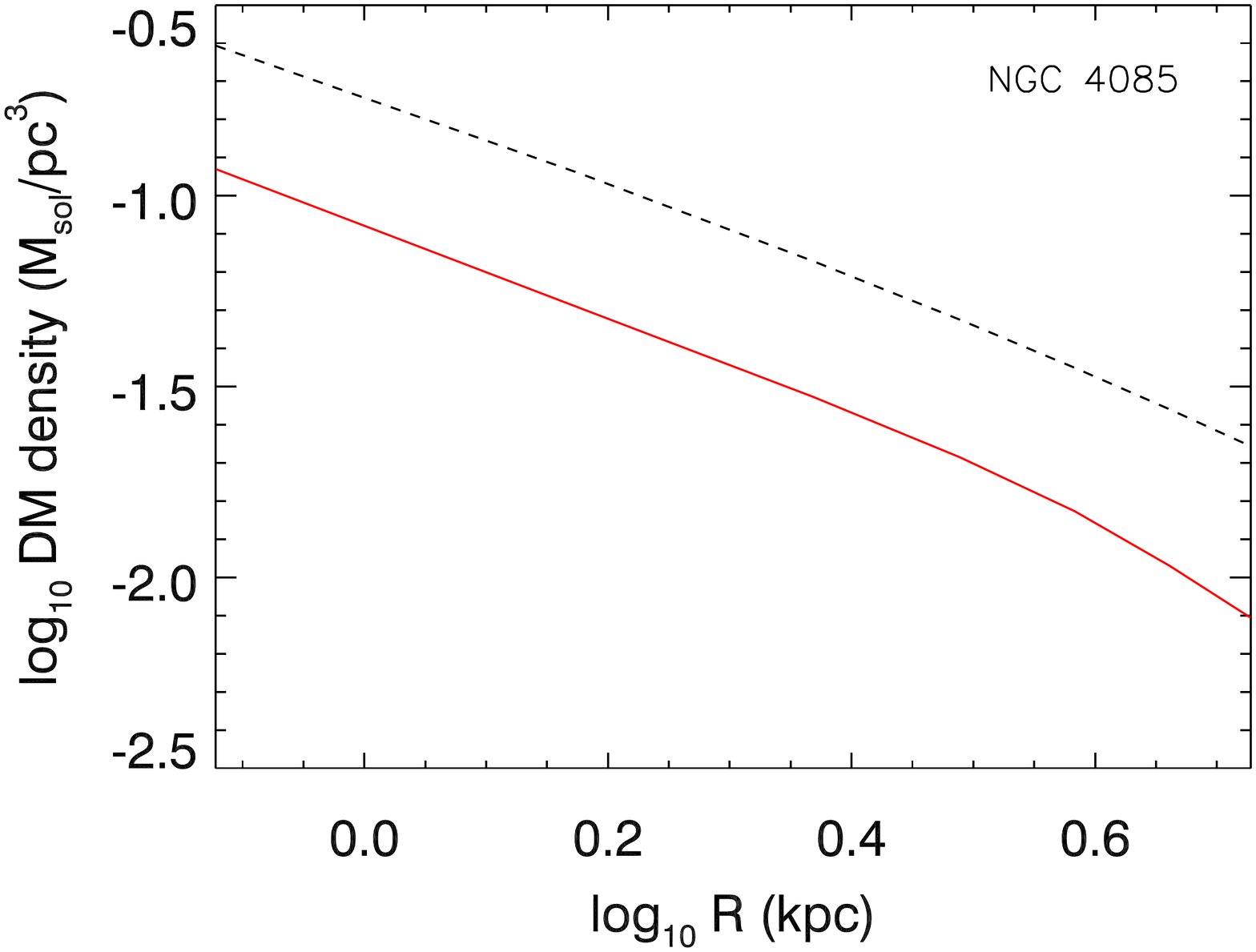}
  \includegraphics[angle=90,width=0.48\textwidth]{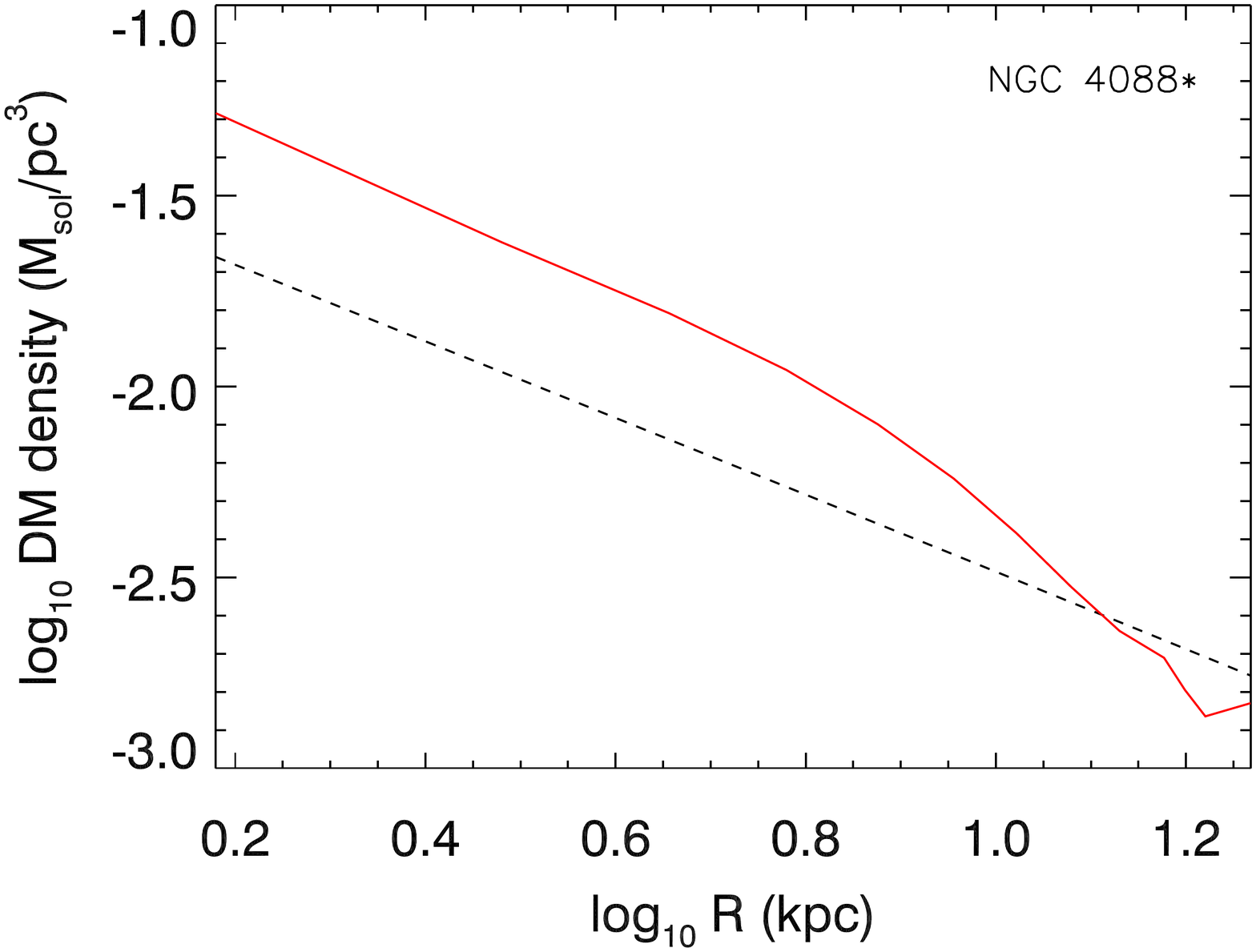}
  \caption{}
\end{figure}

\addtocounter{figure}{-1}

\begin{figure}[!ht]
  \includegraphics[angle=90,width=0.48\textwidth]{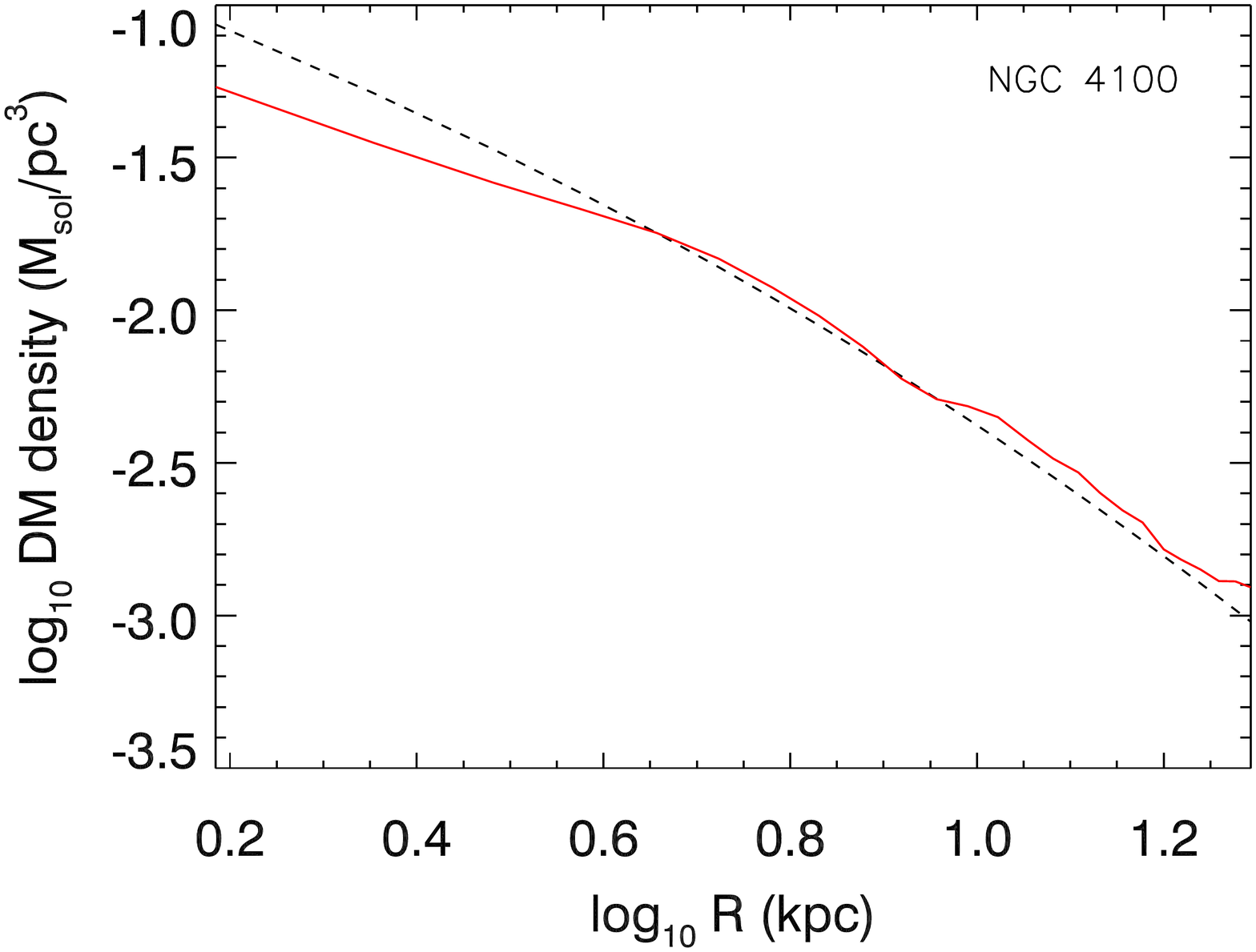}
  \includegraphics[angle=90,width=0.48\textwidth]{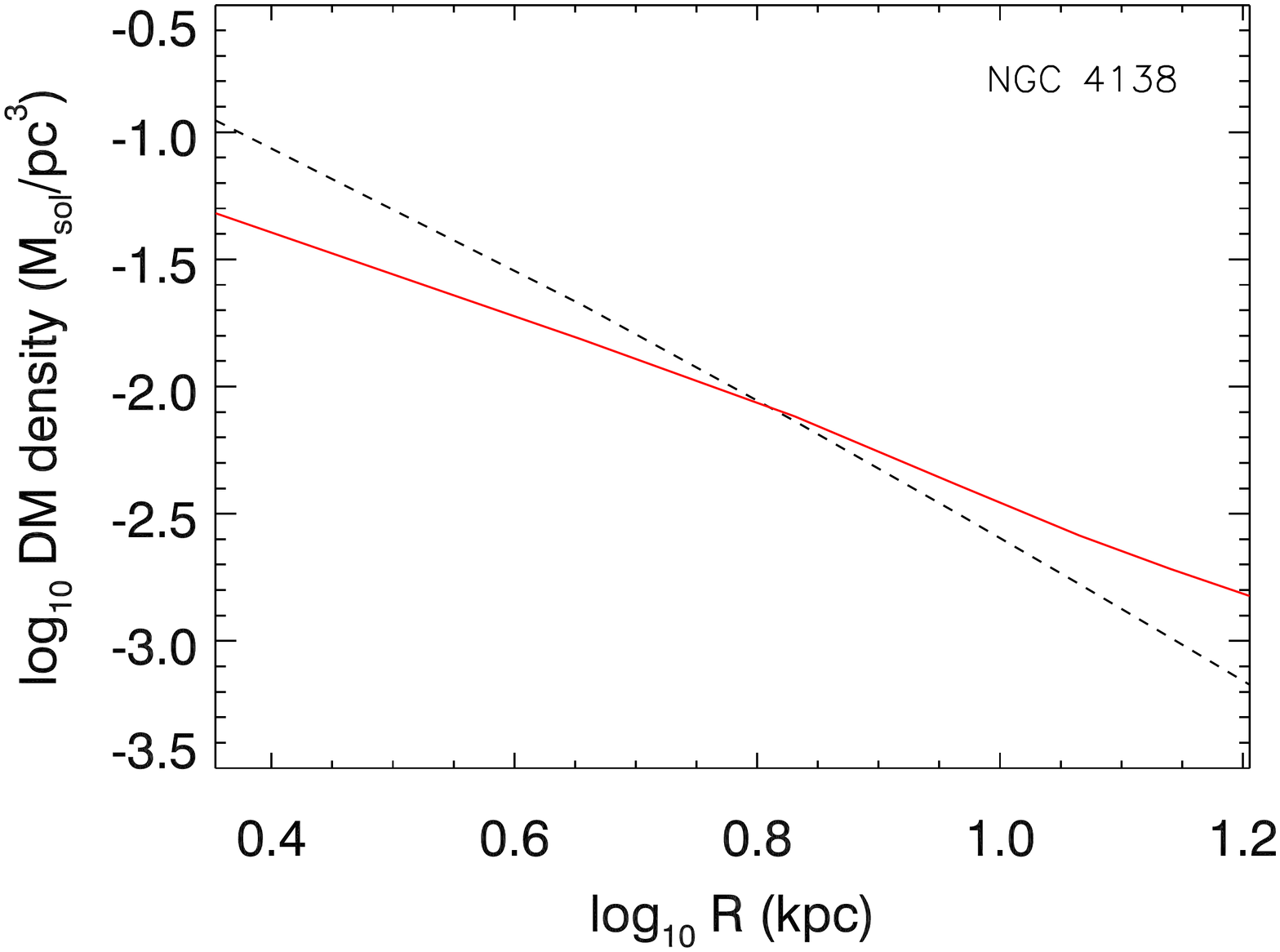}
  \\
  \includegraphics[angle=90,width=0.48\textwidth]{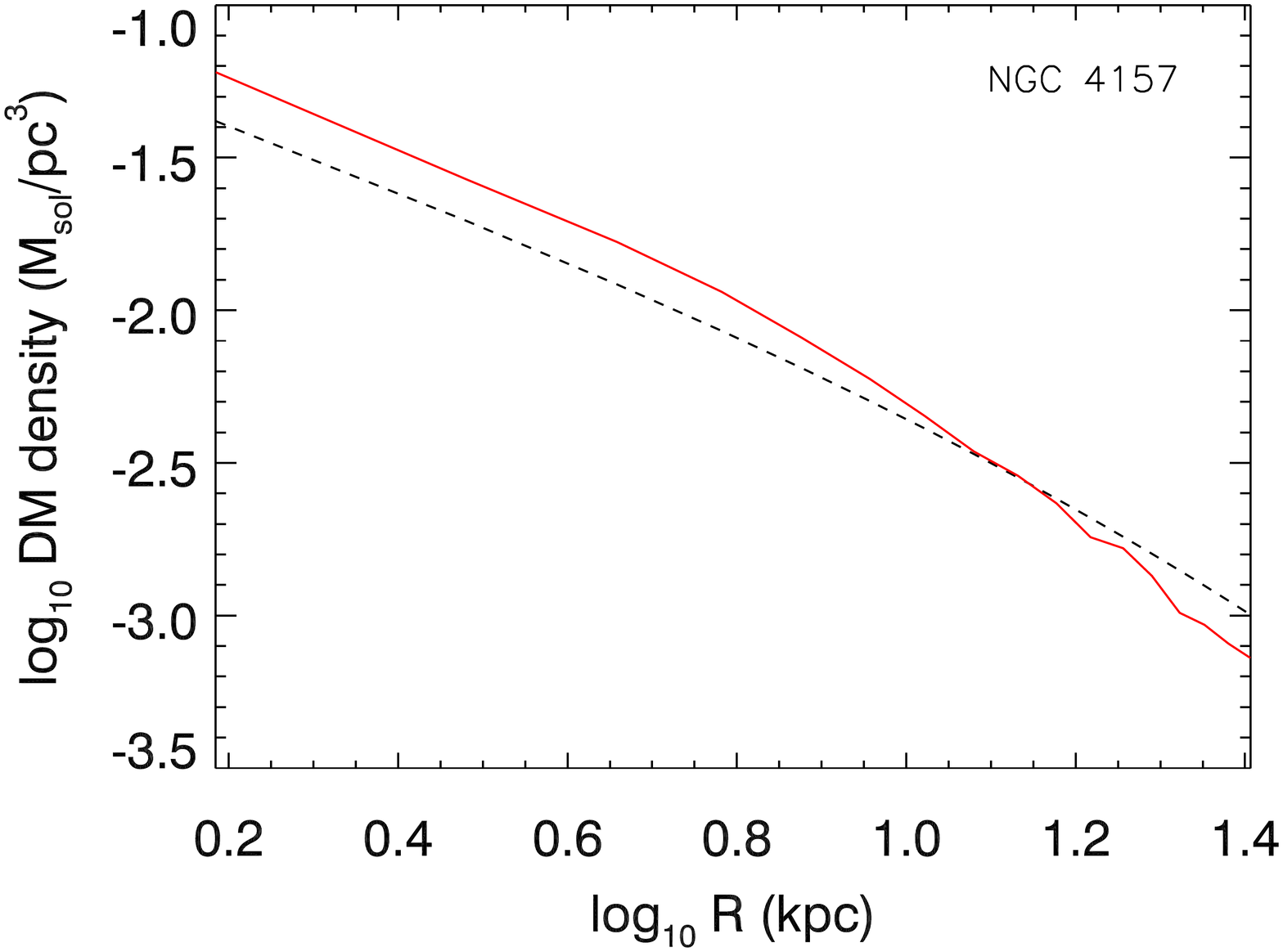}
  \includegraphics[angle=90,width=0.48\textwidth]{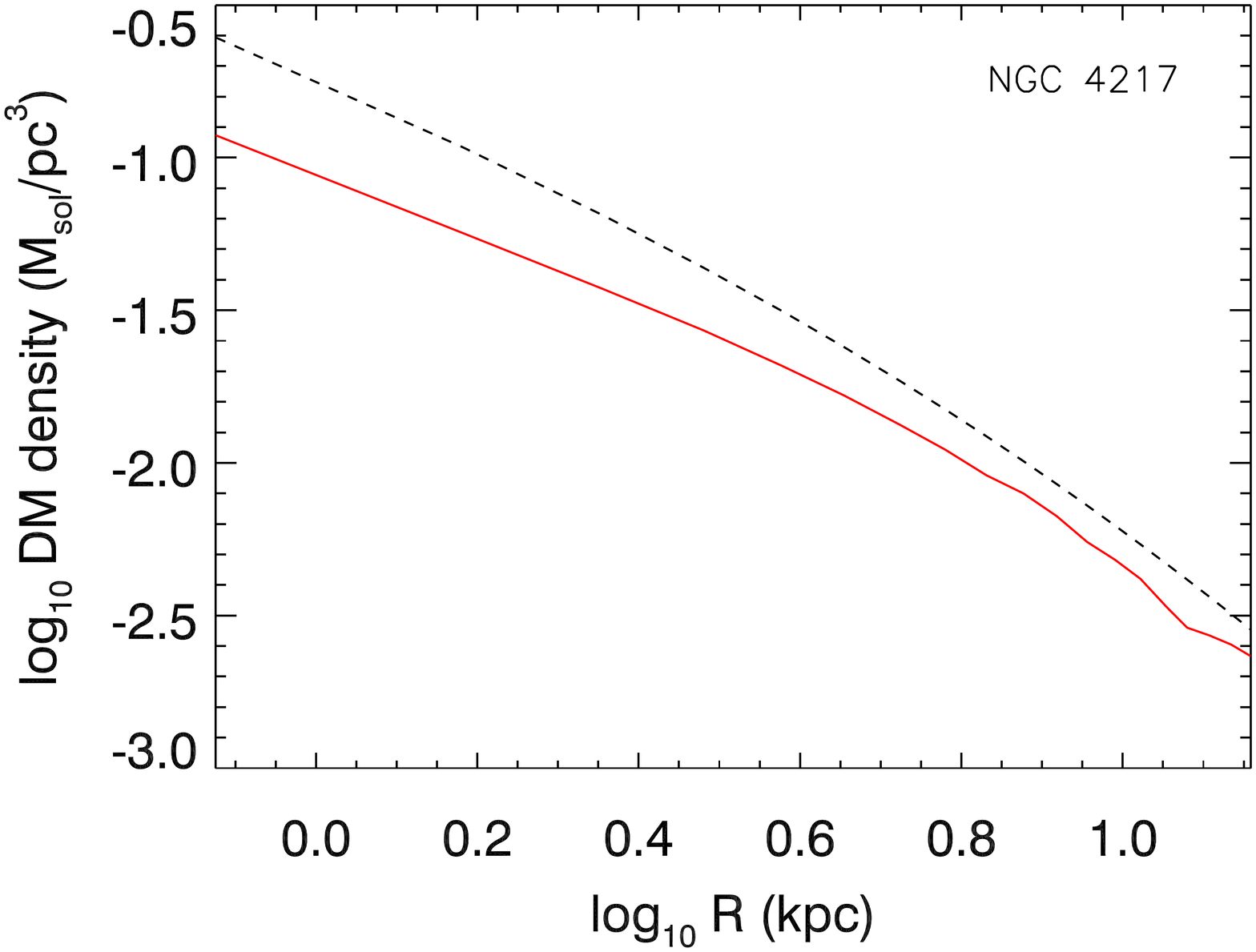}
  \\
  \includegraphics[angle=90,width=0.48\textwidth]{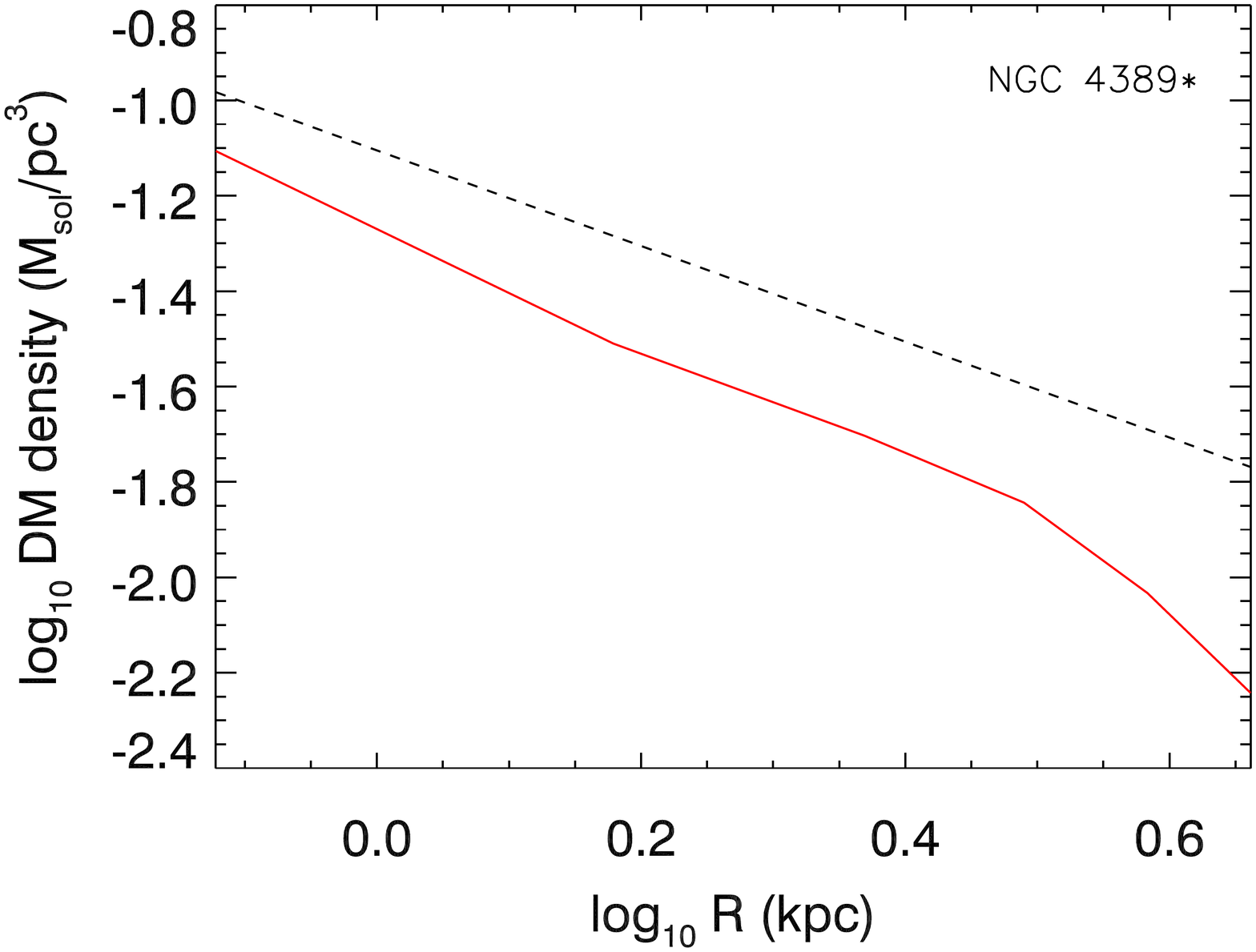}
  \includegraphics[angle=90,width=0.48\textwidth]{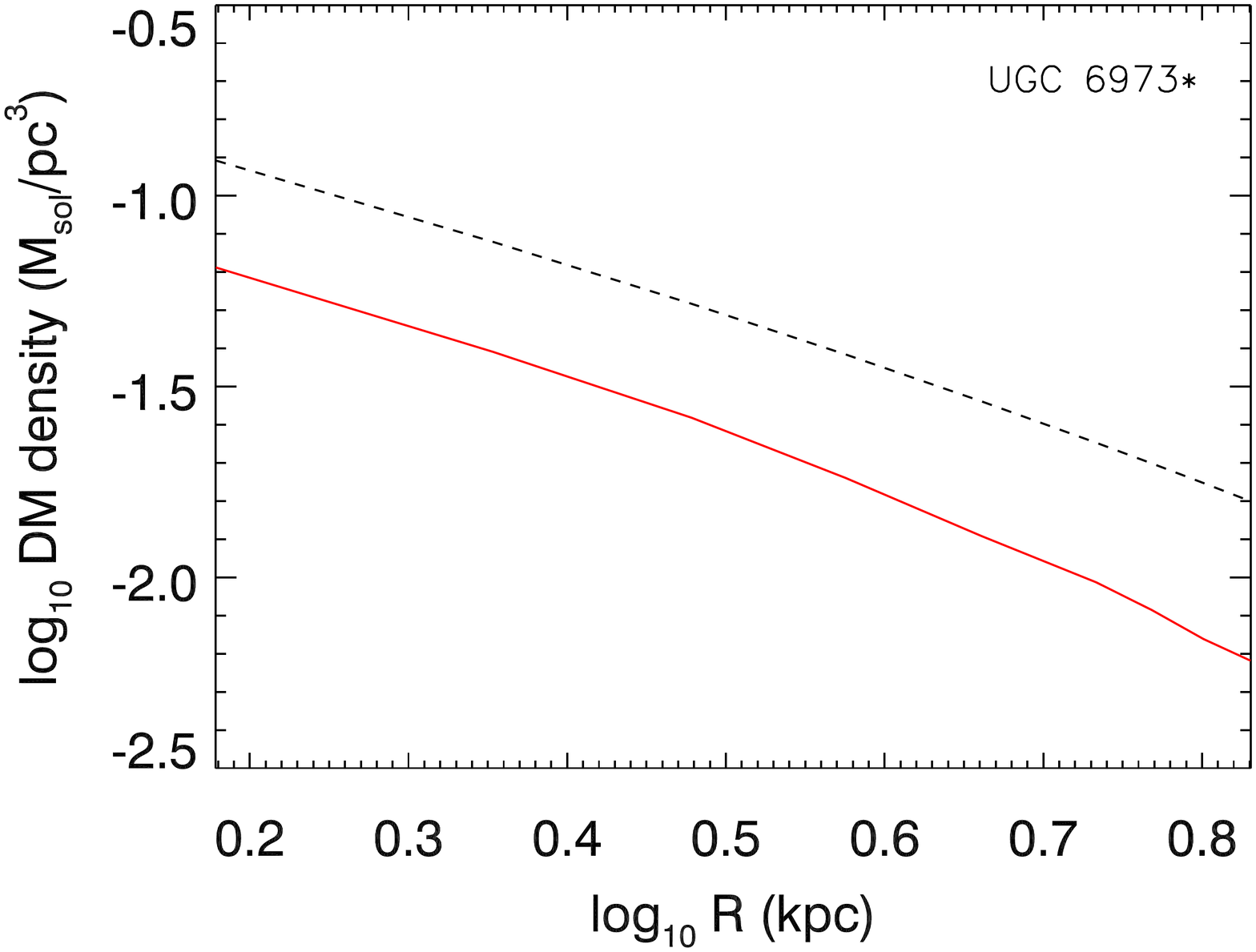}
  \caption{}
\end{figure}

\addtocounter{figure}{-1}

\begin{figure}[!ht]
  \includegraphics[angle=90,width=0.48\textwidth]{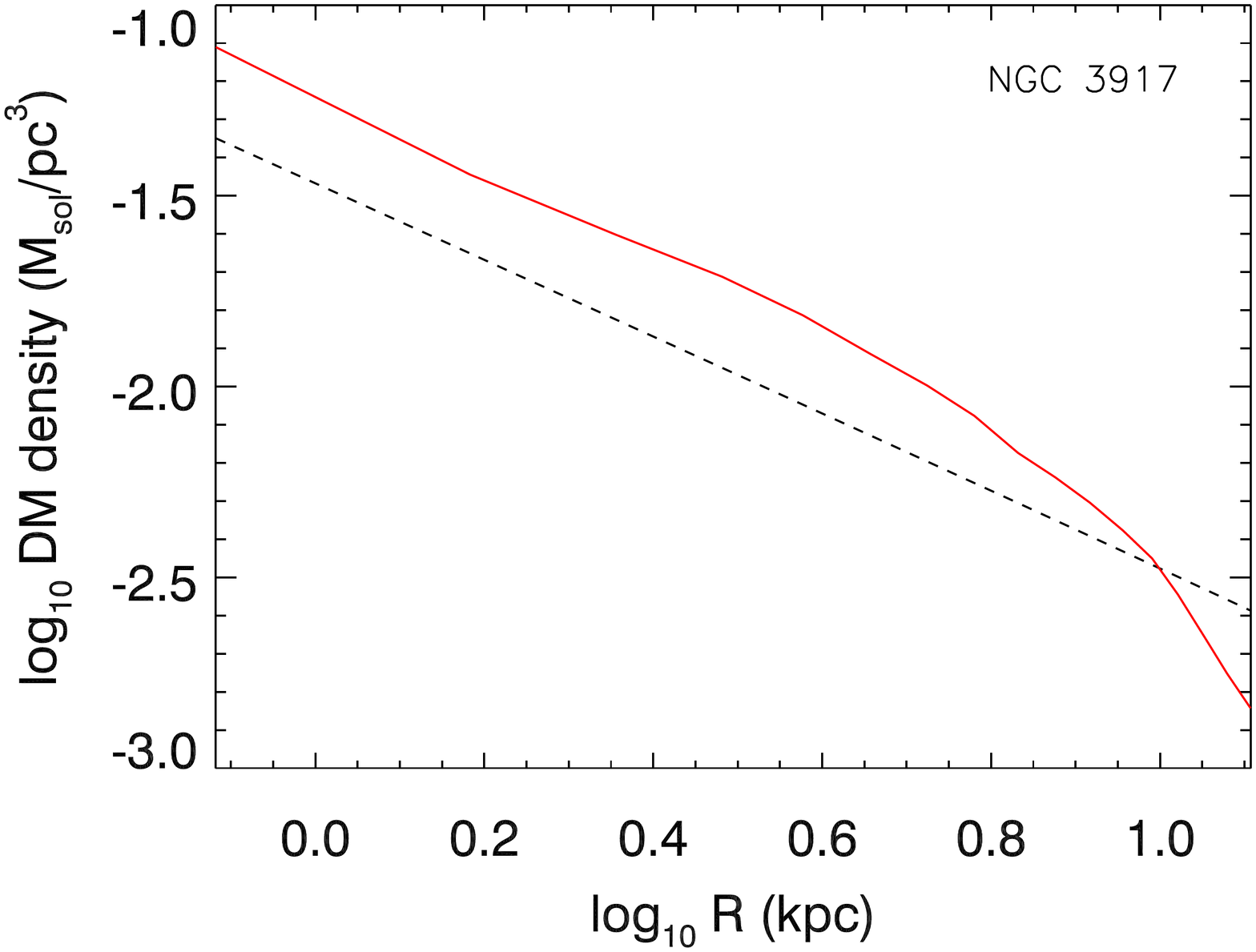}
  \includegraphics[angle=90,width=0.48\textwidth]{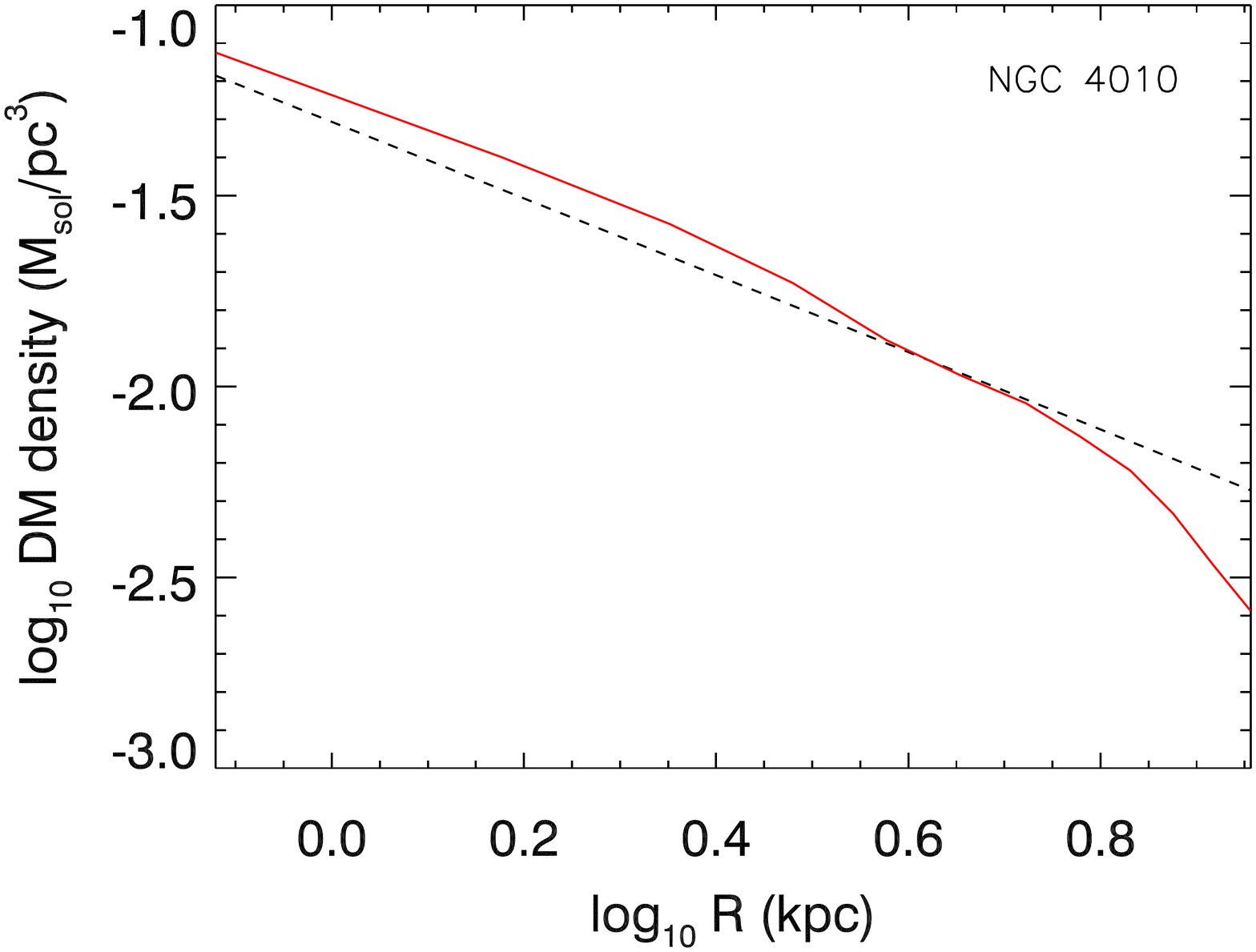}
  \\
  \includegraphics[angle=90,width=0.48\textwidth]{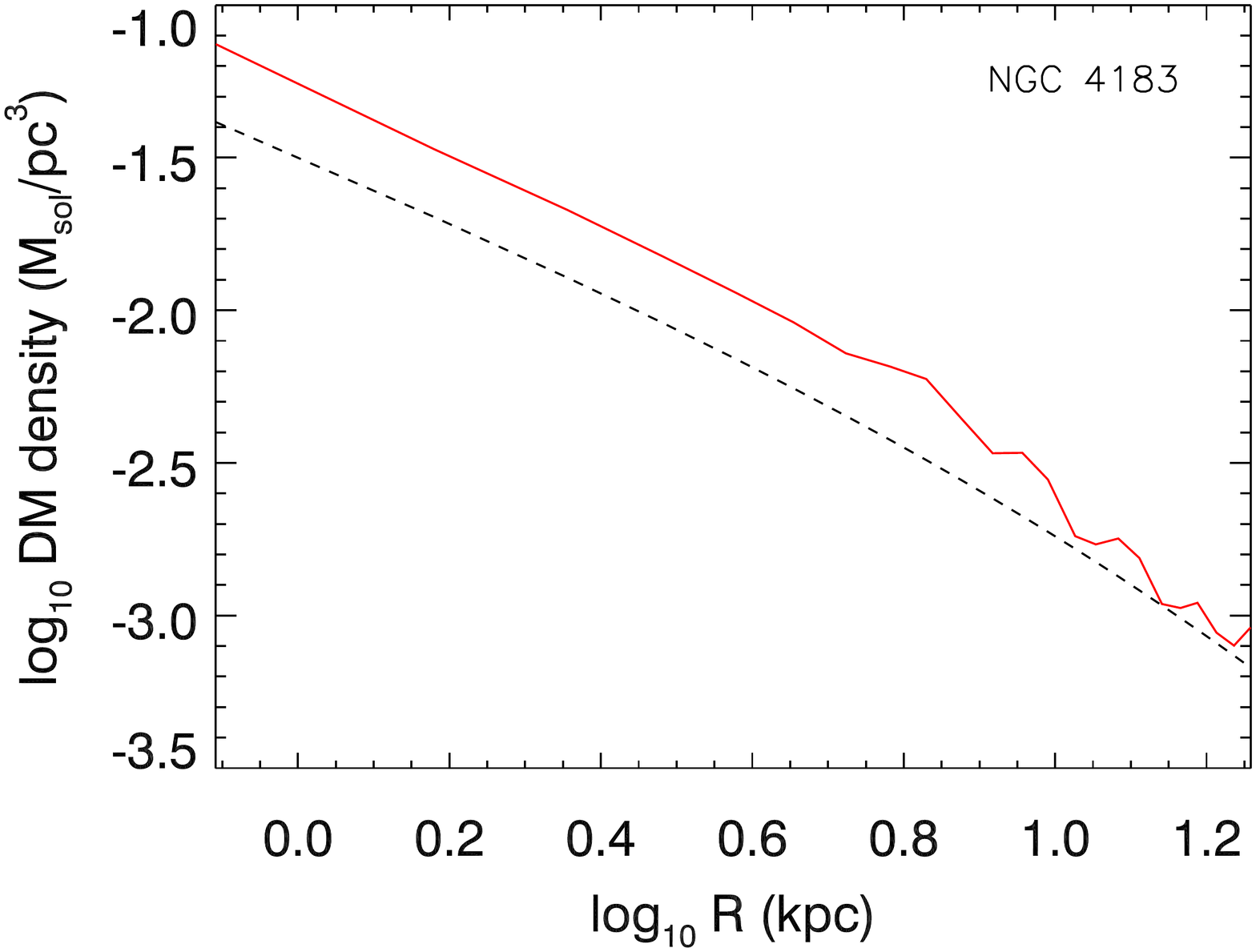}
  \includegraphics[angle=90,width=0.48\textwidth]{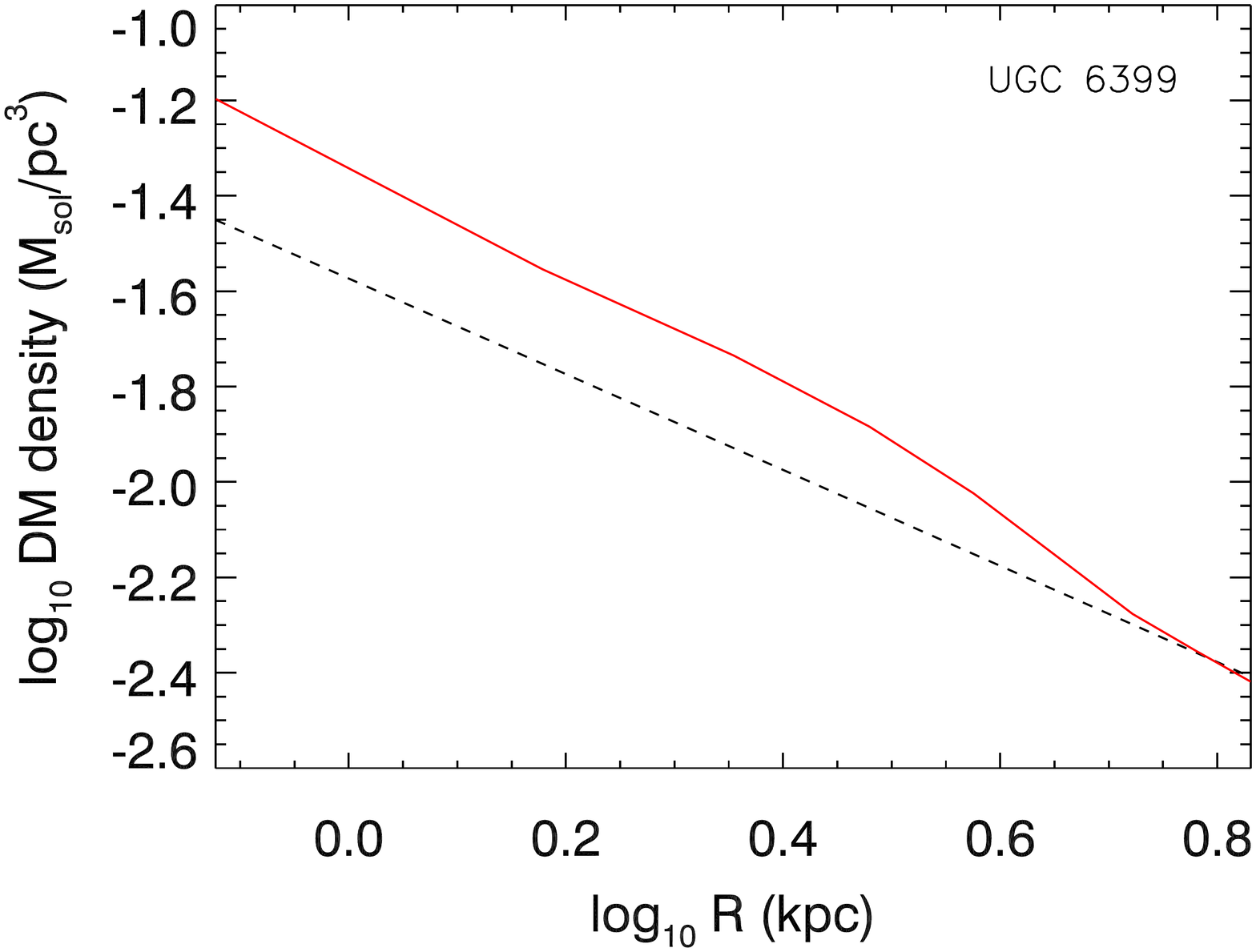}
  \\
  \includegraphics[angle=90,width=0.48\textwidth]{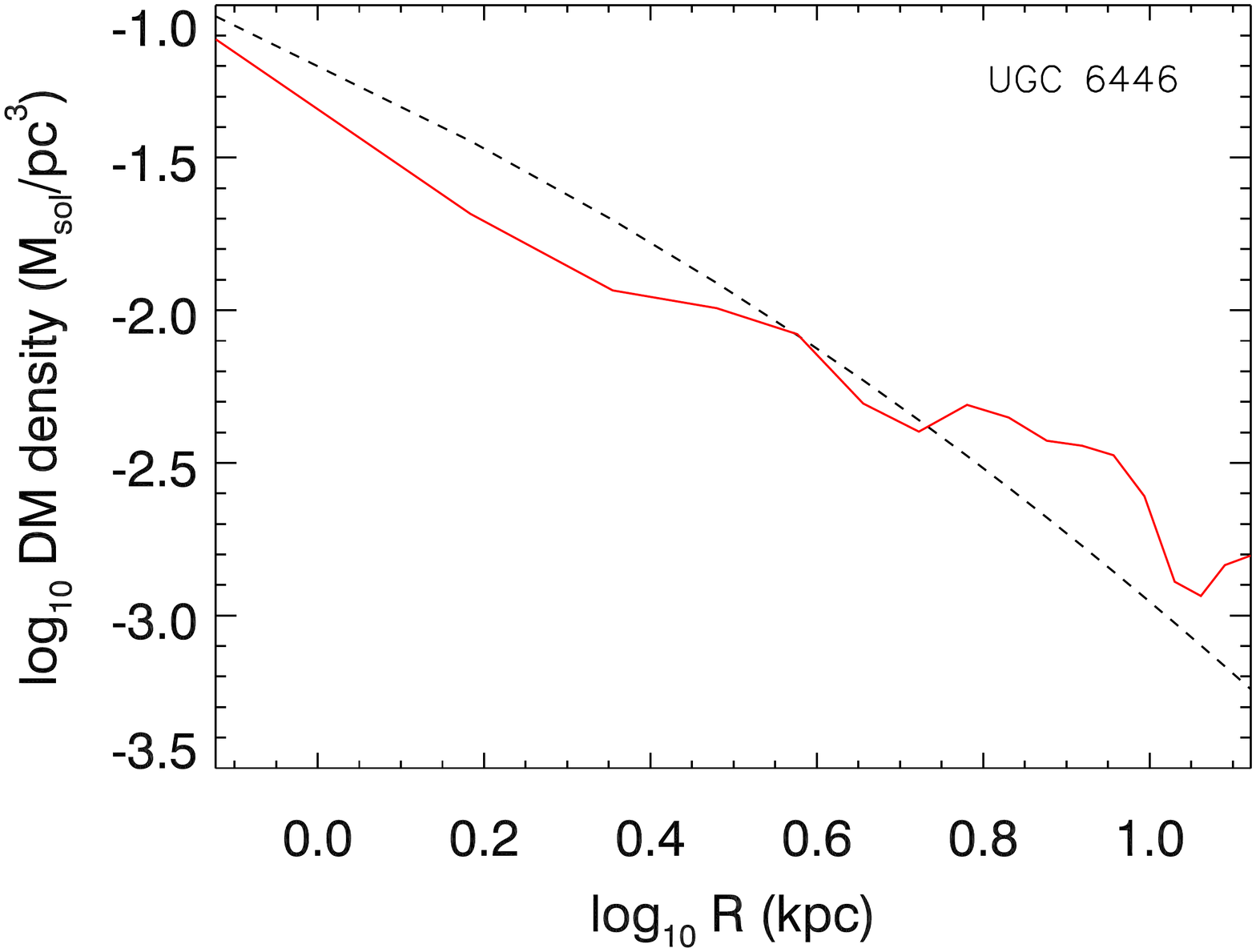}
  \includegraphics[angle=90,width=0.48\textwidth]{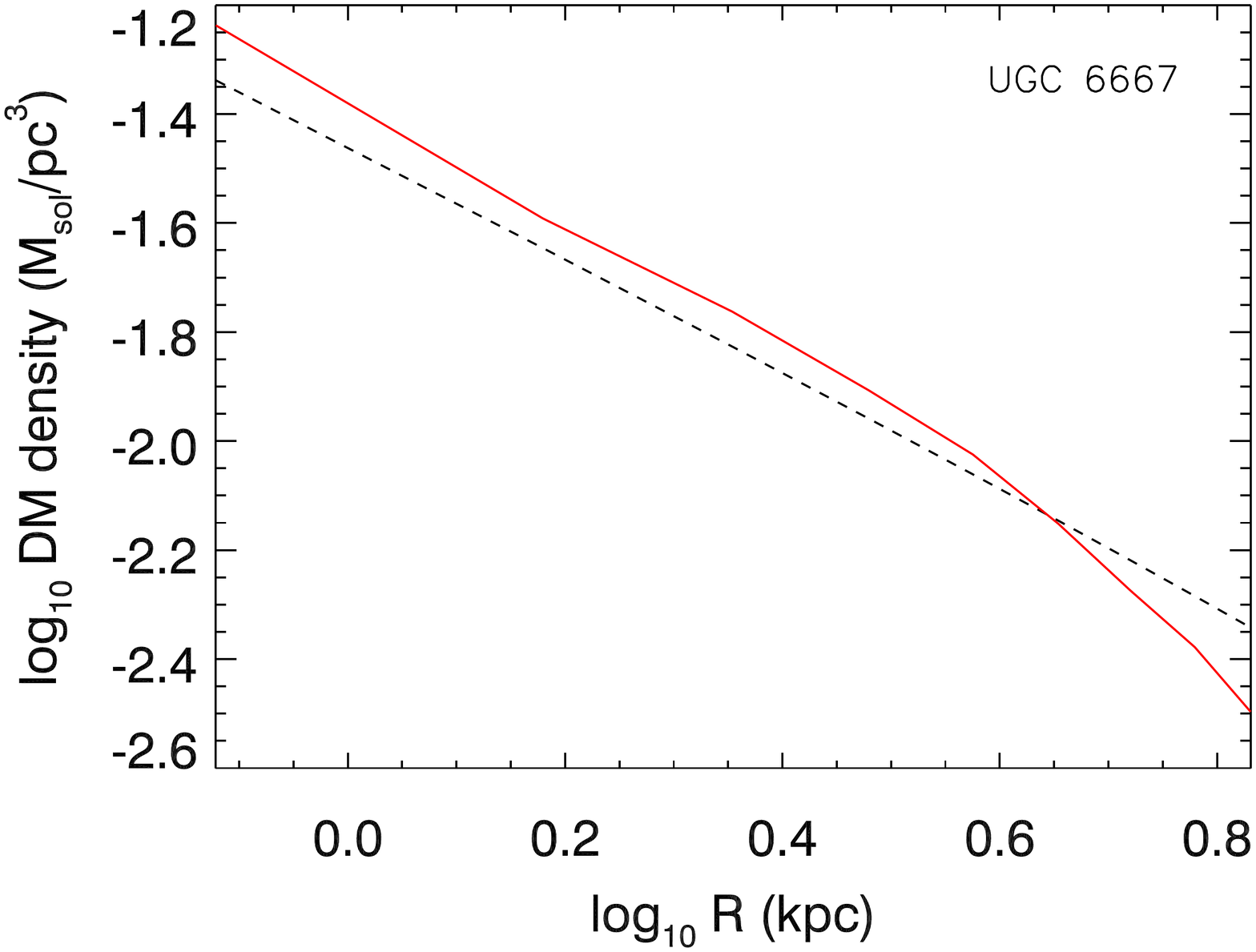}
  \caption{}
 
\end{figure}

\renewcommand{\thefigure}{\arabic{figure} (Cont.)}
\addtocounter{figure}{-1}

\begin{figure}[!ht]
  \includegraphics[angle=90,width=0.48\textwidth]{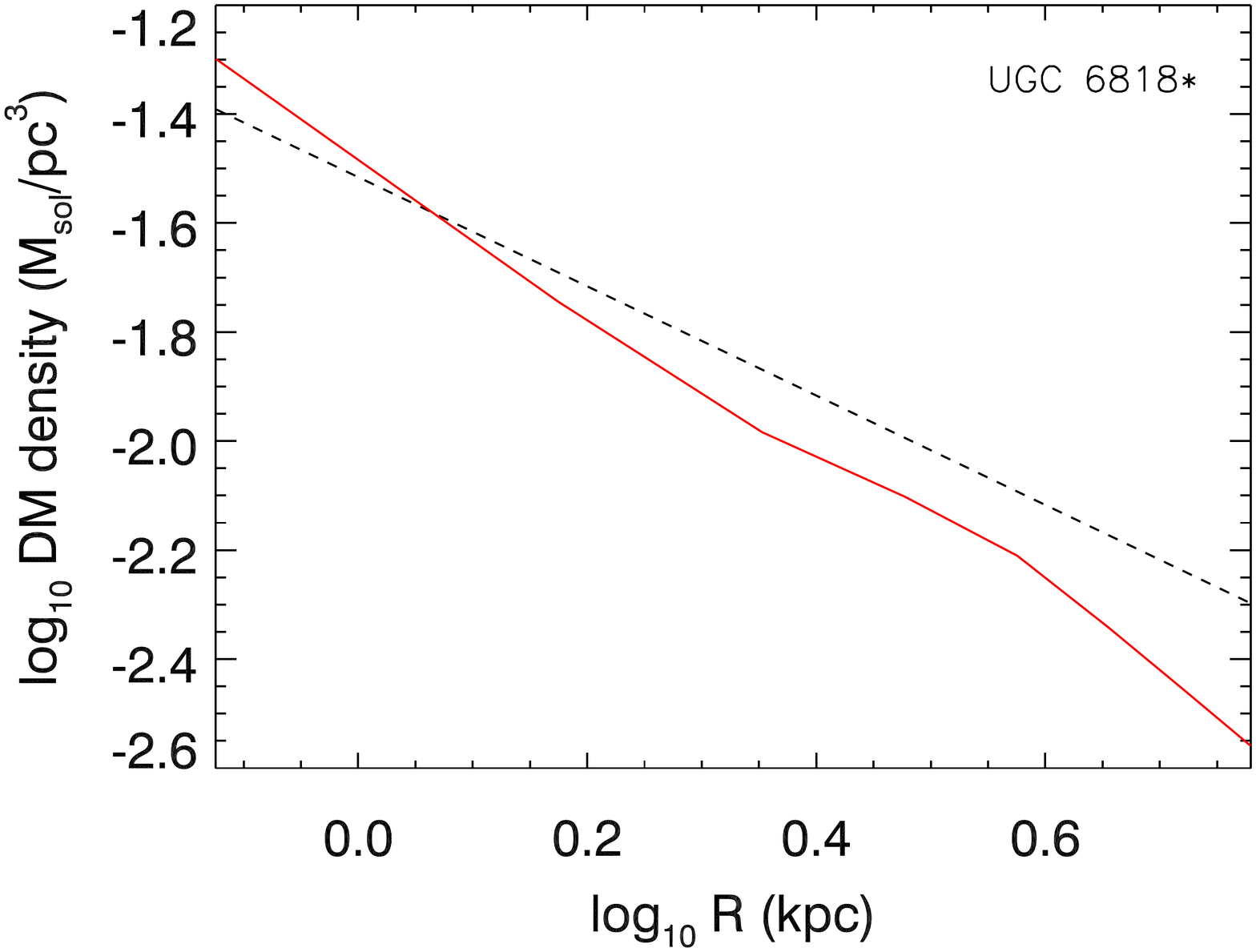}
  \includegraphics[angle=90,width=0.48\textwidth]{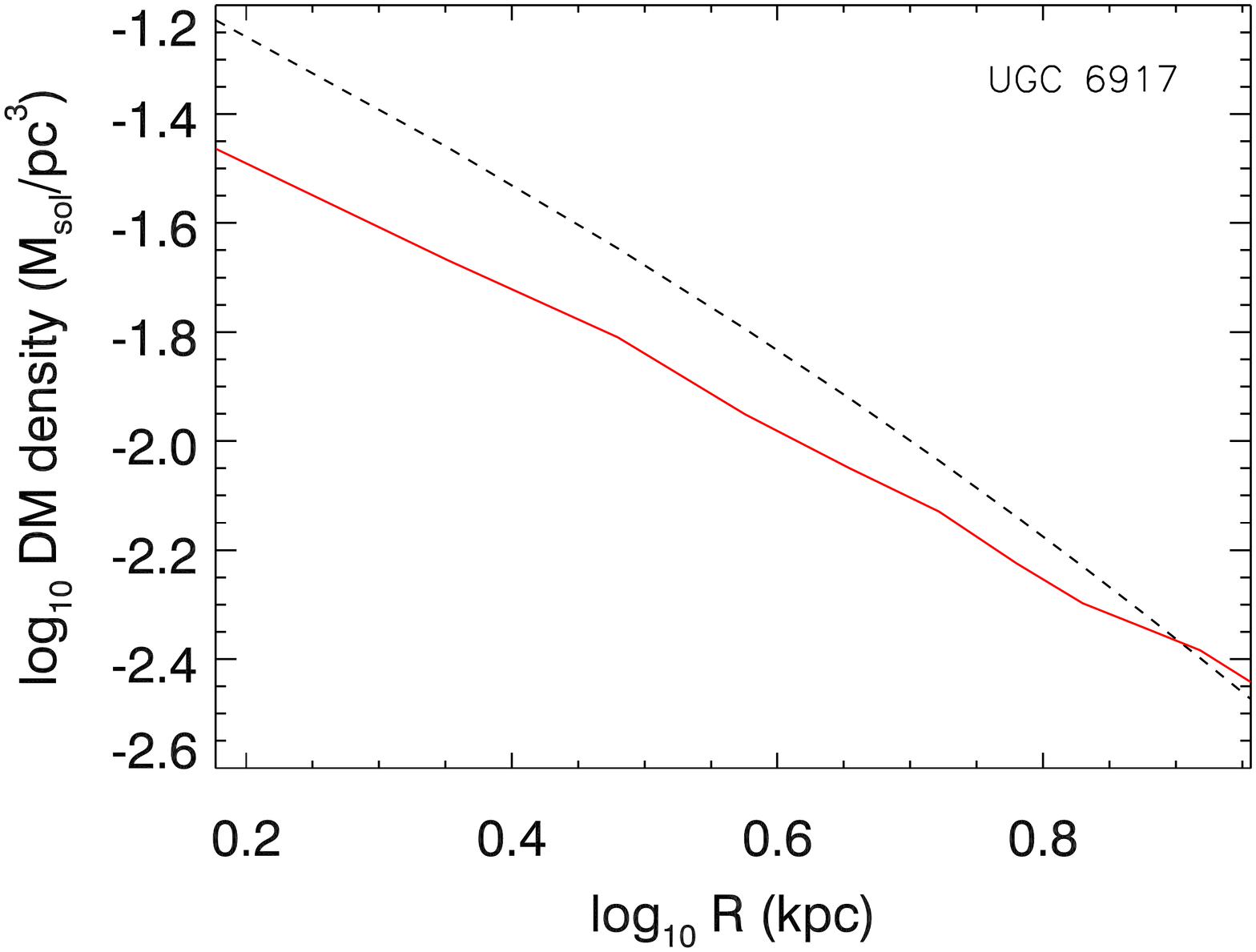}
  \\
  \includegraphics[angle=90,width=0.48\textwidth]{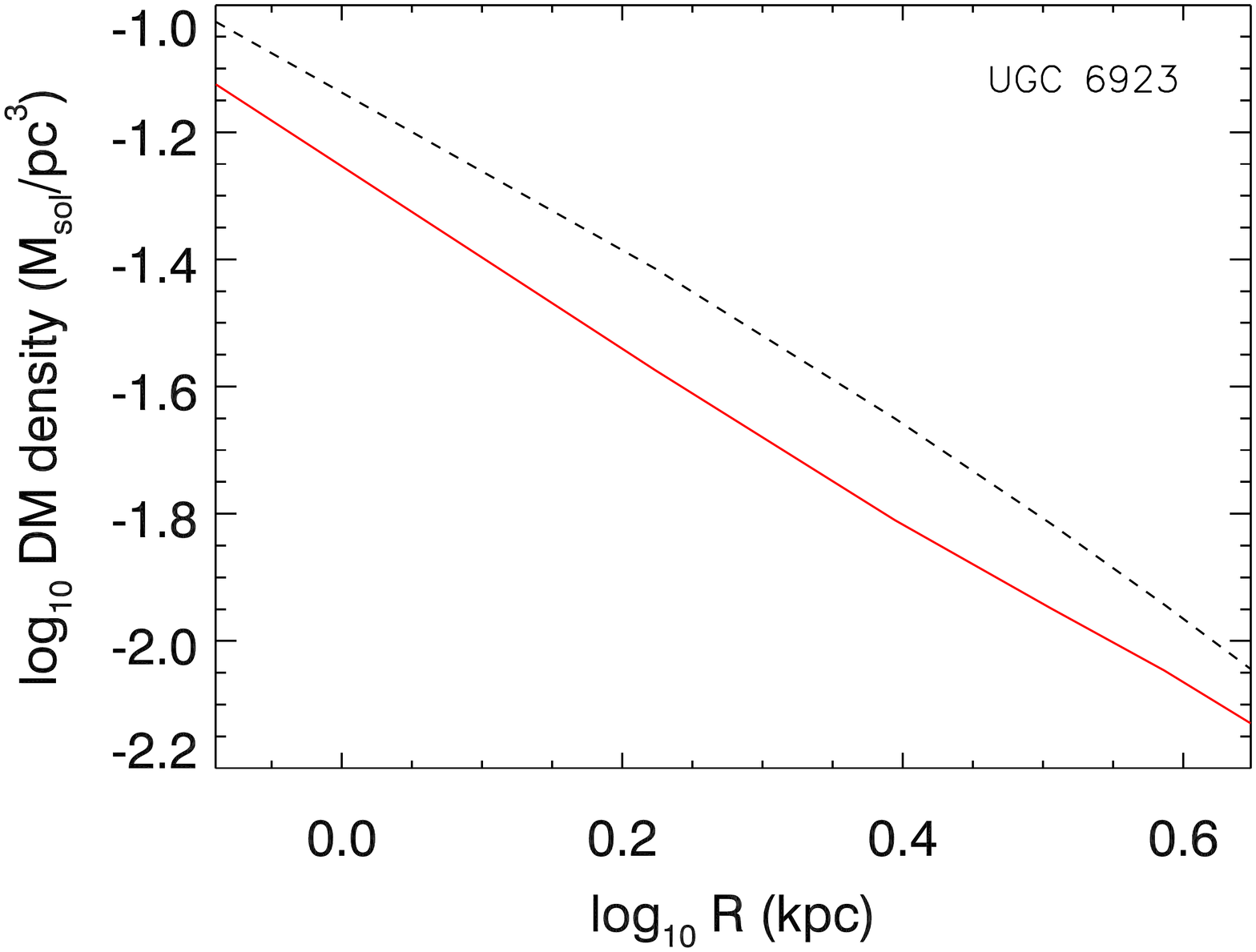}
  \includegraphics[angle=90,width=0.48\textwidth]{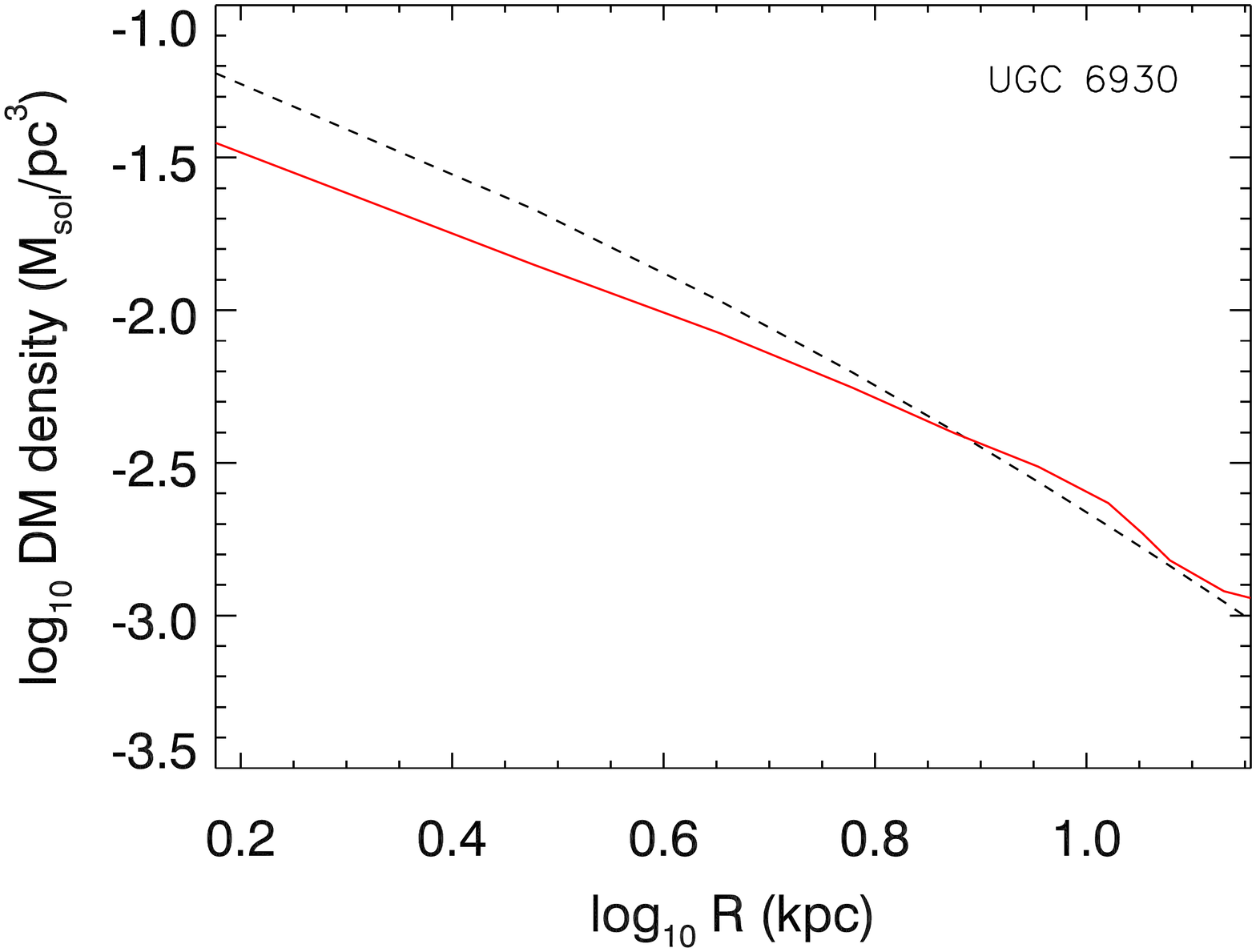}
  \\
  \includegraphics[angle=90,width=0.48\textwidth]{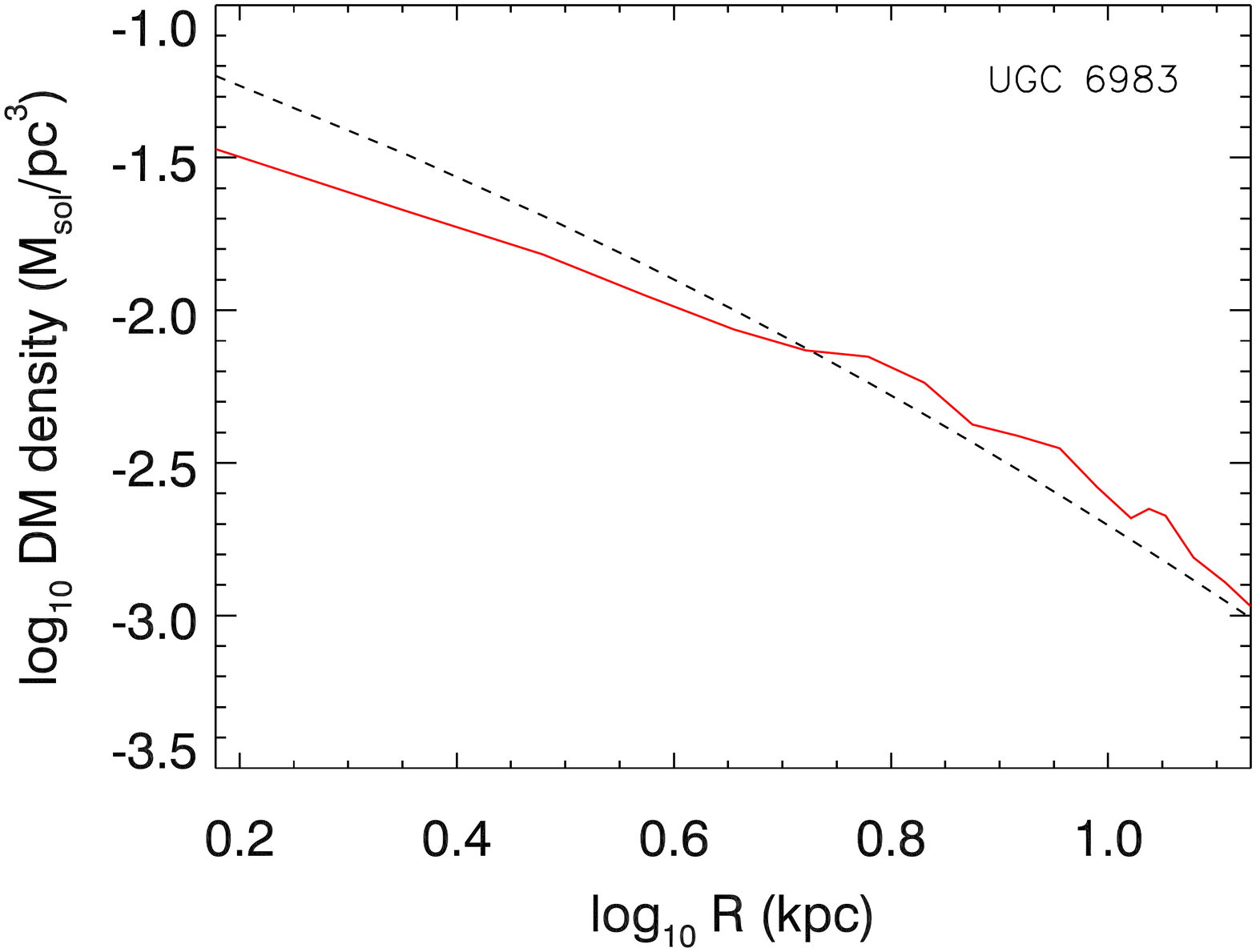}
  \includegraphics[angle=90,width=0.48\textwidth]{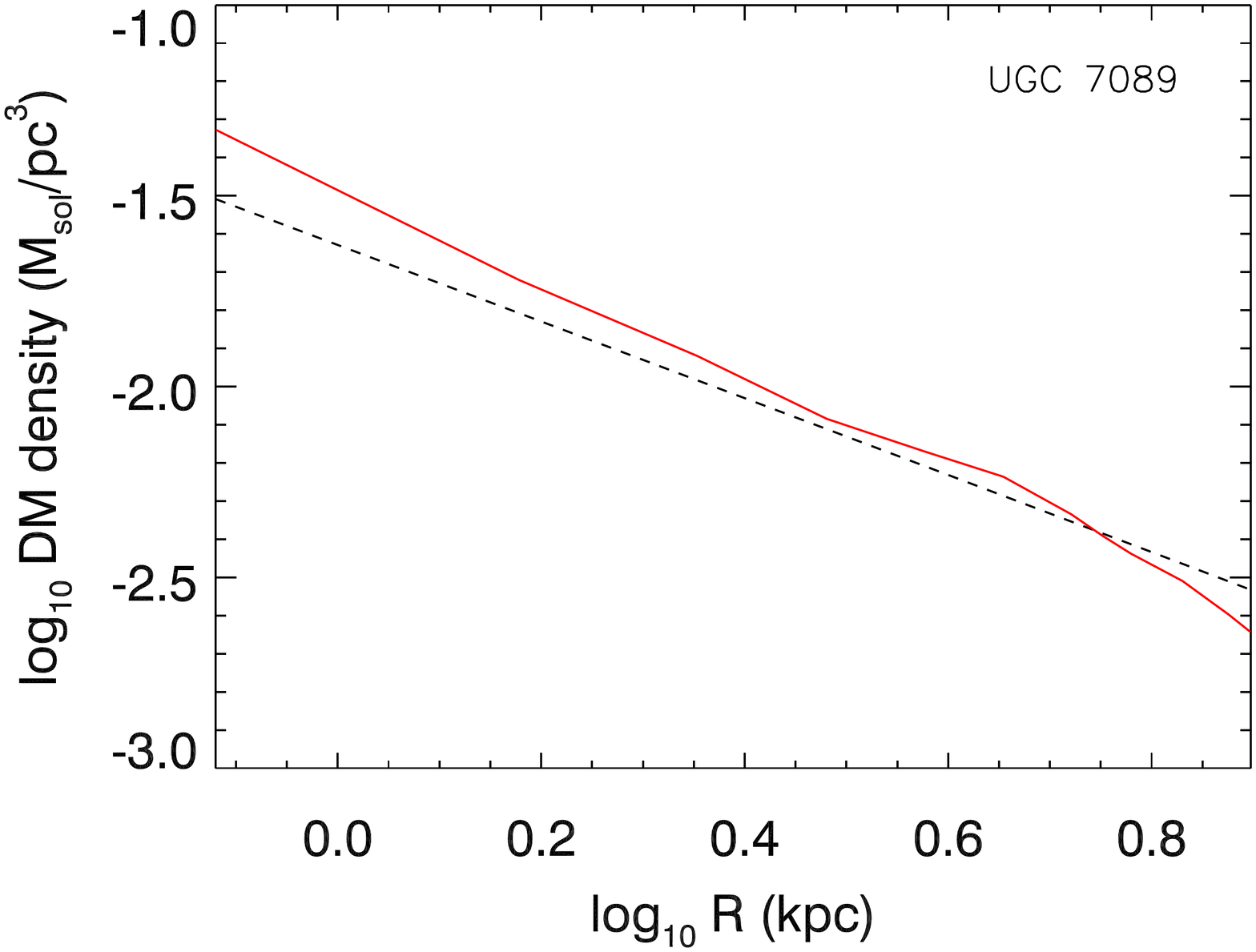}
  \caption{}
\end{figure}

\renewcommand{\thefigure}{\arabic{figure}}



\subsection{Mass Profiles of Galaxy Clusters}
\label{clusters_subsection}

In principle, the above mass profile, Eq.~(\ref{Mprime}) or 
Eq.~(\ref{fMDM2}), 
fixed by the ratio of the corresponding Unruh-Hawking temperatures 
(in the limit of small
$a_0/a$) can be modified due to some well-known physical effects 
associated with a change of scale.
For example, in the presence of gravity, the temperature is not constant
in space at equilibrium.  As a result, it
can be changed due to the so-called Tolman-Ehrenfest 
effect \cite{Tolman:1930zza,Tolman:1930ona} 
\begin{equation}
T \sqrt{g_{00}} \;=\; 2 \tilde{\alpha}\;, 
\quad
\mbox{where}
\quad 
g_{00} \;=\; 1+2\,\Phi / c^2\;,
\end{equation}
with $\Phi$ being the gravitational potential,
and the factor $\tilde{\alpha}$ is determined by the
boundary conditions of the problem.  

But what gravitational 
potential $\Phi$ should be used in our case and 
what sets the value of $\tilde{\alpha}$? 
In general, we should expend the gravitational potential in powers of $r$.
The leading $r$ dependence corresponds to 
a constant background gravitational field, i.e. the linear 
potential (this is indeed the first term in the Taylor expansion for the potential, up to a physically irrelevant constant piece).
Thus we end up with the following modification
of the mass profile :
\begin{eqnarray}
\label{clustermassp}
f_{\mathrm{MDM}}(a/a_0)
\;=\; \dfrac{M'}{M}
& = & \dfrac{\alpha}{\left[\,1+ 
\left(r/r_\textrm{MDM}\right)\,\right]}\left[\dfrac{a_0^2}{(a_\mathrm{obs}+a_0)^2 - a_0^2}\right]
\;,
\end{eqnarray}
where the dimensionless factor $\alpha$ is determined by the ratio of 
$\tilde{\alpha}$ at different scales (in our case, the 
cluster and galactic scales).
\footnote{Interestingly the same prefactor appears in the context of 
conformal gravity when one rewrites the FRW cosmological line element in 
the Schwarzschild coordinate system
(the linear potential also being the direct analogue of the Newtonian 
potential 
for conformal gravity; see Ref.~\citen{Mannheim:2012qw}).}
Note that the prefactor $\alpha/\left[1+ (r/r_\textrm{MDM})\right]$ 
is only the leading term in a more general expression 
that involves higher order terms in $r$.

\begin{figure*}
\includegraphics[angle=0,width=0.48\textwidth]{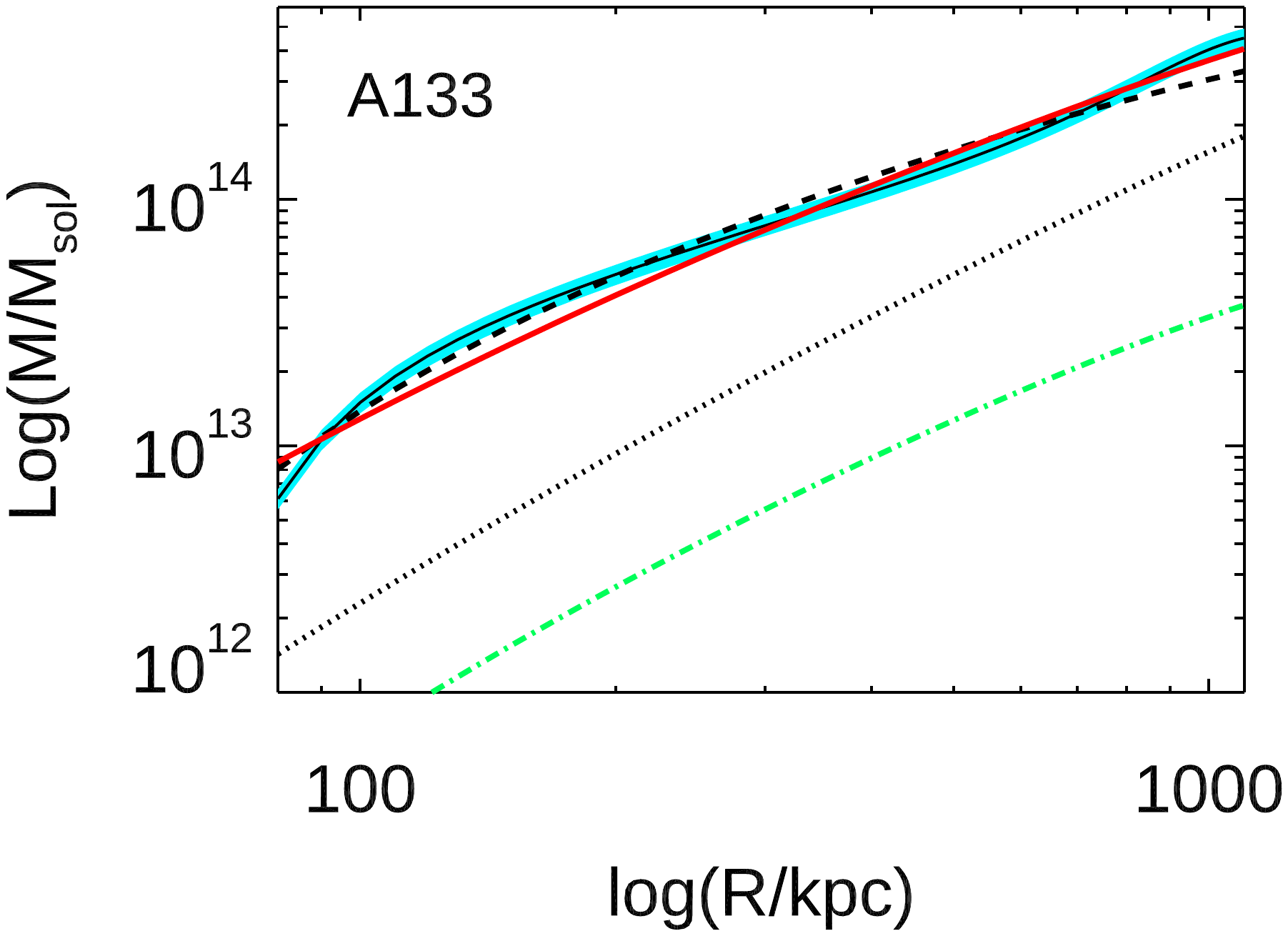}
\includegraphics[angle=0,width=0.48\textwidth]{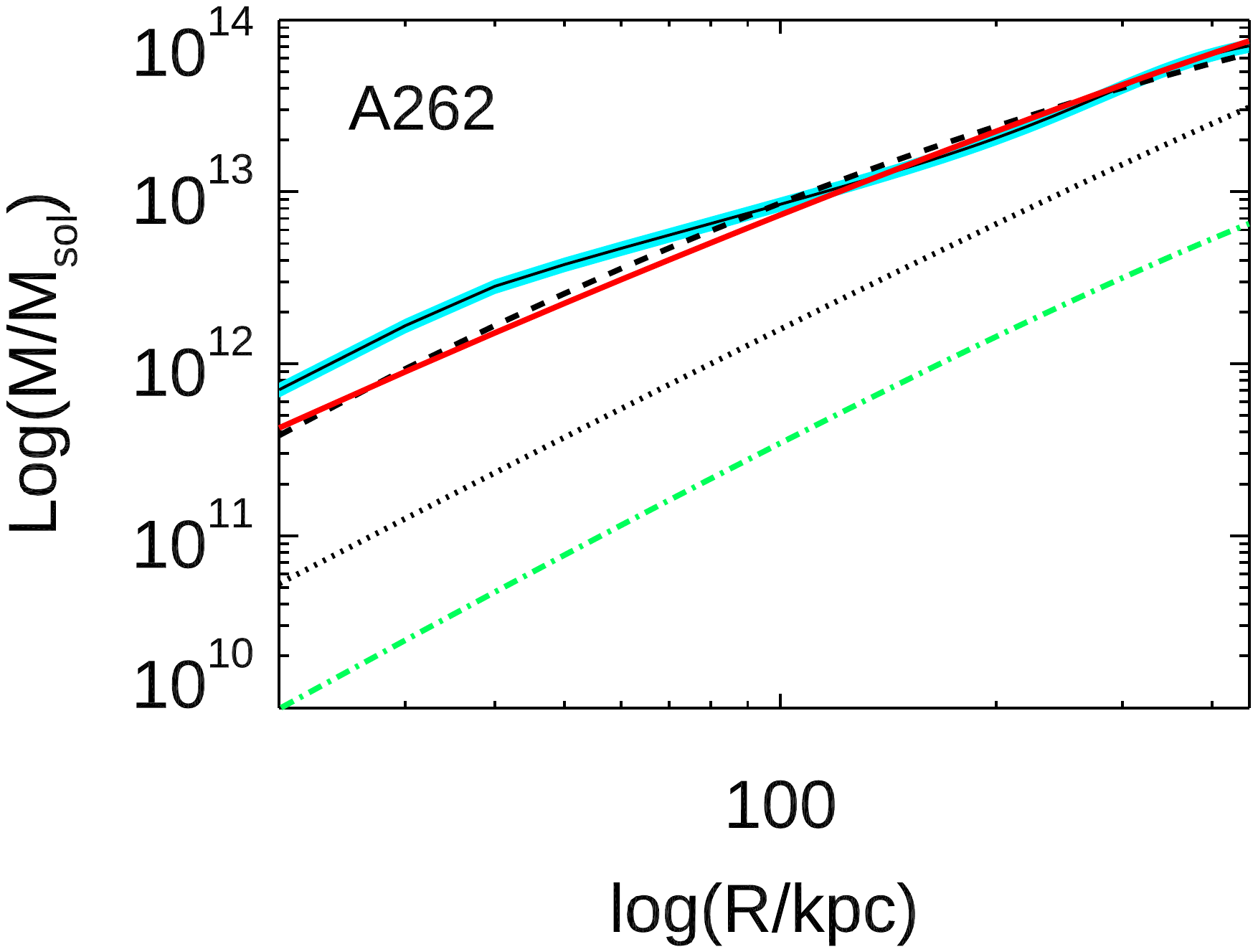}\\
\includegraphics[angle=0,width=0.48\textwidth]{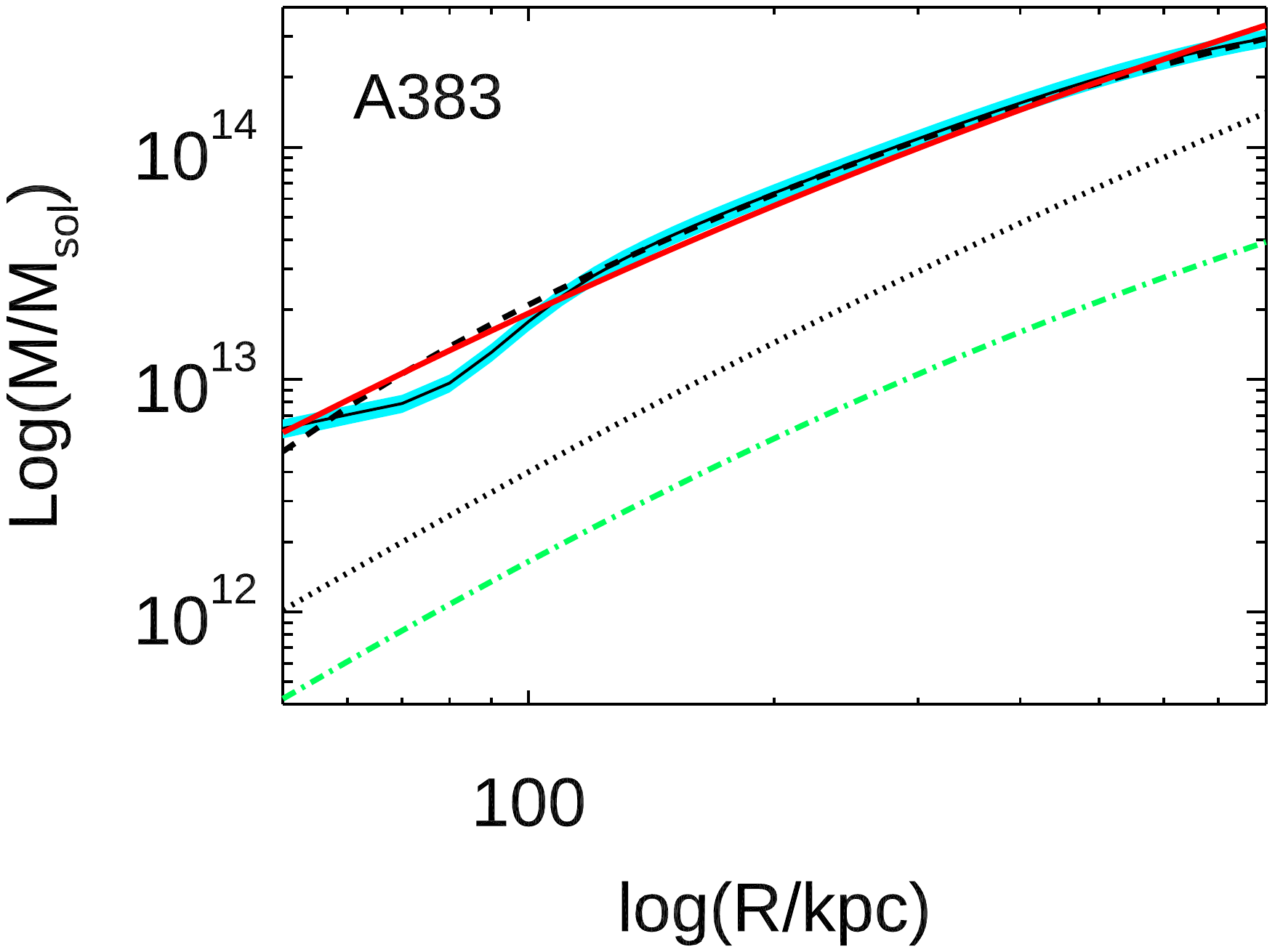}
\includegraphics[angle=0,width=0.48\textwidth]{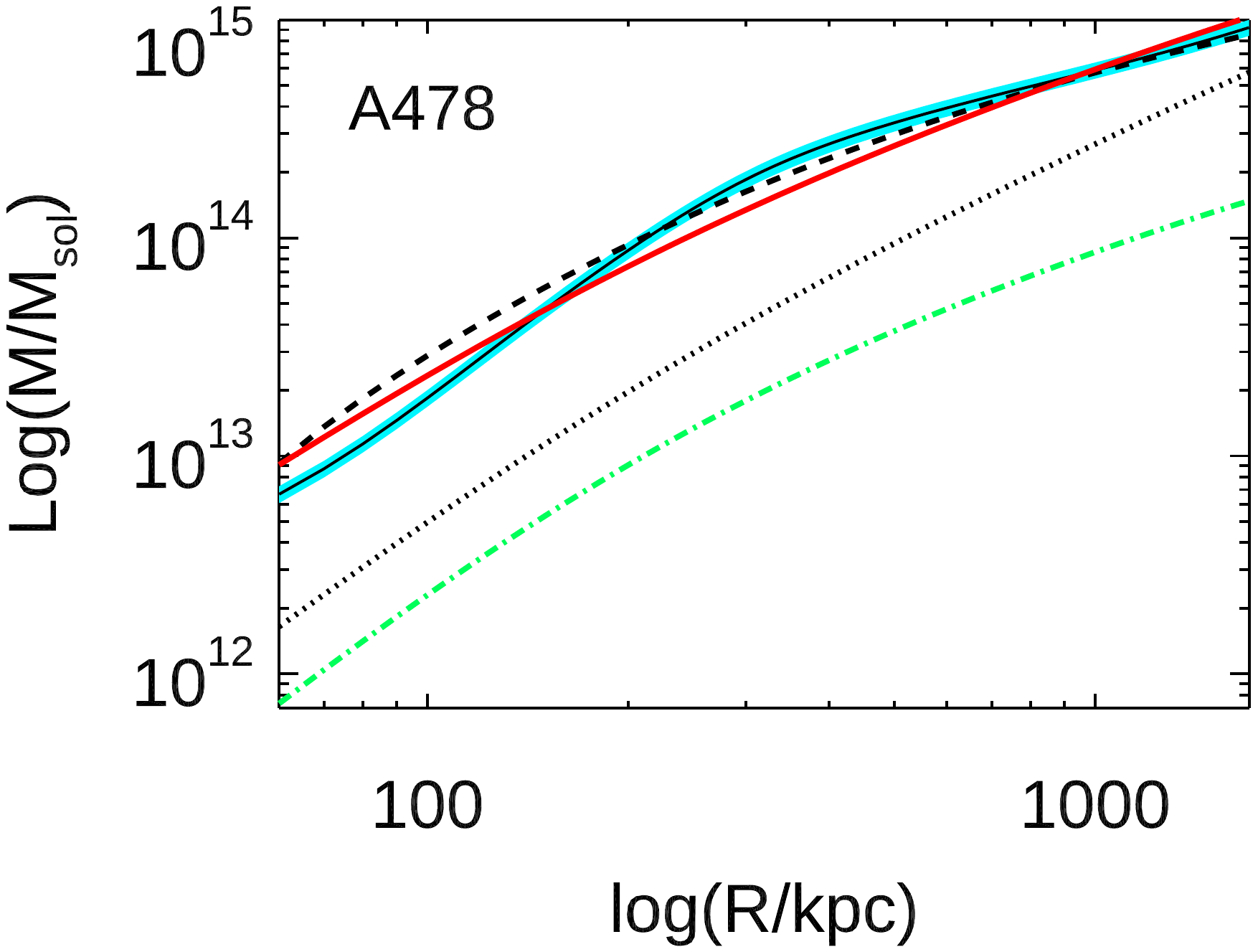}\\
\includegraphics[angle=0,width=0.48\textwidth]{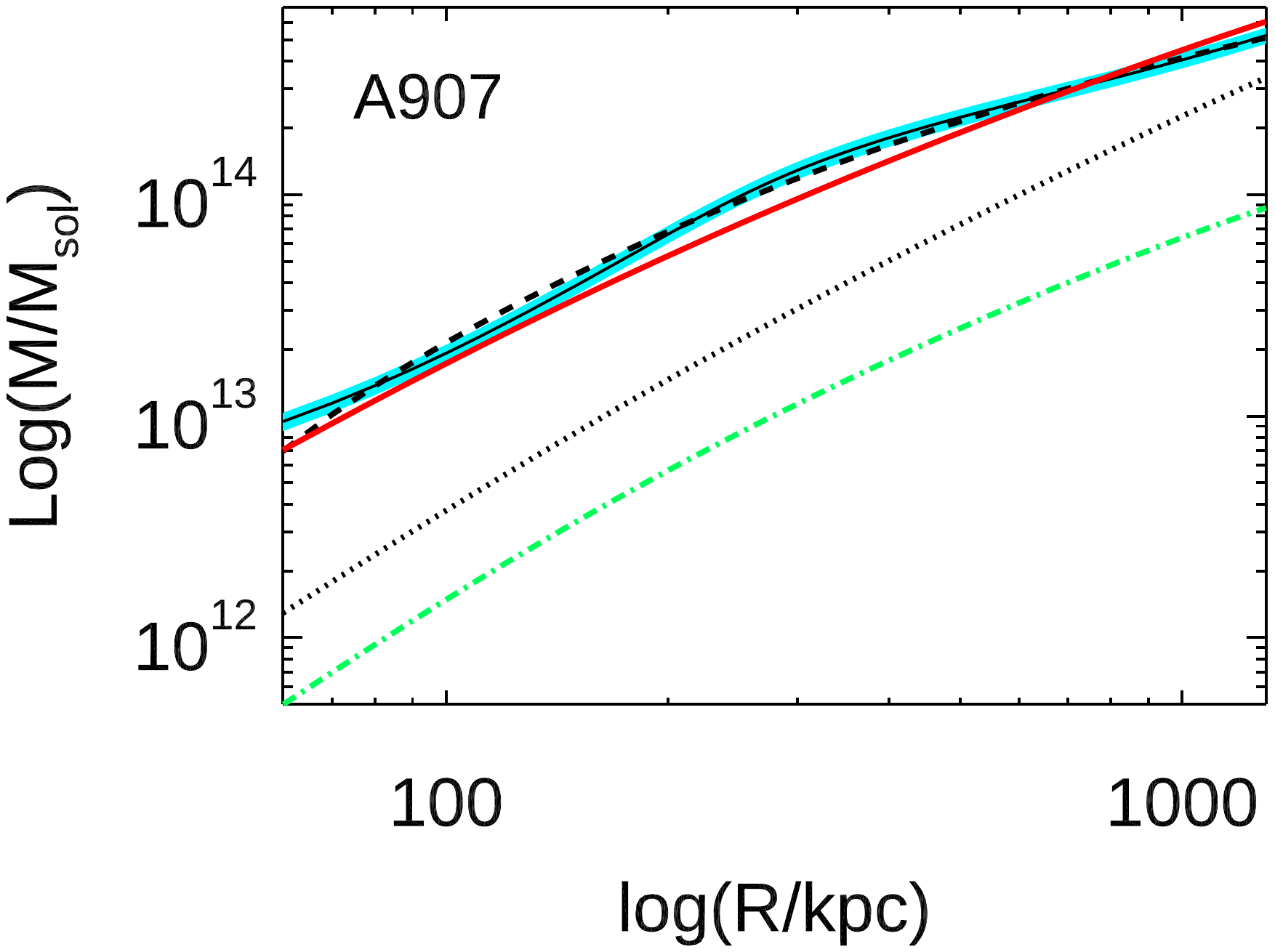}
\includegraphics[angle=0,width=0.48\textwidth]{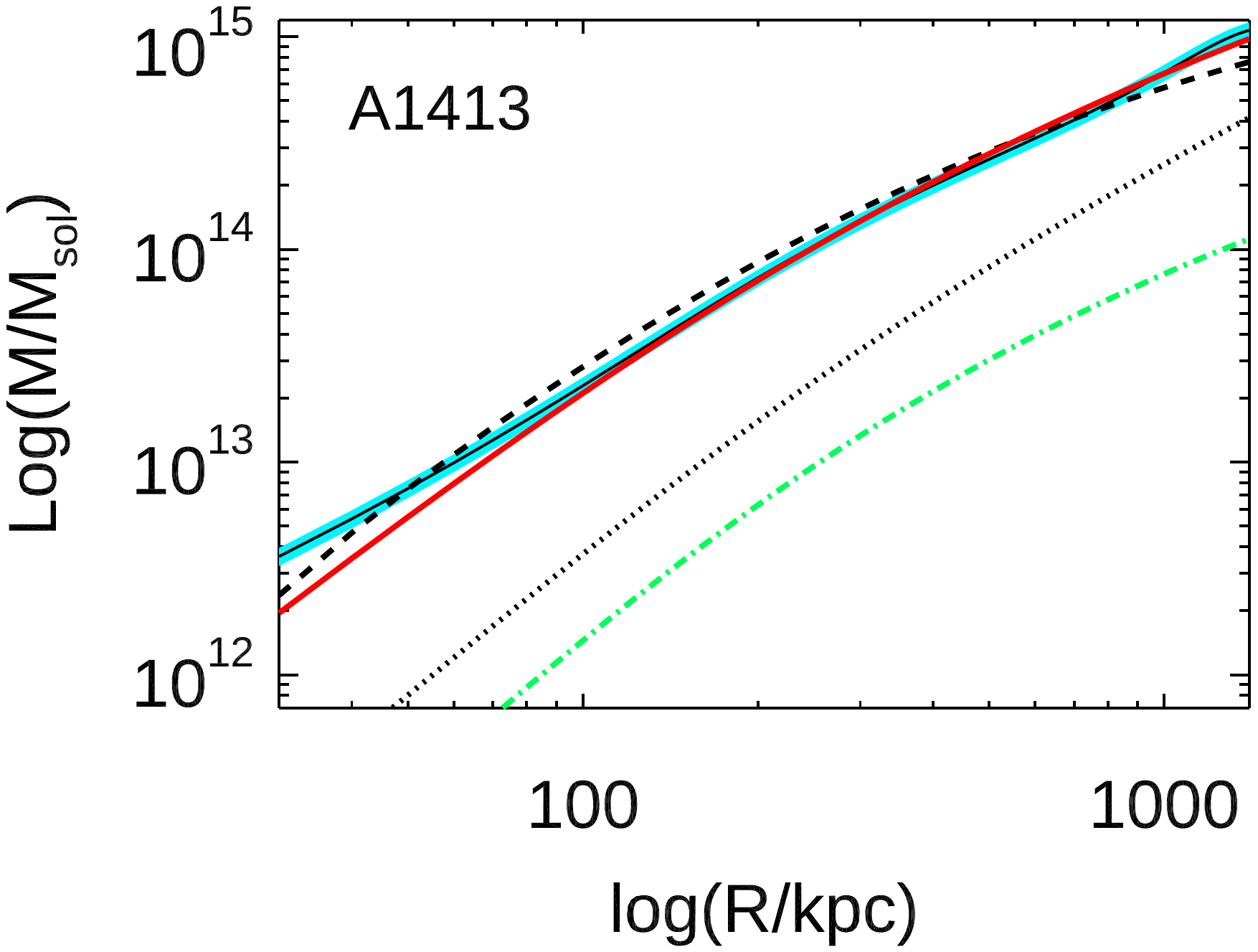}\\
\includegraphics[angle=0,width=0.48\textwidth]{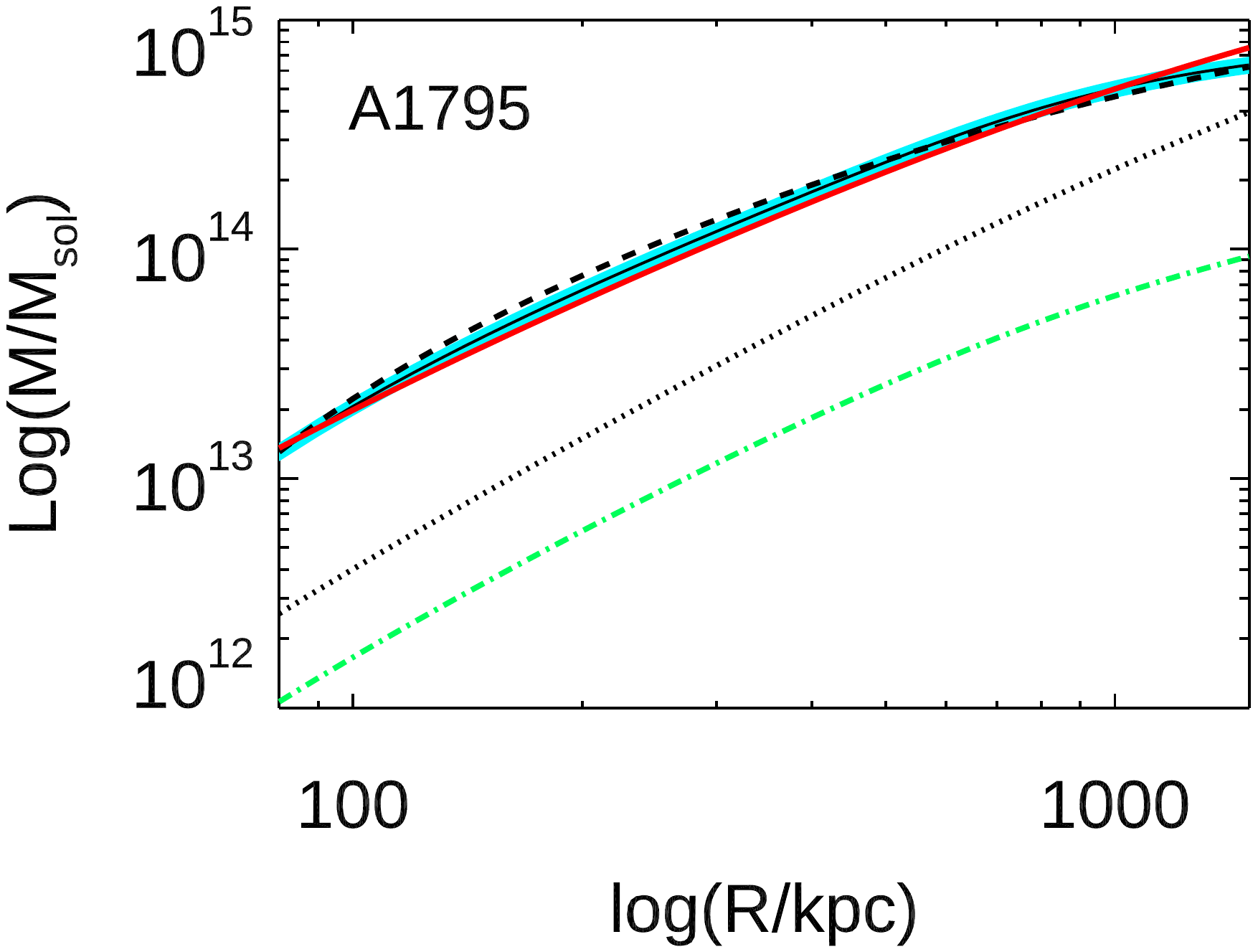}
\includegraphics[angle=0,width=0.48\textwidth]{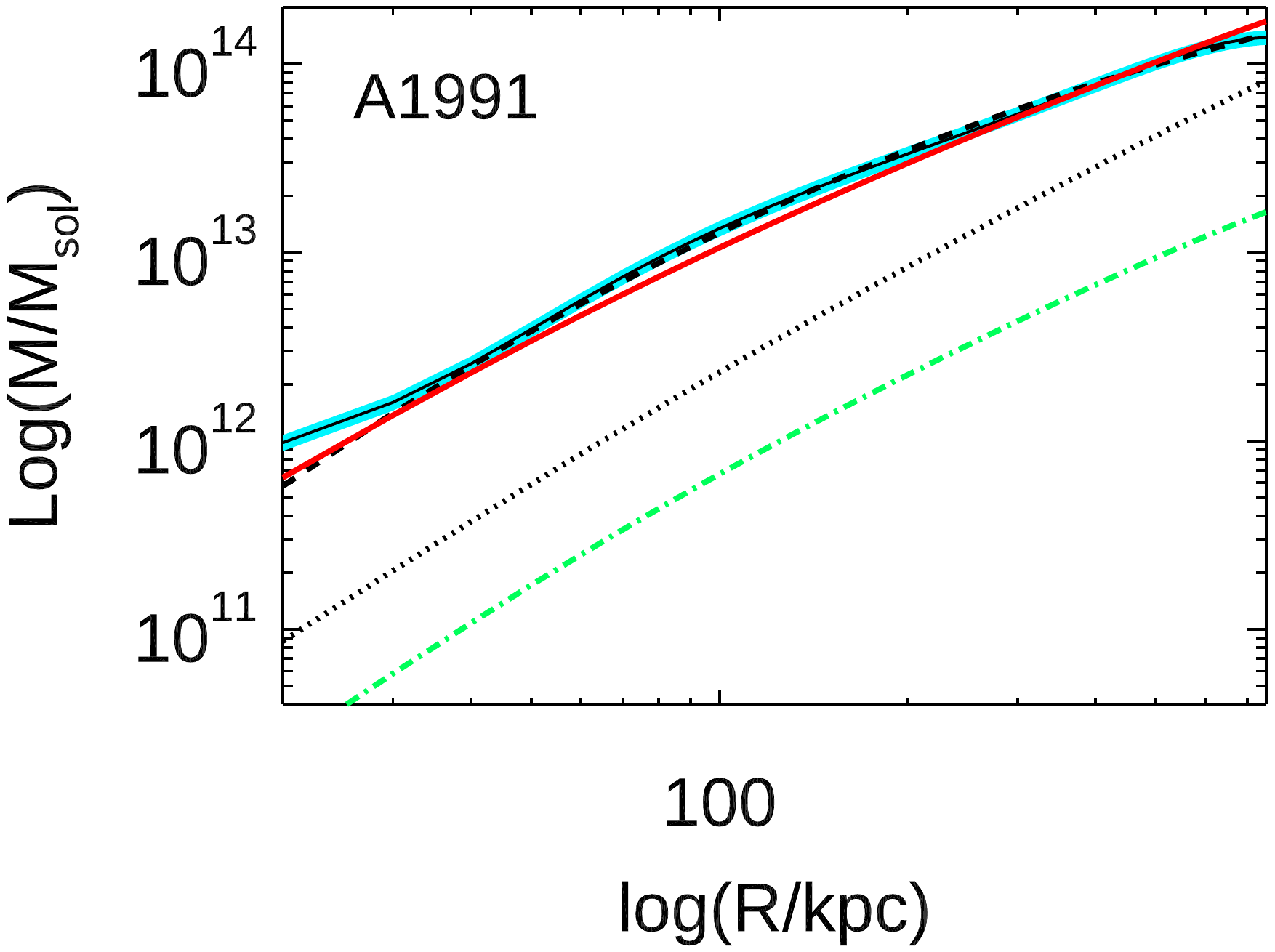}
\caption{Plots of total mass of galactic clusters
within radius R (assuming spherical symmetry). The solid black line is the virial mass and surrounding the black line is a blue shaded region depicting $1-\sigma$ errors; The dot-dashed green line is gas mass; The dotted black line is MOND (effective mass); The dashed black line is CDM; The solid red line is MDM with $\alpha = 100$.}
\label{fig:cluster}
\end{figure*}

In Ref.~\citen{Edmonds:2016tio} we compared MDM mass profiles 
with the observed (virial) mass profiles in a sample of 13 relaxed 
galaxy clusters given in Ref.~\citen{Vikhlinin:2005mp}. 
These authors of Ref.~\citen{Vikhlinin:2005mp}
analyzed all available \textit{Chandra} data which were of
sufficient quality to determine mass profiles
robustly out to large radii ($\sim 0.75\,r_{500}$ and
extended past $r_{500}$ in five clusters).
With the dark matter mass profile predicted by MDM given by 
Eq.~(\ref{clustermassp}), the total mass, 
the sum of dark matter and baryonic matter, is compared
with the virial mass for the sample of galaxy clusters in Fig.~\ref{fig:cluster}.
For comparison, we include dynamical masses predicted by CDM
(Eq.~(\ref{NFWrho})) and MOND (Eq.~(\ref{MpMOND})).

We see in Fig.~\ref{fig:cluster} that, with $\alpha \sim 100$ and
$r_\textrm{MDM} \sim 10$~kpc,
the MDM mass profiles fit the
virial mass data
well.  The fits for MDM mass profiles are as good as those for NFW.
On the other hand, the MOND (effective) mass profiles fail to
reproduce the virial mass profiles in magnitudes and in shape.


\renewcommand{\thefigure}{\arabic{figure} (Cont.)}
\addtocounter{figure}{-1}

\begin{figure*}
\begin{flushleft}
\includegraphics[angle=0,width=0.48\textwidth]{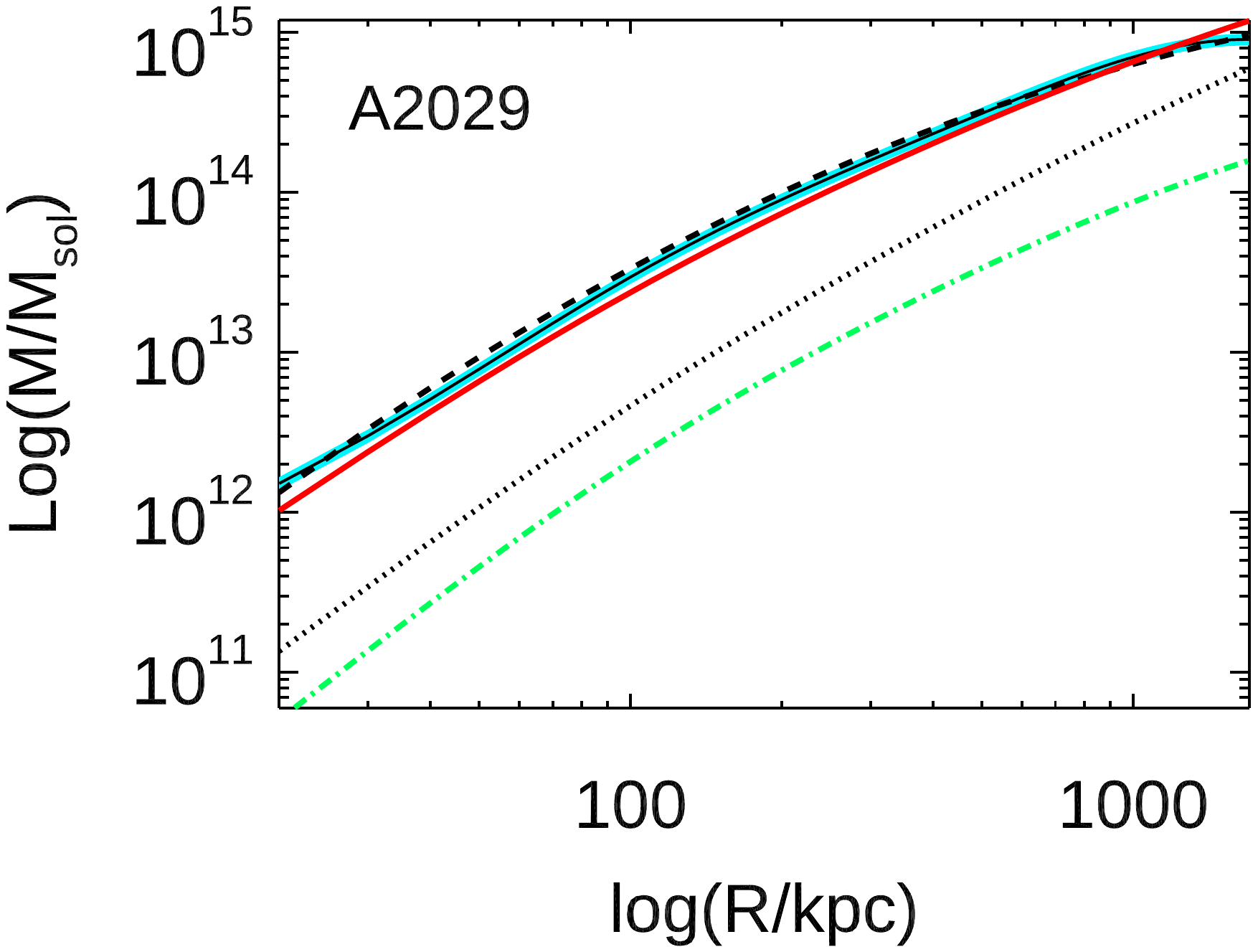}
\includegraphics[angle=0,width=0.48\textwidth]{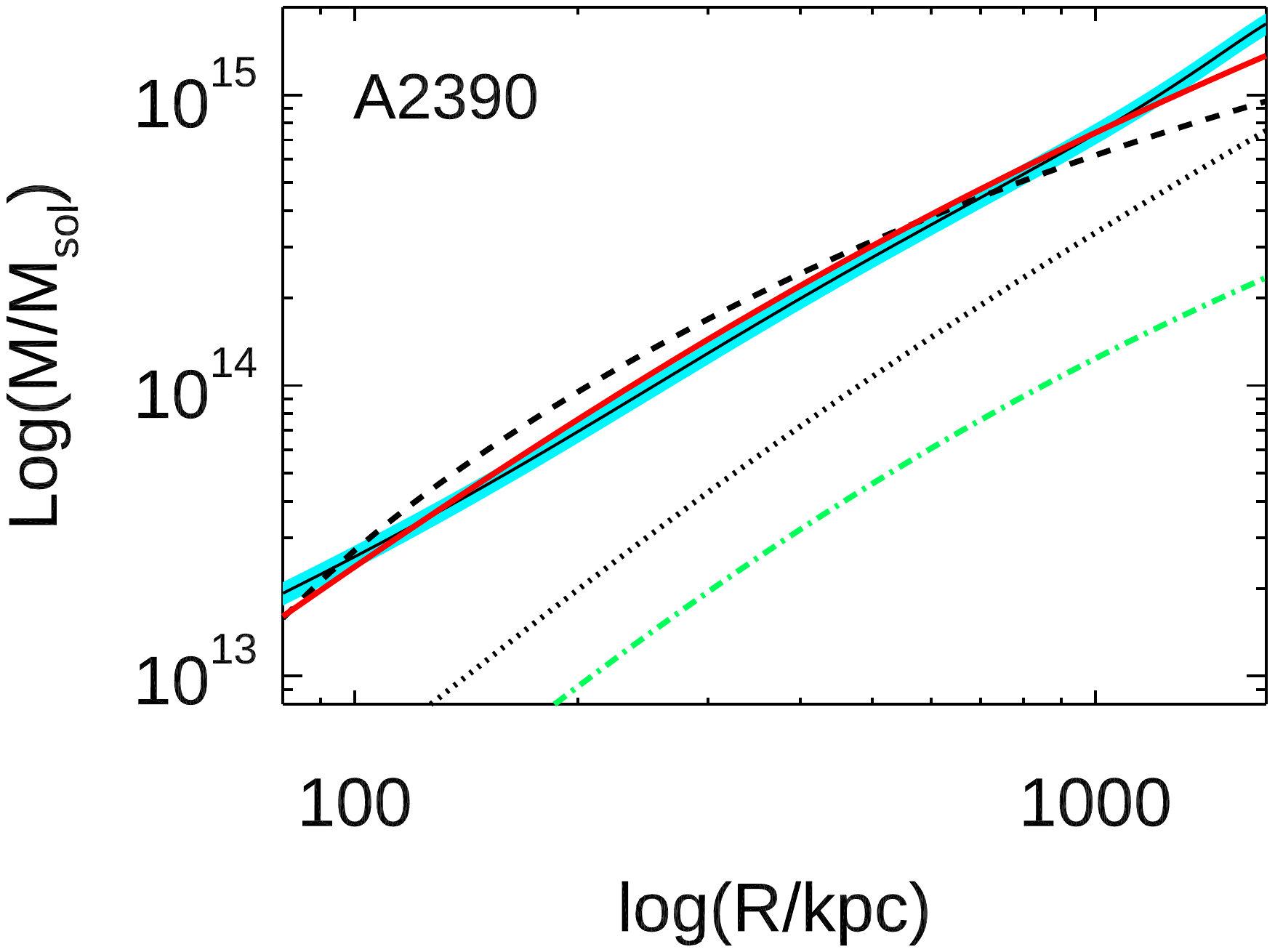}\\
\includegraphics[angle=0,width=0.48\textwidth]{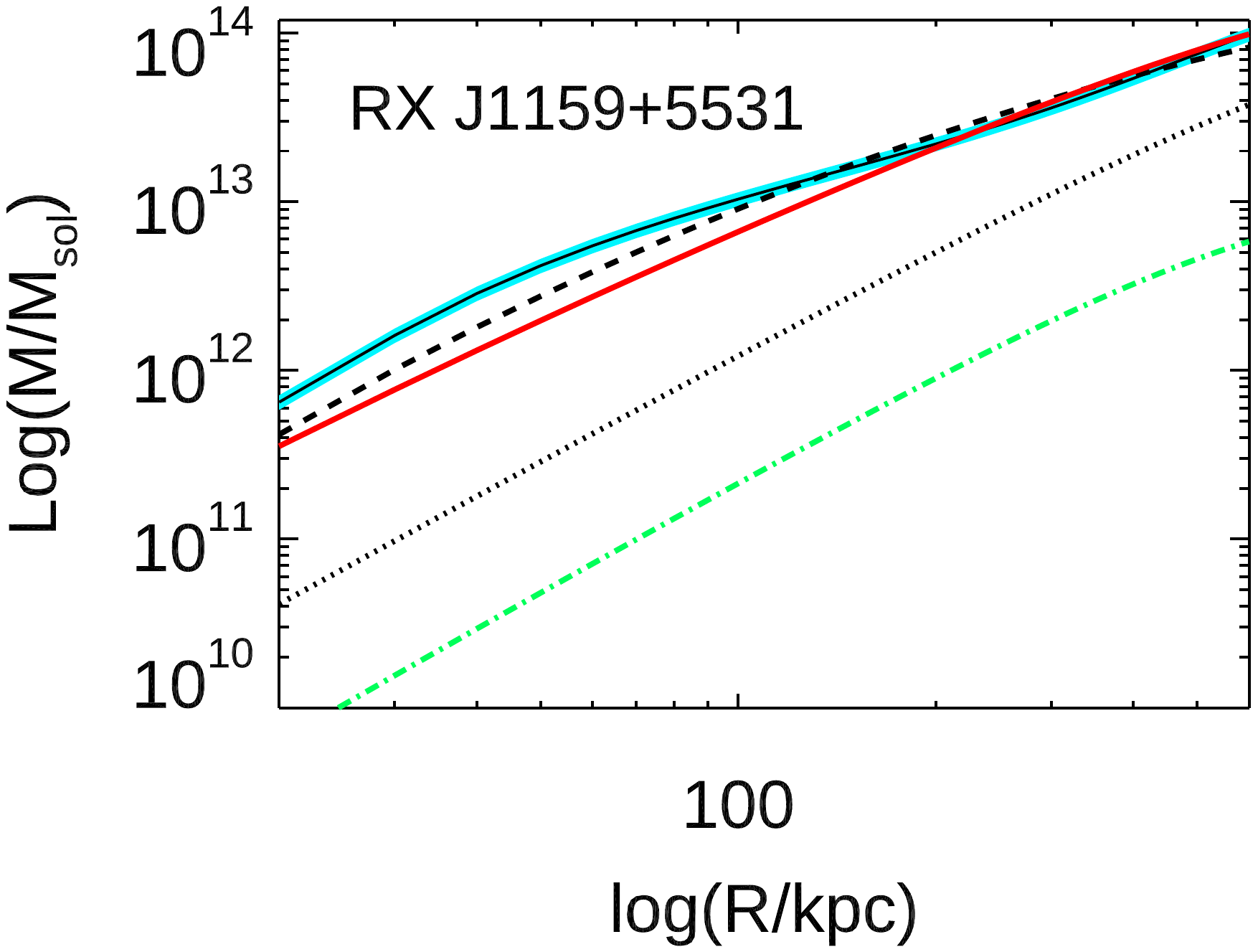}
\includegraphics[angle=0,width=0.48\textwidth]{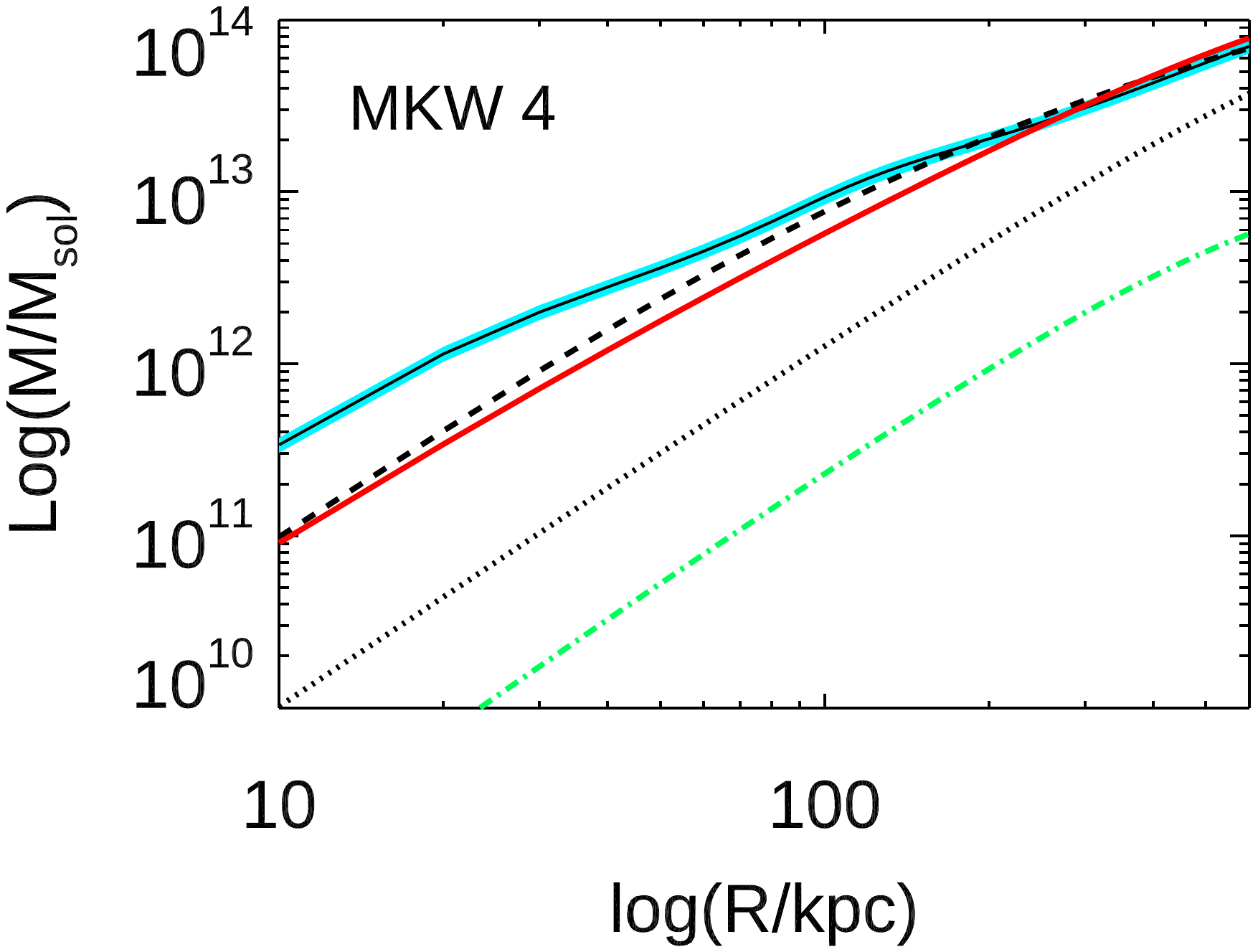}\\
\includegraphics[angle=0,width=0.48\textwidth]{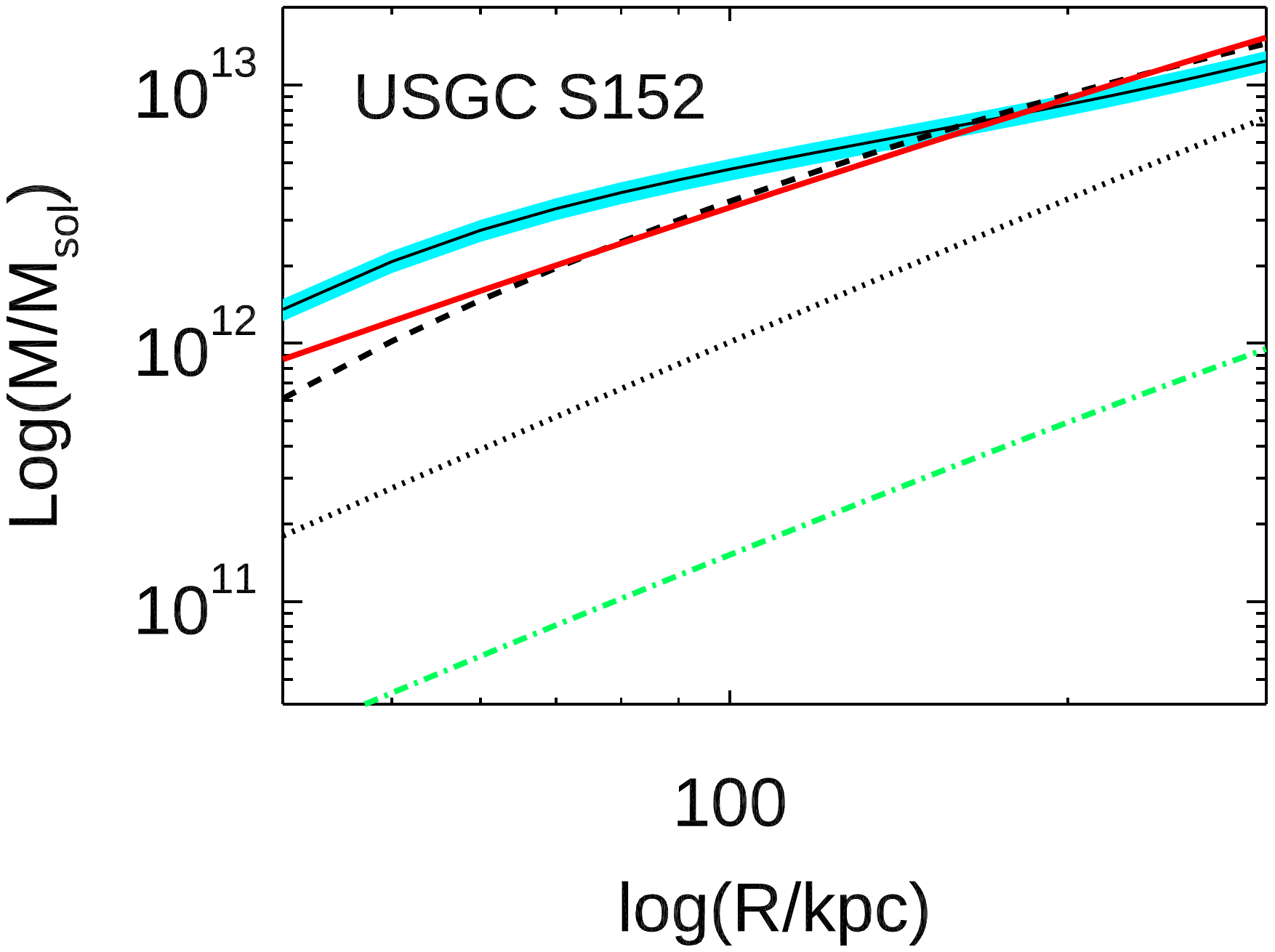}
\end{flushleft}
\caption{}
\end{figure*}

\renewcommand{\thefigure}{\arabic{figure}}


Finally we should recall that,
in Ref.~\citen{Edmonds:2013hba}, we fitted a sample of galactic rotation 
curves using the MDM mass profile
Eq.~(\ref{Mprime}) (see previous sub-section).
We would like 
to see if the new mass profile can work at galactic scales as well as it 
does at cluster scales.
For the galaxy clusters in our sample, we found $\alpha \sim50-100$  for the mass profile given in Eq.~(\ref{clustermassp}) fits the data well, while values $\lessapprox 50$ do not. For the galaxy cluster figures produced in this paper, we used $\alpha=50$. For comparison of data fits using $\alpha=100$, please see Ref.~\citen{Edmonds:2016tio}.
Since the fits to galactic rotation curves presented in 
Ref.~\citen{Edmonds:2013hba} were fit so well with Eq.~(\ref{Mprime}), we 
expect $\alpha \sim 1$ with $r_\textrm{MDM} \gg r$ in the generalized mass 
profile.

One might be disturbed by the appearance of such two radically different values for $\alpha$: $\alpha \sim 1$ for galactic scales and
$\alpha \sim 100$ for the scale of clusters. Here we note that $\alpha$ is essentially a boundary condition for the effective Unruh-Hawking-like
temperature, and thus it can be expected to have radically different values on radically different scales. We note that there is a numerological
coincidence involving the ratio of temperatures of the intracluster medium as compared to the interstellar medium in galaxies that can be around $100$.
Of course, this could be a coincidence and one would have to understand the origin of $\alpha$ better in order to say anything that is more definitive.
Finally, note that the difference in the values of $\alpha$ can be understood as an effective ``renormalizaton'' of $a_0$, needed to fit the cluster data
given the successful fits on galactic scales.
Note however that the rescaling of $a_0$ is not enough to make MOND work on cluster scales. MOND has a hard time reproducing the necessary profiles which are
captured by Tolman-like $r$ dependent factors in the context of our MDM proposal.

%

\subsection{MDM and Cosmology}

To recapitulate: By generalizing the canonical gravitational thermodynamics 
arguments to de-Sitter space, we are led to a new model of dark matter,
which takes into account the observed correlation between dark matter and
baryonic matter on galactic scales. The resulting dark matter mass profile is, by
construction, sensitive to the fundamental acceleration parameter found in the
observed galactic data. Moreover, by taking into account the temperature variation 
at difference scales, and by translating it into the equivalent acceleration variation with scale,
we obtained a successful dark matter profile at cluster scales.  As already emphasized,
our MDM behaves more like MOND at 
galactic scales, but more like CDM at cluster scales.

However, so far we have not discussed the cosmological implications of MDM.
Cosmology provides perhaps the most persuasive evidence for the ``missing mass'' in the form of the
canonical $\Lambda CDM$ model, so we have to answer the
following question:
How does MDM fare at cosmic scales?
While we do not have concrete results yet, qualitative and heuristic
arguments seem to yield an optimistic picture. Let us briefly touch
on the issues of cosmology and gravitational lensing. 
As shown in Ref.~\citen{Ho:2010ca}, at cosmological scales we need to take into account of the fully relativistic sources. 
First, we concentrate on the isotropic and homogeneous situation
described with the FRW metric
\begin{equation}
ds^2 \;=\; -dt^2 + R(t) \left(dr^2 + r^2 d\Omega^2\right)\;,
\end{equation}
and we assume that the matter sources form a perfect fluid with the energy-momentum tensor
\begin{equation}
T_{ab} \;=\; (\rho + p)\, u_a u_b + p\, g_{ab}\;.
\end{equation}
Next, for a fully relativistic description, we need to replace ${M+M'}$  in
%
\begin{equation}
a_\mathrm{obs} \;
\;=\; \frac{G\, (\,M+M'\,) }{\tilde{r}^2} + 4\pi G 
p\,\tilde{r} - \dfrac{\Lambda}{3}\,\tilde{r}\;,
\end{equation}
by the active gravitational mass (Tolman-Komar (TK) mass)
\begin{equation}
M_\mathrm{TK} \;=\; \frac{1}{4 \pi G} \int dV R_{ab} u^a u^b\;,
\end{equation}
which by the Einstein equations of general relativity becomes 
\begin{equation}
M_\mathrm{TK} \;=\; 2 \int dV 
\left(T_{ab} - \dfrac{1}{2} g_{ab} T + \dfrac{\Lambda}{8\pi G} g_{ab}\right) u^a u^b
\;,
\end{equation}
or alternatively
\begin{equation}
M_\mathrm{TK} \;=\; \left(\dfrac{4}{3} \pi r^3 R^3\right)
\left[\left(\rho +3p\right) - \dfrac{\Lambda}{4\pi G}\right]. 
\end{equation}
Then it can be shown \cite{Ho:2010ca} that
\begin{equation}
\frac{1}{R} \dfrac{d^2 R}{dt^2} 
\;=\; -\dfrac{4\pi G}{3}\left(\rho + 3p\right) + \dfrac{\Lambda}{3},
\end{equation}
with $p$ being the pressure. Here $\tilde{r} =r\, R(t)$ denotes the 
physical 
radius, where $r$ and $R(t)$ are the comoving radius and scale factor 
respectively.
At cluster and cosmological scales, either $4\pi G p\tilde{r}$ or 
$\Lambda\tilde{r}/3$ or both could be significant and this 
may explain why MOND does not work well at the cluster and 
cosmological scales.
The main point is that, continuing the argument of Ref.~\citen{Ho:2010ca},  
using the above equation and the continuity equation $\dot{\rho} + 3 H(\rho +p)=0$
one obtains, for the cosmology of MDM, the other canonical 
Friedmann equation 
\begin{equation}
H^2 \;=\; \dfrac{8\pi G}{3}\rho + \dfrac{\Lambda}{3}\;.
\end{equation}
We anticipate that this fact 
will allow MDM to predict the correct cosmic microwave 
background (CMB) spectrum shapes as well as its characteristic alternating peaks.

Next, let us comment on strong gravitational lensing in the context of MDM and
MOND. (Recall that strong lensing refers to the formation of multiple images
of background sources by the central regions of some clusters.)  It is known
that
the critical surface density required for strong lensing is
\begin{equation}
\Sigma_c \;=\; \dfrac{1}{4 \pi} \, \dfrac{c H_0}{\GN} \, F(z_l, z_s)
\;,
\end{equation}
with $F \approx 10$ for typical clusters and background sources at
cosmological distances. 
Sanders \cite{Sanders:1998kv} argued
that, in the deep MOND limit, $\Sigma_\mathrm{MOND} \approx \ac / \GN$.  Recalling
that
numerically $\ac \approx c H_0 / 6$, Sanders concluded that
MOND cannot produce strong lensing on its own: $\Sigma_c \approx 5
\Sigma_\mathrm{MOND}$.  On the other hand, MDM mass distribution appears to be {\it
sufficient} for strong lensing
since the natural scale for the critical acceleration for MDM is $a_0 = c
H_{0} = 2 \pi \ac \approx 6 \ac$, five to six times that for MOND.

\section{Non-local aspects of MDM}

In this section we comment on the crucial feature of MDM contained in
Eq.~(\ref{coincidence}) and Eq.~(\ref{Mprime}) which indicate that modified dark
matter profile is sensitive to the cosmological constant, or equivalently, the Hubble parameter. 
But what is the microscopic basis for this correlation?  
We will now argue that it may have something to do with
the quanta of MDM being non-local (and non-particle-like), or more specifically,
the MDM quanta obeying an exotic statistics
known as infinite statistics or quantum Boltzmann statistics.
The evidence for such non-local nature of MDM quanta is most apparent in the effective 
dark matter mass profile derived from gravitational thermodynamics. 
It is very hard to see why the mass profiles coming from particle-like quanta of
some local effective field theory would be sensitive to a fundamental acceleration parameter, set by the
cosmological scale, even if one assumes some unusual dissipative baryonic dynamics.
Thus, it is not completely outlandish to expect that the MDM dark matter quanta should be
non-local and non-particle like.

In order to get some intuition about
such putative non-local features, without assuming anything about
the microscopic dynamics of MDM quanta,   let us \cite{Ho:2012ar} first reformulate MOND via an
effective gravitational dielectric medium, motivated by the analogy  
\cite{Blanchet:2006yt} between
Coulomb's law in a dielectric medium and Milgrom's law for MOND.
We start with the nonlinear electrostatics embodied in 
the Born-Infeld theory \cite{Gibbons:2001gy},
and write the corresponding gravitational Hamiltonian density as
\begin{equation}
H_g \;=\; \dfrac{b^2}{4\pi}\left(\,\sqrt{1+ \dfrac{D_g^2}{b^2}}-1\,\right)\;,
\end{equation}
where $D$ stands for the electric displacement vector and $b$ is the maximum
field strength in the Born-Infeld theory.
With ${\cal{A}}_0 \equiv b^2$ and $ \vec{{\cal{A}}} \equiv b \, \vec{D_g}$, the Hamiltonian
density becomes
\begin{equation}
H_g \;=\; \dfrac{1}{4\pi}\left(\,\sqrt{{\cal{A}}^2+{\cal{A}}_0^2}-{\cal{A}}_0\,\right)\;.
\end{equation} 
If we invoke energy equipartition
($H_g = \frac{1}{2}k_B T_\mathrm{eff}\,$) and
the Unruh temperature formula ($T_\mathrm{eff} = \dfrac{\hbar}{2\pi k_B c}\,
a_\mathrm{eff}\,$), and apply the equivalence principle (in identifying,
at least locally, the local accelerations $\vec{a}$ and $\vec{a}_0$
with the local gravitational fields
$\vec{{\cal{A}}}$ and $\vec{{\cal{A}}}_0$ respectively), then the effective acceleration
$a_{\mathrm{eff}}$
is identified as $a_\mathrm{eff} \equiv \sqrt{a^2+a_0^2}-a_0$.  
But this, in turn, implies that the Born-Infeld inspired force law takes the form 
$F_\mathrm{BI} = m\left(\sqrt{a^2+a_0^2}-a_0\right)\,$,
for a given test mass $m$,
which is precisely the MONDian force law.

To be a viable cold dark matter candidate, the quanta of
the MDM must
be much heavier than $k_B T_{\mathrm{eff}}$ since $T_{\mathrm{eff}}$, with its
quantum origin (being proportional to $\hbar$), is a very low temperature.
Now
recall that the equipartition theorem in general states that
the average of the Hamiltonian is given by
$\langle H \rangle = - \dfrac{\partial \log{Z(\beta)}}{\partial \beta}\,$,
where $\beta^{-1} = k_B T$.  To obtain
$\langle H \rangle = \dfrac{1}{2}k_B T$ per degree of freedom, even for
very low temperature,
we require the partition function $Z$ to be of the Boltzmann form
$Z = \exp(-\beta H)$.
But this is precisely the case of infinite statistics 
\cite{Greenberg:1989ty,Ho:2012ar}.

What is infinite statistics?  Succinctly, a Fock space realization of infinite
statistics is provided by (a $q$ deformation of) the commutation relations 
of the oscillators:
$a_k^{\phantom{\dagger}} a^{\dagger}_\ell = \delta_{k\ell}$ 
as described by the average of the bosonic and fermionic algebras.
It is known that a theory of particles 
obeying infinite statistics cannot be local\cite{Greenberg:1989ty}.  
For example, the expression for the number operator,
\begin{equation}
n_i \;=\; a_i^{\dagger} a_i^{\phantom{\dagger}} 
+ \sum_k a_k^{\dagger} a_i^{\dagger} a_i^{\phantom{\dagger}} a_k^{\phantom{\dagger}} 
+ \sum_\ell \sum_k a_\ell^{\dagger} a_k^{\dagger} a_i^{\dagger} 
  a_i^{\phantom{\dagger}} a_k^{\phantom{\dagger}} a_\ell^{\phantom{\dagger}} 
+ \cdots\;,
\label{number}
\end{equation}
is both nonlocal and nonpolynomial in the field operators,
and so is the Hamiltonian.  The lack of
locality may make it difficult to formulate a relativistic version of the
theory; but it appears that a non-relativistic theory can be developed.
Lacking locality also means that the familiar spin-statistics relation is 
no longer valid for particles obeying infinite statistics; hence
they can have any spin.  Remarkably, the CPT theorem and cluster
decomposition have been shown to hold despite the lack of locality
\cite{Greenberg:1989ty}. A good example of infinite statistics is the large $N$ (planar) limit
of $SU(N)$  Yang-Mills theory.

This type of statistics suggests that the MDM quanta might be of quasi-one-dimensional
nature, as is natural for string-like excitations found in the context of the planar Yang-Mills theory.
Such non-local excitations are believed to be
an important ingredient in the formulation of quantum gravity, and thus
the lack of locality for theories of infinite statistics may not be a
defect; it can actually be a virtue.  Perhaps it is this lack of
locality that makes it easier to incorporate gravitational interactions in
the theory.  Quantum gravity and infinite statistics appear to fit 
together nicely, and non-locality seems to be a common feature of both of them
\cite{Ho:2010ca,Ho:2012ar}. Conceivably it is the extended nature of 
the MDM quanta that connects them to the cosmological constant/dark energy
and the Hubble parameter, two (related) global aspects of spacetime.
As we will argue in the concluding section, such extended, non-particle excitations may be the generic features of 
quantum theory of gravity, pointing towards a more fundamental origin of the MDM quanta.

\section{Summary and Discussion}

In the conclusion of this review we summarize our main points and address some fundamental underpinnings of our proposal.

At the moment the MDM proposal is not rooted in any fundamental physics, and there is no specific candidate
for the MDM quantum. Here we want to make some comments about the possible fundamental rationale for MDM and the
nature of the MDM quantum.
At this point one might ask: why do we need such unusual quanta? The answer lies, as already mentioned in the introduction, in the comments made by Kaplinghat and Turner \cite{Kaplinghat:2001me}
in the context of CDM regarding the observed Milgrom scaling in the galactic rotations curves.
It is very hard for CDM to reproduce the apparent {\it universality} of this scaling, even though the work on this topic is continuing.
Note that Milgrom's scaling is usually associated with modified Newtonian dynamics (MOND)  which denies dark matter. As emphasized in this review, this is NOT our point. What we have been asking is the following: Given the unknown nature of dark matter, what constraint does the Milgrom scaling impose on its quanta (at all scales)? Can this scaling be accounted for by modifying CDM-like mass profiles? Is this scaling compatible with a non-particle nature of dark matter, thus explaining why we have, so far, not seen the assumed particle quanta of dark matter? If so, what is the ``smoking gun'' signal for such non-local, non-particle quanta of dark matter?

Our central observation is that Milgrom's scaling implies a non-local mixing of physics in the dark matter and dark energy sector. Simply put, the observed value for the cosmological constant can be associated with the acceleration parameter found in Milgrom's scaling. Then the question is, whether a dark matter profile can be found which accounts for the galactic rotation curves and is sensitive to this acceleration parameter. As summarized in this review, we have proposed precisely such a dark matter profile
based on a heuristic viewpoint rooted in gravitational thermodynamics. At
cluster as well as cosmological scales, modified dark matter behaves as CDM, but at the galactic scale, modified dark matter implies the scaling behavior usually associated with MOND.
However, we emphasize once again that MDM is \textit{not} MOND -- MDM is an unconventional, and most probably non-local form of dark matter.
Note that the modified dark matter profile can be related to the classic CDM mass profiles, 
such as the NFW mass profile \cite{Navarro:1995iw,Navarro:1996gj}, 
in certain limits, which might be viewed as another justification for the MDM proposal. Also, the MDM proposal can be viewed as a unification between the dark energy, dark matter and baryonic matter sectors.

We have already made a comment regarding the extended nature of the MDM quanta. This might seem very surprising given the fact that the successful existing theoretical tools of fundamental physics are all rooted in the concept of locality in spacetime. Locality is indeed one of the cornerstones of modern physics. It is one of the key properties underlying effective field theory, which is widely considered as a universal language for describing the fundamental physics and which captures the main features of known particle physics at low energy scales, and thus it is a prominent tool in the standard approaches towards the particle-like CDM quanta. However, it is becoming increasingly clear that non-locality may play a central role in solving some of the most outstanding puzzles in fundamental, such as the vacuum energy problem, the black hole information paradox, as well as the deep and central non-local features of quantum theory and quantum field theory, including the naturalness and hierarchy problems. If this is the case, then the tools of effective field theory are inadequate, and we must develop new ideas and techniques. Similarly, the fundamental concepts of differential geometry constitute the basic mathematical language of general relativity, our deepest theory of space, time and gravity. However, these concepts seem to be just a limiting case of the mathematical language of generalized geometry required to talk about various new non-local phenomena encountered in string theory, such as T-duality (the intrinsic relation between short and long distances).
The concept of non-locality is brought to the forefront in the context of the so-called metastring theory, proposed by Freidel, Leigh and Minic \cite{Freidel:2013zga,Freidel:2014qna,Freidel:2015pka,Freidel:2015uug,Freidel:2016pls, Freidel:2017wst, Freidel:2017nhg}. The metastring formulation of quantum gravity introduces a new concept of quantum spacetime called modular spacetime which, surprisingly, sheds light on the foundational issues in quantum theory and quantum field theory. Also, by pointing out that the effective field theoretic description of strings is generically non-commutative, yet covariant, at long distance, metastring theory sheds light on why effective (Wilsonian) local quantum field theory is so successful in so many domains of physics, and why it is bound to be transcended in more general situations.

Given the successful phenomenology of MDM presented in this review, we expect that the fundamental non-locality and non-commutativity advocated by metastring theory show up in the context of the dark sector, involving dark energy and dark matter. In particular, such fundamental non-commutativity and non-locality of the metastring imply that the spectrum of low energy excitations can be non-particle-like. This happens in non-commutative field theories  \cite{Grosse:2004yu}
which can be viewed as toy models of the metastring \cite{Freidel:2013zga,Freidel:2014qna,Freidel:2015pka,Freidel:2015uug,Freidel:2016pls, Freidel:2017wst, Freidel:2017nhg}.
Such excitations can also have unusual statistics. We expect that such non-local excitations (as part of a covariant non-commutative formulation, as implied by the metastring \cite{ Freidel:2017wst, Freidel:2017nhg}) can serve as unusual dark matter quanta.
Furthermore, the zero mode sector of the metastring, does not look like the classic relativistic particle, but as 
a non-local extension of the usual quanta of free local fields\footnote{The theory of such ``metaparticles'' is currently being developed by Laurent Freidel, Jerzy Kowalski-Glikman, Rob Leigh and Djordje Minic.}. Roughly, the zero mode sector looks like two entangled particles, and it could be viewed as a model for
modified dark matter quanta discussed in this review, implying, perhaps naturally, the observed relation between dark matter and baryonic mass profiles at certain scales.

We can speculate about the extended nature of MDM as follows:
for such extended excitations the change in momentum is proportional to the change in distance (i.e. such excitations expand with an influx of energy) $\delta p \sim \alpha\, \delta x$.
Assuming a non-relativistic situation for which $p \sim \dot{x}$ then we get that change acceleration is proportional to change velocity, or equivalently, momentum, and after using the Heisenberg relation $\delta p\, \delta x \sim \hbar$,
we get that the acceleration is proportional to the inverse of distance, which is exactly what the Milgrom scaling demands. Thus, extended, one-dimensional, excitations could model non-local MDM quanta.

We conclude with our to-do list.
First we aim to understand other static clusters in which the CDM apparently encounters certain
issues (such as Abell 1689). Then we plan to study 
concrete constraints from gravitational lensing 
and the dynamical clusters, such as the famous bullet cluster, on MDM. 
Specifically we would like to answer these questions: 
Can we distinguish MDM from CDM in these physical situations? How strongly
coupled is MDM to baryonic matter?  How does MDM self-interact? We would
also like to test MDM at 
cosmic scales by studying the acoustic peaks in the CMB as well as by doing
simulations of structure formation. 
If the quanta of MDM indeed obey infinite statistics,
as suggested in Ref.~\citen{Ho:2012ar}, it is 
possible that there are (dark) stars made of such quanta. If so, what are some of their observational signatures?

Finally, it is imperative to develop a deeper 
understanding of the fundamental nature of the non-local MDM quanta.  
Recent reformulations of quantum gravity and string theory 
\cite{Freidel:2013zga,Freidel:2014qna,Freidel:2015pka,Freidel:2015uug,Freidel:2016pls} may be helpful in this effort.
A more concrete theory for MDM quanta will allow us to test the
whole scheme of MDM
at colliders, dark matter direct detection experiments and indirect detection 
experiments.  Possibly unusual non-particle phenomenology awaits to be 
discovered.  And from our perspective, this may be regarded as quantum 
gravity phenomenology in disguise. 


\section*{Acknowledgments:}

We are grateful for helpful discussions with J.~Beacom, C.~Frenk, S.~Horiuchi, 
J.~Khoury, P.~Mannheim, S~McGaugh and J.~Moffat.
We would like to especially thank our former collaborator C.~M. Ho
for his tremendous contributions to every aspect of our work on MDM.
DM was supported by the U.S. Department of Energy under contract 
DE-FG02-13ER41917, Task A and by J. Mark Sowers.
YJN was supported in part by the Bahnson Fund and the Kenan Professorship Research Fund of UNC. 


\bibliographystyle{hunsrt}
\bibliography{mdm}

\begin{thebibliography}{100}

\bibitem{Bertone:2016nfn}
Gianfranco Bertone and Dan Hooper.
\newblock {A History of Dark Matter}.
\newblock {\em Submitted to: Rev. Mod. Phys.}, 2016, 1605.04909.

\bibitem{Oort}
J.~H. {Oort}.
\newblock {The force exerted by the stellar system in the direction
  perpendicular to the galactic plane and some related problems}.
\newblock {\em Bull. Astron. Inst. Netherlands}, 6:249, August 1932.

\bibitem{Zwicky:1933gu}
F.~Zwicky.
\newblock {Die Rotverschiebung von extragalaktischen Nebeln}.
\newblock {\em Helv. Phys. Acta}, 6:110--127, 1933.
\newblock [Gen. Rel. Grav.41,207(2009)].

\bibitem{Rubin:1970zza}
Vera~C. Rubin and W.~Kent Ford, Jr.
\newblock {Rotation of the Andromeda Nebula from a Spectroscopic Survey of
  Emission Regions}.
\newblock {\em Astrophys. J.}, 159:379--403, 1970.

\bibitem{Rubin:1978kmz}
Vera~C. Rubin, W.~Kent Ford, Jr., and Norbert Thonnard.
\newblock {Extended rotation curves of high-luminosity spiral galaxies. IV.
  Systematic dynamical properties, Sa through Sc}.
\newblock {\em Astrophys. J.}, 225:L107--L111, 1978.

\bibitem{Rubin:1980zd}
V.~C. Rubin, N.~Thonnard, and W.~K. Ford, Jr.
\newblock {Rotational properties of 21 SC galaxies with a large range of
  luminosities and radii, from NGC 4605 /R = 4kpc/ to UGC 2885 /R = 122 kpc/}.
\newblock {\em Astrophys. J.}, 238:471, 1980.

\bibitem{Peebles:2002gy}
P.~J.~E. Peebles and Bharat Ratra.
\newblock {The Cosmological constant and dark energy}.
\newblock {\em Rev. Mod. Phys.}, 75:559--606, 2003, astro-ph/0207347.

\bibitem{Bekenstein:2004ne}
Jacob~D. Bekenstein.
\newblock {Relativistic gravitation theory for the MOND paradigm}.
\newblock {\em Phys. Rev.}, D70:083509, 2004, astro-ph/0403694.
\newblock [Erratum: Phys. Rev.D71,069901(2005)].

\bibitem{Mannheim:2012qw}
Philip~D. Mannheim and James~G. O'Brien.
\newblock {Galactic rotation curves in conformal gravity}.
\newblock {\em J. Phys. Conf. Ser.}, 437:012002, 2013, 1211.0188.

\bibitem{Moffat:2013}
J.~W. Moffat and S.~Rahvar.
\newblock {The MOG weak field approximation and observational test of galaxy
  rotation curves}.
\newblock {\em Mon. Not. Roy. Astron. Soc.}, 436(2):1439--1451, 2013,
  1306.6383.

\bibitem{Moffat:2014}
J.~W. Moffat and S.~Rahvar.
\newblock {The MOG weak field approximation � II. Observational test of
  Chandra X-ray clusters}.
\newblock {\em Mon. Not. Roy. Astron. Soc.}, 441(4):3724--3732, 2014,
  1309.5077.

\bibitem{alco96}
C.~{Alcock}, R.~A. {Allsman}, T.~S. {Axelrod}, D.~P. {Bennett}, K.~H. {Cook},
  K.~C. {Freeman}, K.~{Griest}, J.~A. {Guern}, M.~J. {Lehner}, S.~L.
  {Marshall}, H.-S. {Park}, S.~{Perlmutter}, B.~A. {Peterson}, M.~R. {Pratt},
  P.~J. {Quinn}, A.~W. {Rodgers}, C.~W. {Stubbs}, and W.~{Sutherland}.
\newblock {The MACHO Project First-Year Large Magellanic Cloud Results: The
  Microlensing Rate and the Nature of the Galactic Dark Halo}.
\newblock {\em ApJ}, 461:84, April 1996, astro-ph/9506113.

\bibitem{alco97}
C.~{Alcock}, R.~A. {Allsman}, D.~{Alves}, T.~S. {Axelrod}, A.~C. {Becker},
  D.~P. {Bennett}, K.~H. {Cook}, K.~C. {Freeman}, K.~{Griest}, J.~{Guern},
  M.~J. {Lehner}, S.~L. {Marshall}, B.~A. {Peterson}, M.~R. {Pratt}, P.~J.
  {Quinn}, A.~W. {Rodgers}, C.~W. {Stubbs}, W.~{Sutherland}, and D.~L. {Welch}.
\newblock {The MACHO Project Large Magellanic Cloud Microlensing Results from
  the First Two Years and the Nature of the Galactic Dark Halo}.
\newblock {\em ApJ}, 486:697--726, September 1997, astro-ph/9606165.

\bibitem{Alcock:2000ph}
C.~Alcock et~al.
\newblock {The MACHO project: Microlensing results from 5.7 years of LMC
  observations}.
\newblock {\em Astrophys. J.}, 542:281--307, 2000, astro-ph/0001272.

\bibitem{calchi05}
S.~{Calchi Novati}, S.~{Paulin-Henriksson}, J.~{An}, P.~{Baillon},
  V.~{Belokurov}, B.~J. {Carr}, M.~{Cr{\'e}z{\'e}}, N.~W. {Evans},
  Y.~{Giraud-H{\'e}raud}, A.~{Gould}, P.~{Hewett}, P.~{Jetzer}, J.~{Kaplan},
  E.~{Kerins}, S.~J. {Smartt}, C.~S. {Stalin}, Y.~{Tsapras}, M.~J. {Weston},
  and {Point-Agape Collaboration}.
\newblock {POINT-AGAPE pixel lensing survey of M 31. Evidence for a MACHO
  contribution to galactic halos}.
\newblock {\em A\&A}, 443:911--928, December 2005, astro-ph/0504188.

\bibitem{tisser07}
P.~{Tisserand}, L.~{Le Guillou}, C.~{Afonso}, J.~N. {Albert}, J.~{Andersen},
  R.~{Ansari}, {\'E}.~{Aubourg}, P.~{Bareyre}, J.~P. {Beaulieu}, X.~{Charlot},
  C.~{Coutures}, R.~{Ferlet}, P.~{Fouqu{\'e}}, J.~F. {Glicenstein},
  B.~{Goldman}, A.~{Gould}, D.~{Graff}, M.~{Gros}, J.~{Haissinski},
  C.~{Hamadache}, J.~{de Kat}, T.~{Lasserre}, {\'E}.~{Lesquoy}, C.~{Loup},
  C.~{Magneville}, J.~B. {Marquette}, {\'E}.~{Maurice}, A.~{Maury},
  A.~{Milsztajn}, M.~{Moniez}, N.~{Palanque-Delabrouille}, O.~{Perdereau},
  Y.~R. {Rahal}, J.~{Rich}, M.~{Spiro}, A.~{Vidal-Madjar}, L.~{Vigroux},
  S.~{Zylberajch}, and {EROS-2 Collaboration}.
\newblock {Limits on the Macho content of the Galactic Halo from the EROS-2
  Survey of the Magellanic Clouds}.
\newblock {\em A\&A}, 469:387--404, July 2007, astro-ph/0607207.

\bibitem{cyburt04}
R.~H. {Cyburt}.
\newblock {Primordial nucleosynthesis for the new cosmology: Determining
  uncertainties and examining concordance}.
\newblock {\em Phys. Rev. D}, 70(2):023505, July 2004, astro-ph/0401091.

\bibitem{smoot92}
G.~F. {Smoot}, C.~L. {Bennett}, A.~{Kogut}, E.~L. {Wright}, J.~{Aymon}, N.~W.
  {Boggess}, E.~S. {Cheng}, G.~{de Amici}, S.~{Gulkis}, M.~G. {Hauser},
  G.~{Hinshaw}, P.~D. {Jackson}, M.~{Janssen}, E.~{Kaita}, T.~{Kelsall},
  P.~{Keegstra}, C.~{Lineweaver}, K.~{Loewenstein}, P.~{Lubin}, J.~{Mather},
  S.~S. {Meyer}, S.~H. {Moseley}, T.~{Murdock}, L.~{Rokke}, R.~F. {Silverberg},
  L.~{Tenorio}, R.~{Weiss}, and D.~T. {Wilkinson}.
\newblock {Structure in the COBE differential microwave radiometer first-year
  maps}.
\newblock {\em ApJL}, 396:L1--L5, September 1992.

\bibitem{debernar00}
P.~{de Bernardis}, P.~A.~R. {Ade}, J.~J. {Bock}, J.~R. {Bond}, J.~{Borrill},
  A.~{Boscaleri}, K.~{Coble}, B.~P. {Crill}, G.~{De Gasperis}, P.~C. {Farese},
  P.~G. {Ferreira}, K.~{Ganga}, M.~{Giacometti}, E.~{Hivon}, V.~V. {Hristov},
  A.~{Iacoangeli}, A.~H. {Jaffe}, A.~E. {Lange}, L.~{Martinis}, S.~{Masi},
  P.~V. {Mason}, P.~D. {Mauskopf}, A.~{Melchiorri}, L.~{Miglio}, T.~{Montroy},
  C.~B. {Netterfield}, E.~{Pascale}, F.~{Piacentini}, D.~{Pogosyan},
  S.~{Prunet}, S.~{Rao}, G.~{Romeo}, J.~E. {Ruhl}, F.~{Scaramuzzi},
  D.~{Sforna}, and N.~{Vittorio}.
\newblock {A flat Universe from high-resolution maps of the cosmic microwave
  background radiation}.
\newblock {\em Nature}, 404:955--959, April 2000, astro-ph/0004404.

\bibitem{spergel07}
D.~N. {Spergel}, R.~{Bean}, O.~{Dor{\'e}}, M.~R. {Nolta}, C.~L. {Bennett},
  J.~{Dunkley}, G.~{Hinshaw}, N.~{Jarosik}, E.~{Komatsu}, L.~{Page}, H.~V.
  {Peiris}, L.~{Verde}, M.~{Halpern}, R.~S. {Hill}, A.~{Kogut}, M.~{Limon},
  S.~S. {Meyer}, N.~{Odegard}, G.~S. {Tucker}, J.~L. {Weiland}, E.~{Wollack},
  and E.~L. {Wright}.
\newblock {Three-Year Wilkinson Microwave Anisotropy Probe (WMAP) Observations:
  Implications for Cosmology}.
\newblock {\em ApJS}, 170:377--408, June 2007, astro-ph/0603449.

\bibitem{komat11}
E.~{Komatsu}, K.~M. {Smith}, J.~{Dunkley}, C.~L. {Bennett}, B.~{Gold},
  G.~{Hinshaw}, N.~{Jarosik}, D.~{Larson}, M.~R. {Nolta}, L.~{Page}, D.~N.
  {Spergel}, M.~{Halpern}, R.~S. {Hill}, A.~{Kogut}, M.~{Limon}, S.~S. {Meyer},
  N.~{Odegard}, G.~S. {Tucker}, J.~L. {Weiland}, E.~{Wollack}, and E.~L.
  {Wright}.
\newblock {Seven-year Wilkinson Microwave Anisotropy Probe (WMAP) Observations:
  Cosmological Interpretation}.
\newblock {\em ApJS}, 192:18, February 2011, 1001.4538.

\bibitem{planck16}
{Planck Collaboration}, P.~A.~R. {Ade}, N.~{Aghanim}, M.~{Arnaud},
  M.~{Ashdown}, J.~{Aumont}, C.~{Baccigalupi}, A.~J. {Banday}, R.~B.
  {Barreiro}, J.~G. {Bartlett}, and et~al.
\newblock {Planck 2015 results. XIII. Cosmological parameters}.
\newblock {\em A\&A}, 594:A13, September 2016, 1502.01589.

\bibitem{davis85}
M.~{Davis}, G.~{Efstathiou}, C.~S. {Frenk}, and S.~D.~M. {White}.
\newblock {The evolution of large-scale structure in a universe dominated by
  cold dark matter}.
\newblock {\em ApJ}, 292:371--394, May 1985.

\bibitem{whfr91}
S.~D.~M. {White} and C.~S. {Frenk}.
\newblock {Galaxy formation through hierarchical clustering}.
\newblock {\em ApJ}, 379:52--79, September 1991.

\bibitem{cole05}
S.~{Cole}, W.~J. {Percival}, J.~A. {Peacock}, P.~{Norberg}, C.~M. {Baugh},
  C.~S. {Frenk}, I.~{Baldry}, J.~{Bland-Hawthorn}, T.~{Bridges}, R.~{Cannon},
  M.~{Colless}, C.~{Collins}, W.~{Couch}, N.~J.~G. {Cross}, G.~{Dalton}, V.~R.
  {Eke}, R.~{De Propris}, S.~P. {Driver}, G.~{Efstathiou}, R.~S. {Ellis},
  K.~{Glazebrook}, C.~{Jackson}, A.~{Jenkins}, O.~{Lahav}, I.~{Lewis},
  S.~{Lumsden}, S.~{Maddox}, D.~{Madgwick}, B.~A. {Peterson}, W.~{Sutherland},
  and K.~{Taylor}.
\newblock {The 2dF Galaxy Redshift Survey: power-spectrum analysis of the final
  data set and cosmological implications}.
\newblock {\em MNRAS }, 362:505--534, September 2005, astro-ph/0501174.

\bibitem{eisen05}
D.~J. {Eisenstein}, I.~{Zehavi}, D.~W. {Hogg}, R.~{Scoccimarro}, M.~R.
  {Blanton}, R.~C. {Nichol}, R.~{Scranton}, H.-J. {Seo}, M.~{Tegmark},
  Z.~{Zheng}, S.~F. {Anderson}, J.~{Annis}, N.~{Bahcall}, J.~{Brinkmann},
  S.~{Burles}, F.~J. {Castander}, A.~{Connolly}, I.~{Csabai}, M.~{Doi},
  M.~{Fukugita}, J.~A. {Frieman}, K.~{Glazebrook}, J.~E. {Gunn}, J.~S.
  {Hendry}, G.~{Hennessy}, Z.~{Ivezi{\'c}}, S.~{Kent}, G.~R. {Knapp}, H.~{Lin},
  Y.-S. {Loh}, R.~H. {Lupton}, B.~{Margon}, T.~A. {McKay}, A.~{Meiksin}, J.~A.
  {Munn}, A.~{Pope}, M.~W. {Richmond}, D.~{Schlegel}, D.~P. {Schneider},
  K.~{Shimasaku}, C.~{Stoughton}, M.~A. {Strauss}, M.~{SubbaRao}, A.~S.
  {Szalay}, I.~{Szapudi}, D.~L. {Tucker}, B.~{Yanny}, and D.~G. {York}.
\newblock {Detection of the Baryon Acoustic Peak in the Large-Scale Correlation
  Function of SDSS Luminous Red Galaxies}.
\newblock {\em ApJ}, 633:560--574, November 2005, astro-ph/0501171.

\bibitem{sprin05}
V.~{Springel}, S.~D.~M. {White}, A.~{Jenkins}, C.~S. {Frenk}, N.~{Yoshida},
  L.~{Gao}, J.~{Navarro}, R.~{Thacker}, D.~{Croton}, J.~{Helly}, J.~A.
  {Peacock}, S.~{Cole}, P.~{Thomas}, H.~{Couchman}, A.~{Evrard}, J.~{Colberg},
  and F.~{Pearce}.
\newblock {Simulations of the formation, evolution and clustering of galaxies
  and quasars}.
\newblock {\em Nature}, 435:629--636, June 2005, astro-ph/0504097.

\bibitem{clowe06}
D.~{Clowe}, M.~{Brada{\v c}}, A.~H. {Gonzalez}, M.~{Markevitch}, S.~W.
  {Randall}, C.~{Jones}, and D.~{Zaritsky}.
\newblock {A Direct Empirical Proof of the Existence of Dark Matter}.
\newblock {\em ApJL}, 648:L109--L113, September 2006, astro-ph/0608407.

\bibitem{angus06}
G.~W. {Angus}, B.~{Famaey}, and H.~S. {Zhao}.
\newblock {Can MOND take a bullet? Analytical comparisons of three versions of
  MOND beyond spherical symmetry}.
\newblock {\em MNRAS }, 371:138--146, September 2006, astro-ph/0606216.

\bibitem{lililin13}
X.~{Li}, M.-H. {Li}, H.-N. {Lin}, and Z.~{Chang}.
\newblock {Finslerian MOND versus observations of Bullet Cluster 1E 0657-558}.
\newblock {\em MNRAS }, 428:2939--2948, February 2013, 1209.3086.

\bibitem{Peebles:1982ff}
P.~J.~E. Peebles.
\newblock {Large scale background temperature and mass fluctuations due to
  scale invariant primeval perturbations}.
\newblock {\em Astrophys. J.}, 263:L1--L5, 1982.

\bibitem{Bond:1982uy}
J.~R. Bond, A.~S. Szalay, and Michael~S. Turner.
\newblock {Formation of Galaxies in a Gravitino Dominated Universe}.
\newblock {\em Phys. Rev. Lett.}, 48:1636, 1982.

\bibitem{Blumenthal:1982mv}
George~R. Blumenthal, Heinz Pagels, and Joel~R. Primack.
\newblock {Galaxy Formation by Dissipationless Particles Heavier Than
  Neutrinos}.
\newblock {\em Nature}, 299:37--38, 1982.

\bibitem{Blumenthal:1984bp}
George~R. Blumenthal, S.~M. Faber, Joel~R. Primack, and Martin~J. Rees.
\newblock {Formation of Galaxies and Large Scale Structure with Cold Dark
  Matter}.
\newblock {\em Nature}, 311:517--525, 1984.

\bibitem{Jungman:1995df}
Gerard Jungman, Marc Kamionkowski, and Kim Griest.
\newblock {Supersymmetric dark matter}.
\newblock {\em Phys. Rept.}, 267:195--373, 1996, hep-ph/9506380.

\bibitem{Bode:2000gq}
Paul Bode, Jeremiah~P. Ostriker, and Neil Turok.
\newblock {Halo formation in warm dark matter models}.
\newblock {\em Astrophys. J.}, 556:93--107, 2001, astro-ph/0010389.

\bibitem{Asztalos:2009yp}
S.~J. Asztalos et~al.
\newblock {A SQUID-based microwave cavity search for dark-matter axions}.
\newblock {\em Phys. Rev. Lett.}, 104:041301, 2010, 0910.5914.

\bibitem{Strassler:2006im}
Matthew~J. Strassler and Kathryn~M. Zurek.
\newblock {Echoes of a hidden valley at hadron colliders}.
\newblock {\em Phys. Lett.}, B651:374--379, 2007, hep-ph/0604261.

\bibitem{Goodman:1984dc}
Mark~W. Goodman and Edward Witten.
\newblock {Detectability of Certain Dark Matter Candidates}.
\newblock {\em Phys. Rev.}, D31:3059, 1985.

\bibitem{Klasen:2015uma}
Michael Klasen, Martin Pohl, and Gunter Sigl.
\newblock {Indirect and direct search for dark matter}.
\newblock {\em Prog. Part. Nucl. Phys.}, 85:1--32, 2015, 1507.03800.

\bibitem{Moore:1998zn}
Ben Moore, George Lake, Thomas~R. Quinn, and Joachim Stadel.
\newblock {On the survival and destruction of spiral galaxies in clusters}.
\newblock {\em Mon. Not. Roy. Astron. Soc.}, 304:465--474, 1999,
  astro-ph/9811127.

\bibitem{Moore:1999jv}
Ben Moore, George Lake, Joachim Stadel, and Thomas~R. Quinn.
\newblock {The fate of lsb galaxies in clusters and the origin of the diffuse
  intra-cluster light}.
\newblock {\em ASP Conf. Ser.}, 170:229, 1999, astro-ph/9903064.

\bibitem{Klypin:1999uc}
Anatoly~A. Klypin, Andrey~V. Kravtsov, Octavio Valenzuela, and Francisco Prada.
\newblock {Where are the missing Galactic satellites?}
\newblock {\em Astrophys. J.}, 522:82--92, 1999, astro-ph/9901240.

\bibitem{Tollerud:2008ze}
Erik~J. Tollerud, James~S. Bullock, Louis~E. Strigari, and Beth Willman.
\newblock {Hundreds of Milky Way Satellites? Luminosity Bias in the Satellite
  Luminosity Function}.
\newblock {\em Astrophys. J.}, 688:277--289, 2008, 0806.4381.

\bibitem{Springel:2008cc}
Volker Springel, Jie Wang, Mark Vogelsberger, Aaron Ludlow, Adrian Jenkins,
  Amina Helmi, Julio~F. Navarro, Carlos~S. Frenk, and Simon D.~M. White.
\newblock {The Aquarius Project: the subhalos of galactic halos}.
\newblock {\em Mon. Not. Roy. Astron. Soc.}, 391:1685--1711, 2008, 0809.0898.

\bibitem{Kravtsov:2004cm}
Andrey~V. Kravtsov, Oleg~Y. Gnedin, and Anatoly~A. Klypin.
\newblock {The Tumultuous lives of Galactic dwarfs and the missing satellites
  problem}.
\newblock {\em Astrophys. J.}, 609:482--497, 2004, astro-ph/0401088.

\bibitem{Geha:2008zr}
Marla Geha, Beth Willman, Josh~D. Simon, Louis~E. Strigari, Evan~N. Kirby,
  David~R. Law, and Jay Strader.
\newblock {The Least Luminous Galaxy: Spectroscopy of the Milky Way Satellite
  Segue 1}.
\newblock {\em Astrophys. J.}, 692:1464--1475, 2009, 0809.2781.

\bibitem{Crain:2006sb}
Robert~A. Crain, Vincent~R. Eke, Carlos~S. Frenk, Adrian Jenkins, Ian~G.
  McCarthy, Julio~F. Navarro, and Frazer~R. Pearce.
\newblock {The baryon fraction of Lambda-CDM haloes}.
\newblock {\em Mon. Not. Roy. Astron. Soc.}, 377:41--49, 2007,
  astro-ph/0610602.

\bibitem{Tassis:2006zt}
Konstantinos Tassis, Andrey~V. Kravtsov, and Nickolay~Y. Gnedin.
\newblock {Scaling Relations of Dwarf Galaxies without Supernova-Driven Winds}.
\newblock {\em Astrophys. J.}, 672:888--903, 2008, astro-ph/0609763.

\bibitem{wadepuh11}
M.~{Wadepuhl} and V.~{Springel}.
\newblock {Satellite galaxies in hydrodynamical simulations of Milky Way sized
  galaxies}.
\newblock {\em MNRAS }, 410:1975--1992, January 2011, 1004.3217.

\bibitem{brooks13}
A.~M. {Brooks}, M.~{Kuhlen}, A.~{Zolotov}, and D.~{Hooper}.
\newblock {A Baryonic Solution to the Missing Satellites Problem}.
\newblock {\em ApJ}, 765:22, March 2013, 1209.5394.

\bibitem{Maccio:2009aek}
Andrea~V. Maccio', Xi~Kang, Fabio Fontanot, Rachel~S. Somerville, Sergey~E.
  Koposov, and Pierluigi Monaco.
\newblock {Luminosity function and radial distribution of Milky Way satellites
  in a $\Lambda$CDM Universe}.
\newblock {\em Mon. Not. Roy. Astron. Soc.}, 402:1995, 2010, 0903.4681.

\bibitem{Wetzel:2016wro}
Andrew~R. Wetzel, Philip~F. Hopkins, Ji-hoon Kim, Claude-Andre Faucher-Giguere,
  Dusan Keres, and Eliot Quataert.
\newblock {Reconciling dwarf galaxies with $\Lambda$CDM cosmology: Simulating a
  realistic population of satellites around a Milky Way-mass galaxy}.
\newblock {\em Astrophys. J.}, 827(2):L23, 2016, 1602.05957.

\bibitem{Dubinski:1991bm}
John Dubinski and R.~G. Carlberg.
\newblock {The Structure of cold dark matter halos}.
\newblock {\em Astrophys. J.}, 378:496, 1991.

\bibitem{Governato:2009bg}
Fabio Governato et~al.
\newblock {At the heart of the matter: the origin of bulgeless dwarf galaxies
  and Dark Matter cores}.
\newblock {\em Nature}, 463:203--206, 2010, 0911.2237.

\bibitem{Walker:2011zu}
Matthew~G. Walker and Jorge Penarrubia.
\newblock {A Method for Measuring (Slopes of) the Mass Profiles of Dwarf
  Spheroidal Galaxies}.
\newblock {\em Astrophys. J.}, 742:20, 2011, 1108.2404.

\bibitem{Oh:2015xoa}
Se-Heon Oh et~al.
\newblock {High-resolution mass models of dwarf galaxies from LITTLE THINGS}.
\newblock {\em Astron. J.}, 149:180, 2015, 1502.01281.

\bibitem{Navarro:2016bfs}
Julio~F. Navarro, Alejandro Ben�tez-Llambay, Azadeh Fattahi, Carlos~S. Frenk,
  Aaron~D. Ludlow, Kyle~A. Oman, Matthieu Schaller, and Tom Theuns.
\newblock {The origin of the mass discrepancy-acceleration relation in
  $\Lambda$CDM}.
\newblock 2016, 1612.06329.

\bibitem{Kroupa:2010hf}
P.~Kroupa, B.~Famaey, K.~S. de~Boer, J.~Dabringhausen, M.~S. Pawlowski, C.~M.
  Boily, H.~Jerjen, D.~Forbes, G.~Hensler, and M.~Metz.
\newblock {Local-Group tests of dark-matter Concordance Cosmology: Towards a
  new paradigm for structure formation?}
\newblock {\em Astron. Astrophys.}, 523:A32, 2010, 1006.1647.

\bibitem{BoylanKolchin:2011de}
Michael Boylan-Kolchin, James~S. Bullock, and Manoj Kaplinghat.
\newblock {Too big to fail? The puzzling darkness of massive Milky Way
  subhaloes}.
\newblock {\em Mon. Not. Roy. Astron. Soc.}, 415:L40, 2011, 1103.0007.

\bibitem{BoylanKolchin:2011dk}
Michael Boylan-Kolchin, James~S. Bullock, and Manoj Kaplinghat.
\newblock {The Milky Way's bright satellites as an apparent failure of LCDM}.
\newblock {\em Mon. Not. Roy. Astron. Soc.}, 422:1203--1218, 2012, 1111.2048.

\bibitem{Tollerud:2014zha}
Erik~J. Tollerud, Michael Boylan-Kolchin, and James~S. Bullock.
\newblock {M31 Satellite Masses Compared to LCDM Subhaloes}.
\newblock {\em Mon. Not. Roy. Astron. Soc.}, 440(4):3511--3519, 2014,
  1403.6469.

\bibitem{Papastergis:2014aba}
Emmanouil Papastergis, Riccardo Giovanelli, Martha~P. Haynes, and Francesco
  Shankar.
\newblock {Is there a �too big to fail� problem in the field?}
\newblock {\em Astron. Astrophys.}, 574:A113, 2015, 1407.4665.

\bibitem{Kroupa:2004pt}
Pavel Kroupa, Christian Theis, and Christian~M. Boily.
\newblock {The Great disk of Milky Way satellites and cosmological
  sub-structures}.
\newblock {\em Astron. Astrophys.}, 431:517--521, 2005, astro-ph/0410421.

\bibitem{Metz:2006zc}
Manuel Metz, P.~Kroupa, and H.~Jerjen.
\newblock {The spatial distribution of the Milky Way and Andromeda satellite
  galaxies}.
\newblock {\em Mon. Not. Roy. Astron. Soc.}, 374:1125--1145, 2007,
  astro-ph/0610933.

\bibitem{Ibata:2014pja}
Rodrigo~A. Ibata, Neil~G. Ibata, Geraint~F. Lewis, Nicolas~F. Martin, Anthony
  Conn, Pascal Elahi, Veronica Arias, and Nuwanthika Fernando.
\newblock {A thousand shadows of Andromeda: rotating planes of satellites in
  the Millennium-II cosmological simulation}.
\newblock {\em Astrophys. J.}, 784:L6, 2014, 1403.2389.

\bibitem{Gillet:2015}
Nicolas Gillet, Pierre Ocvirk, Dominique Aubert, Alexander Knebe, Noam
  Libeskind, Gustavo Yepes, Stefan Gottl{\"o}ber, and Yehuda Hoffman.
\newblock {Vast planes of satellites in a high resolution simulation of the
  Local Group: comparison to Andromeda}.
\newblock {\em Astrophys. J.}, 800:1, 2015, 1412.3110.

\bibitem{Kroupa:2012qj}
Pavel Kroupa.
\newblock {The dark matter crisis: falsification of the current standard model
  of cosmology}.
\newblock {\em Publ. Astron. Soc. Austral.}, 29:395--433, 2012, 1204.2546.

\bibitem{Famaey:2013ty}
Benoit Famaey and Stacy McGaugh.
\newblock {Challenges for Lambda-CDM and MOND}.
\newblock {\em J. Phys. Conf. Ser.}, 437:012001, 2013, 1301.0623.

\bibitem{McGaugh:2014nsa}
Stacy~S. McGaugh.
\newblock {A tale of two paradigms: the mutual incommensurability of
  $\Lambda$CDM and MOND}.
\newblock {\em Can. J. Phys.}, 93(2):250--259, 2015, 1404.7525.

\bibitem{Weinberg:2013aya}
David~H. Weinberg, James~S. Bullock, Fabio Governato, Rachel Kuzio~de Naray,
  and Annika H.~G. Peter.
\newblock {Cold dark matter: controversies on small scales}.
\newblock {\em Proc. Nat. Acad. Sci.}, 112:12249--12255, 2014, 1306.0913.

\bibitem{Walker:2014isa}
Matthew~G. Walker and Abraham Loeb.
\newblock {Is the universe simpler than ?CDM?}
\newblock {\em Contemp. Phys.}, 55(3):198--211, 2014, 1401.1146.

\bibitem{Pawlowski:2015qta}
Marcel~S. Pawlowski, Benoit Famaey, David Merritt, and Pavel Kroupa.
\newblock {On the persistence of two small-scale problems in $\Lambda$CDM}.
\newblock {\em Astrophys. J.}, 815(1):19, 2015, 1510.08060.

\bibitem{Kroupa:2014ria}
Pavel Kroupa.
\newblock {Galaxies as simple dynamical systems: observational data disfavor
  dark matter and stochastic star formation}.
\newblock {\em Can. J. Phys.}, 93(2):169--202, 2015, 1406.4860.

\bibitem{Schaye:2014tpa}
Joop Schaye et~al.
\newblock {The EAGLE project: Simulating the evolution and assembly of galaxies
  and their environments}.
\newblock {\em Mon. Not. Roy. Astron. Soc.}, 446:521--554, 2015, 1407.7040.

\bibitem{Koposov:2009ru}
Sergey~E. Koposov, Jaiyul Yoo, Hans-Walter Rix, David~H. Weinberg, Andrea~V.
  Maccio, and Jordi Miralda-Escude.
\newblock {A quantitative explanation of the observed population of Milky Way
  satellite galaxies}.
\newblock {\em Astrophys. J.}, 696:2179--2194, 2009, 0901.2116.

\bibitem{Pontzen:2011ty}
Andrew Pontzen and Fabio Governato.
\newblock {How supernova feedback turns dark matter cusps into cores}.
\newblock {\em Mon. Not. Roy. Astron. Soc.}, 421:3464, 2012, 1106.0499.

\bibitem{Governato:2012fa}
F.~Governato, A.~Zolotov, A.~Pontzen, C.~Christensen, S.~H. Oh, A.~M. Brooks,
  T.~Quinn, S.~Shen, and J.~Wadsley.
\newblock {Cuspy No More: How Outflows Affect the Central Dark Matter and
  Baryon Distribution in Lambda CDM Galaxies}.
\newblock {\em Mon. Not. Roy. Astron. Soc.}, 422:1231--1240, 2012, 1202.0554.

\bibitem{Ogiya:2012jq}
Go~Ogiya and Masao Mori.
\newblock {The core-cusp problem in cold dark matter halos and supernova
  feedback: Effects of Oscillation}.
\newblock {\em Astrophys. J.}, 793:46, 2014, 1206.5412.

\bibitem{Garrison-Kimmel:2017zes}
Shea Garrison-Kimmel et~al.
\newblock {Not so lumpy after all: modeling the depletion of dark matter
  subhalos by Milky Way-like galaxies}.
\newblock 2017, 1701.03792.

\bibitem{Elbert:2014bma}
Oliver~D. Elbert, James~S. Bullock, Shea Garrison-Kimmel, Miguel Rocha, Jose
  O�orbe, and Annika H.~G. Peter.
\newblock {Core formation in dwarf haloes with self-interacting dark matter: no
  fine-tuning necessary}.
\newblock {\em Mon. Not. Roy. Astron. Soc.}, 453(1):29--37, 2015, 1412.1477.

\bibitem{Milgrom:2007br}
Mordehai Milgrom and Robert~H. Sanders.
\newblock {Rings and shells of dark matter as MOND artifacts}.
\newblock {\em Astrophys. J.}, 678:131--143, 2008, 0709.2561.

\bibitem{McGaugh:2016leg}
Stacy McGaugh, Federico Lelli, and Jim Schombert.
\newblock {Radial Acceleration Relation in Rotationally Supported Galaxies}.
\newblock {\em Phys. Rev. Lett.}, 117(20):201101, 2016, 1609.05917.

\bibitem{Lelli:2017vgz}
Federico Lelli, Stacy~S. McGaugh, James~M. Schombert, and Marcel~S. Pawlowski.
\newblock {One Law to Rule Them All: The Radial Acceleration Relation of
  Galaxies}.
\newblock {\em Astrophys. J.}, 836(2):152, 2017.

\bibitem{Ade:2015xua}
P.~A.~R. Ade et~al.
\newblock {Planck 2015 results. XIII. Cosmological parameters}.
\newblock {\em Astron. Astrophys.}, 594:A13, 2016, 1502.01589.

\bibitem{Milgrom:1998sy}
Mordehai Milgrom.
\newblock {The modified dynamics as a vacuum effect}.
\newblock {\em Phys. Lett.}, A253:273--279, 1999, astro-ph/9805346.

\bibitem{Ludlow:2016qzh}
Aaron~D. Ludlow et~al.
\newblock {The Mass-Discrepancy Acceleration Relation: a Natural Outcome of
  Galaxy Formation in Cold Dark Matter halos}.
\newblock {\em Phys. Rev. Lett.}, 118(16):161103, 2017, 1610.07663.

\bibitem{Tully:1977fu}
R.~B. Tully and J.~R. Fisher.
\newblock {A New method of determining distances to galaxies}.
\newblock {\em Astron. Astrophys.}, 54:661--673, 1977.

\bibitem{Milgrom:1983ca}
Mordehai Milgrom.
\newblock {A Modification of the Newtonian dynamics as a possible alternative
  to the hidden mass hypothesis}.
\newblock {\em Astrophys. J.}, 270:365--370, 1983.

\bibitem{Milgrom:1983pn}
Mordehai Milgrom.
\newblock {A Modification of the Newtonian dynamics: Implications for
  galaxies}.
\newblock {\em Astrophys. J.}, 270:371--383, 1983.

\bibitem{Milgrom:1983zz}
Mordehai Milgrom.
\newblock {A modification of the Newtonian dynamics: implications for galaxy
  systems}.
\newblock {\em Astrophys. J.}, 270:384--389, 1983.

\bibitem{Persic:1991}
Massimo Persic and Paolo Salucci.
\newblock {The universal galaxy rotation curve}.
\newblock {\em Astrophys. J.}, 368:60--65, 1991.

\bibitem{Famaey:2011kh}
Benoit Famaey and Stacy McGaugh.
\newblock {Modified Newtonian Dynamics (MOND): Observational Phenomenology and
  Relativistic Extensions}.
\newblock {\em Living Rev. Rel.}, 15:10, 2012, 1112.3960.

\bibitem{Sanders:1996ua}
R.~H. Sanders.
\newblock {The published extended rotation curves of spiral galaxies:
  confrontation with modified dynamics}.
\newblock {\em Astrophys. J.}, 473:117, 1996, astro-ph/9606089.

\bibitem{Sanders:1998gr}
R.~H. Sanders and M.~A.~W. Verheijen.
\newblock {Rotation curves of uma galaxies in the context of modified newtonian
  dynamics}.
\newblock {\em Astrophys. J.}, 503:97, 1998, astro-ph/9802240.

\bibitem{Kaplinghat:2001me}
Manoj Kaplinghat and Michael~S. Turner.
\newblock {How cold dark matter theory explains Milgrom's law}.
\newblock {\em Astrophys. J.}, 569:L19, 2002, astro-ph/0107284.

\bibitem{Smolin:2017kkb}
Lee Smolin.
\newblock {MOND as a regime of quantum gravity}.
\newblock 2017, 1704.00780.

\bibitem{Jacobson:1995ab}
Ted Jacobson.
\newblock {Thermodynamics of space-time: The Einstein equation of state}.
\newblock {\em Phys. Rev. Lett.}, 75:1260--1263, 1995, gr-qc/9504004.

\bibitem{Verlinde:2010hp}
Erik~P. Verlinde.
\newblock {On the Origin of Gravity and the Laws of Newton}.
\newblock {\em JHEP}, 04:029, 2011, 1001.0785.

\bibitem{Verlinde:2016toy}
Erik~P. Verlinde.
\newblock {Emergent Gravity and the Dark Universe}.
\newblock 2016, 1611.02269.

\bibitem{Ho:2010ca}
Chiu~Man Ho, Djordje Minic, and Y.~Jack Ng.
\newblock {Cold Dark Matter with MOND Scaling}.
\newblock {\em Phys. Lett.}, B693:567--570, 2010, 1005.3537.

\bibitem{Ho:2011xc}
Chiu~Man Ho, Djordje Minic, and Y.~Jack Ng.
\newblock {Quantum Gravity and Dark Matter}.
\newblock {\em Gen. Rel. Grav.}, 43:2567--2573, 2011, 1105.2916.
\newblock [Int. J. Mod. Phys.D20,2887(2011)].

\bibitem{Ho:2012ar}
Chiu~Man Ho, Djordje Minic, and Y.~Jack Ng.
\newblock {Dark Matter, Infinite Statistics and Quantum Gravity}.
\newblock {\em Phys. Rev.}, D85:104033, 2012, 1201.2365.

\bibitem{Edmonds:2013hba}
Doug Edmonds, Duncan Farrah, Chiu~Man Ho, Djordje Minic, Y.~Jack Ng, and Tatsu
  Takeuchi.
\newblock {Testing MONDian Dark Matter with Galactic Rotation Curves}.
\newblock {\em Astrophys. J.}, 793:41, 2014, 1308.3252.

\bibitem{Edmonds:2016tio}
Doug Edmonds, Duncan Farrah, Chi~Man Ho, Djordje Minic, Y.~Jack Ng, and Tatsu
  Takeuchi.
\newblock {Testing Modified Dark Matter with Galaxy Clusters: Does Dark Matter
  know about the Cosmological Constant?}
\newblock {\em International Journal of Modern Physics A}, 32, 2017, 1750108.

\bibitem{Davies:1974th}
P.~C.~W. Davies.
\newblock {Scalar particle production in Schwarzschild and Rindler metrics}.
\newblock {\em J. Phys.}, A8:609--616, 1975.

\bibitem{Unruh:1976db}
W.~G. Unruh.
\newblock {Notes on black hole evaporation}.
\newblock {\em Phys. Rev.}, D14:870, 1976.

\bibitem{tHooft:1993dmi}
Gerard 't~Hooft.
\newblock {Dimensional reduction in quantum gravity}.
\newblock In {\em {Salamfest 1993:0284-296}}, pages 0284--296, 1993,
  gr-qc/9310026.

\bibitem{Susskind:1994vu}
Leonard Susskind.
\newblock {The World as a hologram}.
\newblock {\em J. Math. Phys.}, 36:6377--6396, 1995, hep-th/9409089.

\bibitem{Deser:1997ri}
Stanley Deser and Orit Levin.
\newblock {Accelerated detectors and temperature in (anti)-de Sitter spaces}.
\newblock {\em Class. Quant. Grav.}, 14:L163--L168, 1997, gr-qc/9706018.

\bibitem{Jacobson:1997ux}
Ted Jacobson.
\newblock {Comment on `Accelerated detectors and temperature in anti-de Sitter
  spaces'}.
\newblock {\em Class. Quant. Grav.}, 15:251--253, 1998, gr-qc/9709048.

\bibitem{Bekenstein:1973ur}
Jacob~D. Bekenstein.
\newblock {Black holes and entropy}.
\newblock {\em Phys. Rev.}, D7:2333--2346, 1973.

\bibitem{Hawking:1974sw}
S.~W. Hawking.
\newblock {Particle Creation by Black Holes}.
\newblock {\em Commun. Math. Phys.}, 43:199--220, 1975.
\newblock [,167(1975)].

\bibitem{Khoury:2014tka}
Justin Khoury.
\newblock {Alternative to particle dark matter}.
\newblock {\em Phys. Rev.}, D91(2):024022, 2015, 1409.0012.

\bibitem{Berezhiani:2015bqa}
Lasha Berezhiani and Justin Khoury.
\newblock {Theory of dark matter superfluidity}.
\newblock {\em Phys. Rev.}, D92:103510, 2015, 1507.01019.

\bibitem{Berezhiani:2015pia}
Lasha Berezhiani and Justin Khoury.
\newblock {Dark Matter Superfluidity and Galactic Dynamics}.
\newblock {\em Phys. Lett.}, B753:639--643, 2016, 1506.07877.

\bibitem{Navarro:1995iw}
Julio~F. Navarro, Carlos~S. Frenk, and Simon D.~M. White.
\newblock {The Structure of cold dark matter halos}.
\newblock {\em Astrophys. J.}, 462:563--575, 1996, astro-ph/9508025.

\bibitem{Navarro:1996gj}
Julio~F. Navarro, Carlos~S. Frenk, and Simon D.~M. White.
\newblock {A Universal density profile from hierarchical clustering}.
\newblock {\em Astrophys. J.}, 490:493--508, 1997, astro-ph/9611107.

\bibitem{Tolman:1930zza}
Richard~C. Tolman.
\newblock {On the Weight of Heat and Thermal Equilibrium in General
  Relativity}.
\newblock {\em Phys. Rev.}, 35:904--924, 1930.

\bibitem{Tolman:1930ona}
Richard Tolman and Paul Ehrenfest.
\newblock {Temperature Equilibrium in a Static Gravitational Field}.
\newblock {\em Phys. Rev.}, 36(12):1791--1798, 1930.

\bibitem{Vikhlinin:2005mp}
Alexey Vikhlinin, A.~Kravtsov, W.~Forman, C.~Jones, M.~Markevitch, S.~S.
  Murray, and L.~Van~Speybroeck.
\newblock {Chandra sample of nearby relaxed galaxy clusters: Mass, gas
  fraction, and mass-temperature relation}.
\newblock {\em Astrophys. J.}, 640:691--709, 2006, astro-ph/0507092.

\bibitem{Sanders:1998kv}
R.~H. Sanders.
\newblock {Resolving the virial discrepancy in clusters of galaxies with
  modified newtonian dynamics}.
\newblock {\em Astrophys. J.}, 512:L23, 1999, astro-ph/9807023.

\bibitem{Blanchet:2006yt}
Luc Blanchet.
\newblock {Gravitational polarization and the phenomenology of MOND}.
\newblock {\em Class. Quant. Grav.}, 24:3529--3540, 2007, astro-ph/0605637.

\bibitem{Gibbons:2001gy}
G~W Gibbons.
\newblock {Aspects of Born-Infeld theory and string / M theory}.
\newblock {\em Rev. Mex. Fis.}, 49S1:19--29, 2003, hep-th/0106059.
\newblock [AIP Conf. Proc.589,324(2001)].

\bibitem{Greenberg:1989ty}
O.~W. Greenberg.
\newblock {Example of Infinite Statistics}.
\newblock {\em Phys. Rev. Lett.}, 64:705, 1990.

\bibitem{Freidel:2013zga}
Laurent Freidel, Robert~G. Leigh, and Djordje Minic.
\newblock {Born Reciprocity in String Theory and the Nature of Spacetime}.
\newblock {\em Phys. Lett.}, B730:302--306, 2014, 1307.7080.

\bibitem{Freidel:2014qna}
Laurent Freidel, Robert~G. Leigh, and Djordje Minic.
\newblock {Quantum Gravity, Dynamical Phase Space and String Theory}.
\newblock {\em Int. J. Mod. Phys.}, D23(12):1442006, 2014, 1405.3949.

\bibitem{Freidel:2015pka}
Laurent Freidel, Robert~G. Leigh, and Djordje Minic.
\newblock {Metastring Theory and Modular Space-time}.
\newblock {\em JHEP}, 06:006, 2015, 1502.08005.

\bibitem{Freidel:2015uug}
Laurent Freidel, Robert~G. Leigh, and Djordje Minic.
\newblock {Modular spacetime}.
\newblock {\em Int. J. Mod. Phys.}, D24(12):1544028, 2015.

\bibitem{Freidel:2016pls}
Laurent Freidel, Robert~G. Leigh, and Djordje Minic.
\newblock {Quantum Spaces are Modular}.
\newblock {\em Phys. Rev.}, D94(10):104052, 2016, 1606.01829.

\bibitem{Freidel:2017wst}
Laurent Freidel, Robert~G. Leigh, and Djordje Minic.
\newblock {Intrinsic non-commutativity of closed string theory}.
\newblock 2017, 1706.03305.

\bibitem{Freidel:2017nhg}
Laurent Freidel, Robert~G. Leigh, and Djordje Minic.
\newblock {On the Non-commutativity of Closed String Zero Modes}.
\newblock 2017, 1707.00312.

\bibitem{Grosse:2004yu}
Harald Grosse and Raimar Wulkenhaar.
\newblock {Renormalization of phi**4 theory on noncommutative R**4 in the
  matrix base}.
\newblock {\em Commun. Math. Phys.}, 256:305--374, 2005, hep-th/0401128.

\end{thebibliography}


\end{document}